\documentclass[preprint2]{aastex62}

\shorttitle{Members in Coma Berenices}
\shortauthors{Tang et al.}

\begin{document}

\title{ Characterization of Stellar and Substellar 
Members in the Coma Berenices Star Cluster 
        }
	
\author{Shih-Yun Tang} 
\affiliation{Department of Physics, National Central University, 
300 Zhongda Road, Zhongli, Taoyuan 32001, Taiwan}

\author{W.~P. Chen}
\affiliation{Graduate Institute of Astronomy, National Central University, 
300 Zhongda Road, Zhongli, Taoyuan 32001, Taiwan}
\affiliation{Department of Physics, National Central University, 
300 Zhongda Road, Zhongli, Taoyuan 32001, Taiwan}

\author{P.~S. Chiang}
\affiliation{Graduate Institute of Astronomy, National Central University, 
300 Zhongda Road, Zhongli, Taoyuan 32001, Taiwan}

\author{Jessy Jose}
\affiliation{Kavli Institute for Astronomy and Astrophysics, Peking University,
Yi He Yuan Lu 5, Haidian District, Beijing 100871, China }
\affiliation{Department of Physics, Indian Institute of Science Education and Research, 
Rami Reddy Nagar, Karakambadi Road, Mangalam (P.O.) Tirupati 517507, India }

\author{Gregory J. Herczeg}
\affiliation{Kavli Institute for Astronomy and Astrophysics, Peking University, 
Yi He Yuan Lu 5, Haidian District, Beijing 100871, China }

\author{Bertrand Goldman}
\affiliation{Max Planck Institute for Astronomy, Konigstuhl 17, D-69117 Heidelberg,
Germany}
\affiliation{Universite de Strasbourg, CNRS, Observatoire astronomique de Strasbourg, 
UMR 7550, F-67000 Strasbourg, France}

\accepted{June 4, 2018}
\submitjournal{Astrophysical Journal}

\begin{abstract}
We have identified stellar and substellar members in the nearby star cluster Coma 
Berenices, using photometry, proper motions, and distances of a combination of 
2MASS, UKIDSS, URAT1, and {\it Gaia}/DR2 data.  Those with {\it Gaia}/DR2 
parallax measurements provide the most reliable sample to constrain the distance, 
averaging 86.7~pc with a dispersion 7.1~pc, and the age $\sim800$~Myr, of the cluster.  
This age is older than the 400--600~Myr commonly adopted in the literature. 
Our analysis, complete within $5\degr$ of the cluster radius, leads to identification of 
192 candidates, among which, after field contamination is considered, about 148 are 
true members.  The members have $J\sim3$~mag to $\sim17.5$~mag, 
corresponding to stellar masses 2.3--0.06~$M_\sun$.  The mass function of the cluster 
peaks around 0.3~$M_\sun$ and, in the sense of $dN/dm = m^{-\alpha}$, 
where $N$ is the number of members and $m$ is stellar mass, has a 
slope $\alpha\approx 0.49\pm0.03$ in the mass range 0.3--2.3~$M_\sun$.  This is 
much shallower than that of the field population in the solar neighborhood.  
The slope $\alpha=-1.69\pm0.14$ from 0.3~$M_\sun$ to 0.06~$M_\sun$, 
the lowest mass in our sample.  The cluster is mass segregated and has a shape elongated 
toward the Galactic plane.  Our list contains nine substellar members, including three 
new discoveries of an M8, an L1 and an L4 brown dwarfs, extending from the previously known 
coolest members of late-M types to even cooler types.  
\end{abstract}


\section{Introduction}

Stars are formed in groups out of interstellar molecular clouds.   Those clusters that remain 
gravitational bound, 
i.e., surviving internal dynamics and external disturbances, appear as star clusters.  Stellar aggregates 
provide the early evolutionary environments for a star, in which planets and moons are formed. 
Since members in a star cluster are formed essentially at the same time, and share similar compositions, 
space motions, and spatial locations in space, star clusters have been used extensively for studies 
of stellar evolution, tests and calibrations of stellar atmospheric models, clustered star formation, 
starburst processes, or stellar dynamics \citep{bra08,roc10,gen11}.

An embedded or infrared cluster may not remain gravitationally 
bound as the turbulent parental cloud disperses \citep{lad03}.  Later on through mutual gravitational 
interaction, higher-mass members lose kinetic energy and sink to the center, whereas lower-mass members 
gain speed and occupy a progressively larger volume of space.  Those that exceed the escape velocity 
of the system, notably the least-massive members at the time, would be most vulnerable of being thrown 
out (``stellar evaporation'') to supply as field stars \citep[e.g.,][]{mat84}, 
with an evaporation timescale $\tau_{\rm evap}\approx 100\,(D/v)\,(0.1 N/ \ln N) $, for  
a system of $N$ equal-mass stars, of a size scale $D$ and typical velocity $v$ \citep{shu82,bin87,bha17}. 

It is not clear if there is an ``initial mass function'' for star clusters; namely, if more massive systems 
are favored or less preferred in formation.  It is plausible that most star clusters we witness now 
are remnants of massive systems such as those super star clusters seen near the Galactic center 
\citep[e.g.,][]{bra08}, the Orion Nebular cluster \citep[e.g.,][]{hil97}, and the pristine globular 
clusters.  In addition to internal 
stellar dynamics, Galactic disturbances also act to disintegrate a star cluster, with effects such as the 
tidal disruption from nearby giant molecular clouds or star clusters, passages through spiral arms or disks, 
or shear forces arising from Galactic differential rotation.  While the youngest systems are shaped by 
the parental cloud structure \citep{che04}, tidal distortion is evidenced in many open clusters or even 
in globular clusters \citep{che04,che10,bha17}. Only a recently dissolved star cluster in the solar 
neighborhood may be recognized as a star moving group, if the then-members still share common space 
position and kinematics \citep{zuc04}.  

While low-mass stars are susceptible to ejection, their total mass plays a decisive role in the 
survival of a star cluster \citep{deg07}; a cluster must have a sufficient number of low-mass stars to have  
longevity ($\ga 1$~Gyr) against external stirring \citep{deg09}.  Nearby young systems such as 
Hyades ($\sim47$~pc, 625~Myr), Praesepe ($\sim170$~pc, 757~Myr, \citep{gas09,van09}) and the Coma 
Berenices star cluster ($\sim90$~pc, 600~Myr, \citep{tsv89,van99}) are particularly suitable targets 
to identify the low-mass stellar or even substellar members in the context of cluster disintegration.  

The Coma Berenices star cluster (Melotte\,111, hereafter Coma Ber, R.A.=$12^{\rm h} 25^{\rm m}$, 
Decl.=$26\degr 06^{\prime}$, J2000) was first listed by \citet{mel15}, with \citet{tru38} 
pioneering in characterization of its stellar members.  Despite its proximity, 
the cluster has been relatively poorly studied due to its large sky coverage ($>5\degr$), 
hence the difficulty in distinguishing members against field stars. 
\citet{cas06} combined 2MASS (Two Micron All Sky Survey) and USNO-B1.0 (by United States Naval 
Observatory) data to identify some 100 possible cluster members.  
Using optical and 2MASS photometric data, \citet{mel12} 
identified very low-mass candidates to the limit of $ I < 20.1$~mag in an area of 
22.5~deg$^2$, with no proper motion constraints except removal of high proper-motion stars. 
Five of their candidates have luminosities and colors consistent 
with being brown dwarfs.  \citet{ter14} included SDSS/APOGEE (Sloan Digital Sky Surve, 
Apache Point Observatory Galactic Evolution Experiment) radial velocity data in 
membership determination, and found a few K and early-M members that were previously unknown.  

\citet{kra07} conducted a comparative study between Praesepe and Coma Ber, using 
the 2MASS, SDSS, USNOB1.0 and UCAC-2.0 (USNO CCD Astrograph Catalog) surveys for photometric and 
astrometric member selection.  They found a clear mass segregation in 
Praesepe, i.e., with massive stars being concentrated toward the central region whereas 
lower-mass member occupying progressively large volume in space, but not in 
Coma Ber, which has a similar linear size but was thought to be somewhat younger.  
\citet{wan14} confirmed the mass segregation in Praesepe, and 
concluded that the lowest-mass members they detected, $\sim0.1$~M$_\sun$, are 
being stripped away.  

Here we present a comprehensive characterization of the stellar and substellar member candidates 
of Coma Ber.  We first summarize the archival data on photometry, astrometry, and 
distance used in this work, and then report how membership is determined.  
A set of bright candidates with parallax distances serve as the high-confidence sample to constrain 
the cluster parameters, such as the distance, age, and size, etc., which in turn guide  
the identification of faint stellar and substellar candidates.  We then present the 
infrared spectroscopy that confirm the brown dwarf nature of these members.   With a sample of stellar and 
substellar members, we derive the luminosity function, mass function, shape, and dynamical status 
of the cluster.  For a star with no parallax measurement available, we derive the distance by 
first estimating its spectral type from photometric colors, and then comparing the observed flux to 
the expected luminosity for that spectral type.   We describe the method in Appendix.  

\section{Data and Analysis}\label{sec:data}

In this work, stellar membership is diagnosed by grouping of stars in position in space 
and in kinematics.  For bright stars, we use 2MASS photometry 
and URAT1 (USNO Robotic Astrometric Telescope) proper motions, whereas for faint stars, 
we analyze both the photometry and proper motions from the UKIRT Infrared Deep Sky 
Survey (UKIDSS) Galactic Clusters Survey \citep[GCS,][]{law12}.  Distance information 
comes from parallax measurements by {\it Gaia}/DR2, or estimated by the spectral type.

\subsection{Archival Data for Distance}  

Distance determination is based on parallax measurements whenever available by {\it Gaia}/DR2 \citep{bro18}.
{\it Gaia} is a space mission designed for astrometry by the European Space Agency,   
launched on 19 December 2013.  The latest data release (DR2) including 
the first 22 months of the nominal mission lifetime, contains celestial positions and 
apparent brightness for $\sim1.7$ billion sources, among which 1.3 billion also have 
parallaxes and proper motions available \citep{lin18}.  For our study the empirical limit for the 
{\it Gaia}/DR2 is $J\sim 15$~mag, and only measurements with $ \varpi/ \Delta \varpi > 10$ are 
considered in the analysis, where $\varpi$ is the parallax and $\Delta\varpi$ is the error 
\citep{lin18}. 

For an object with no parallax data, we estimate its spectral type via multi-band photometry, 
from which the distance is derived.  Photometric data in optical 
wavelengths include those of SDSS/DR12 \citep{ala15} and PS1 \citep[Panoramic Survey 
Telescope and Rapid Response System, ][]{cha16}.  In a few cases we utilize the SDSS flags 
to distinguish a star from a galaxy.  For photometry extending to mid-infrared wavelengths, 
``ALLWISE'' \citep{cut13} has been used, which combines the data of {\it WISE} 
\citep[{\it Wide-field Infrared Survey Explorer}, ][]{wri10} in the cryogenic phase of 
the mission, and NEOWISE \citep{mai11} in the first post-cryogenic phase.

\subsection{Archival Data for Proper Motion and Photometry} 

Proper motions are taken from {\it Gaia}/DR2 when available, as long as the measurements are 
reliable, again with $\varpi/ \Delta \varpi > 10$.  Alternatively, proper
motions are extracted from URAT1 \citep{zac15}, which is an astrometric catalog as a follow-up project of UCAC. 
In addition to proper motions, with typical errors 5--8~mas~yr$^{-1}$, URAT1 provides photometry in one 
single ``$f$'' band (between $R$ and $I$).  URAT1 covers almost the entire northern sky and extends 
down to Declination $-15\degr$ in some areas, cataloguing over 228 million objects at a mean epoch 
around May 2013.  A large fraction (83\%) of the URAT1 entries ($3\arcsec$ matching radius) 
list 2MASS $J$, $H$, and $K_s$ magnitudes.  Some 16\% of URAT1 sources are supplemented with five-band 
photometry (BVgri) from the AAVSO Photometric All-Sky Survey (APASS).  

In our analysis, photometry is taken from 2MASS whenever available.  The 2MASS Point Source Catalog \citep{skr06} 
has the 10$\sigma$ detection limits of $J\sim15.8$~mag, $H\sim15.1$~mag, and $K_s\sim14.3$~mag, and \
saturates around $J\sim9$~mag, $H\sim8.5$~mag, and $K_s\sim8$~mag.  

\begin{figure}[t!]
	\centering
	\includegraphics[width=0.85\columnwidth,angle=0]{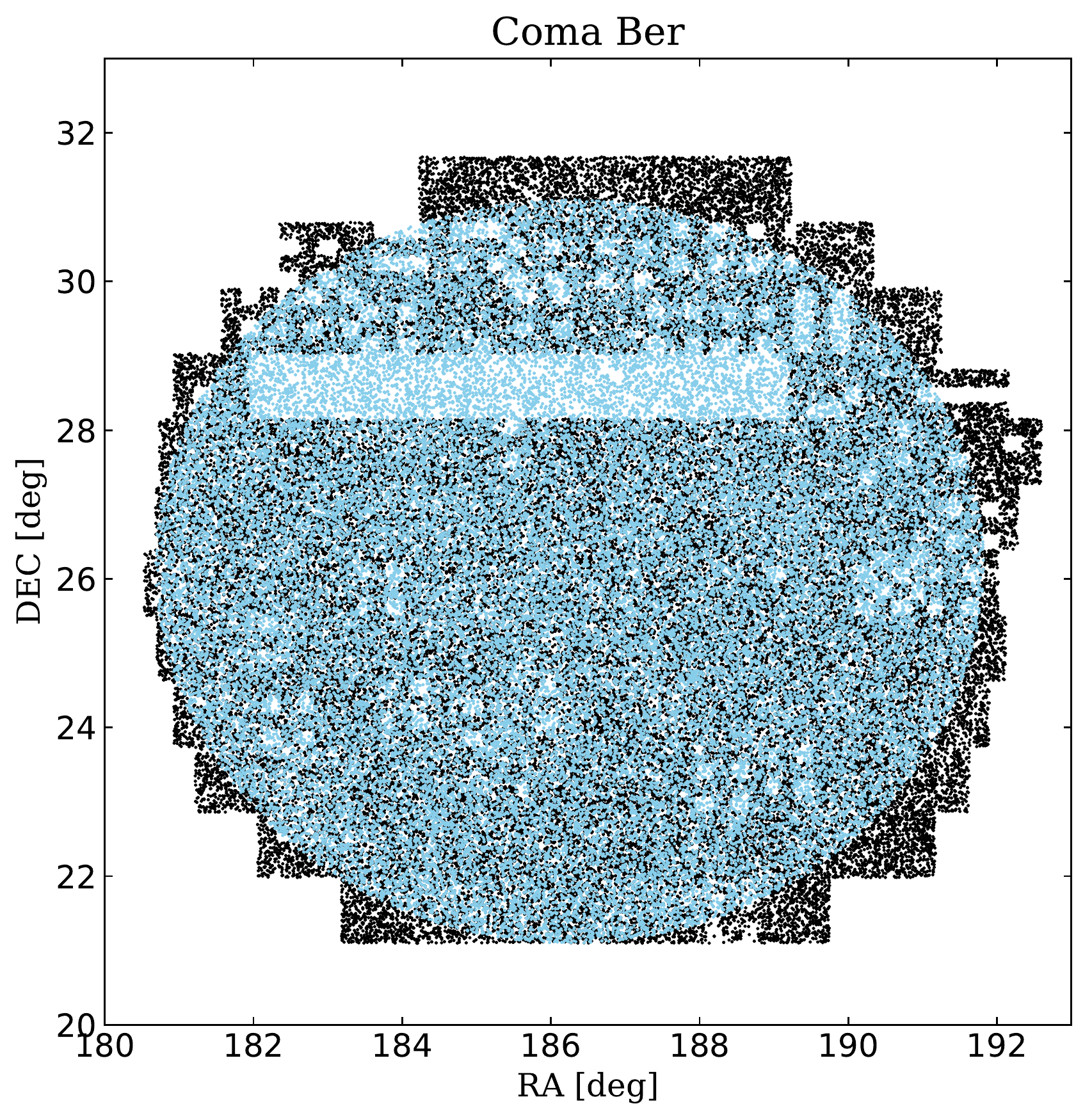}
	\caption{UKIDSS/GCS sources (in black) toward Coma Ber with $ZYJKs$ photometric 
		measurements.  No data are available in the blank regions due to 
		poor image quality flagged by the UKIDSS quality control.
        	 Also shown are the 2MASS sources (in light blue) within the 5$\degr$ radius of the 
		 ``cluster region'' of our study.  
		 For display clarity, only one in two sources, selected randomly, is shown.
			}
  \label{fig:comaCGS}
\end{figure}
\begin{figure}[htb!]
	\centering
  \includegraphics[width=\columnwidth,angle=0]{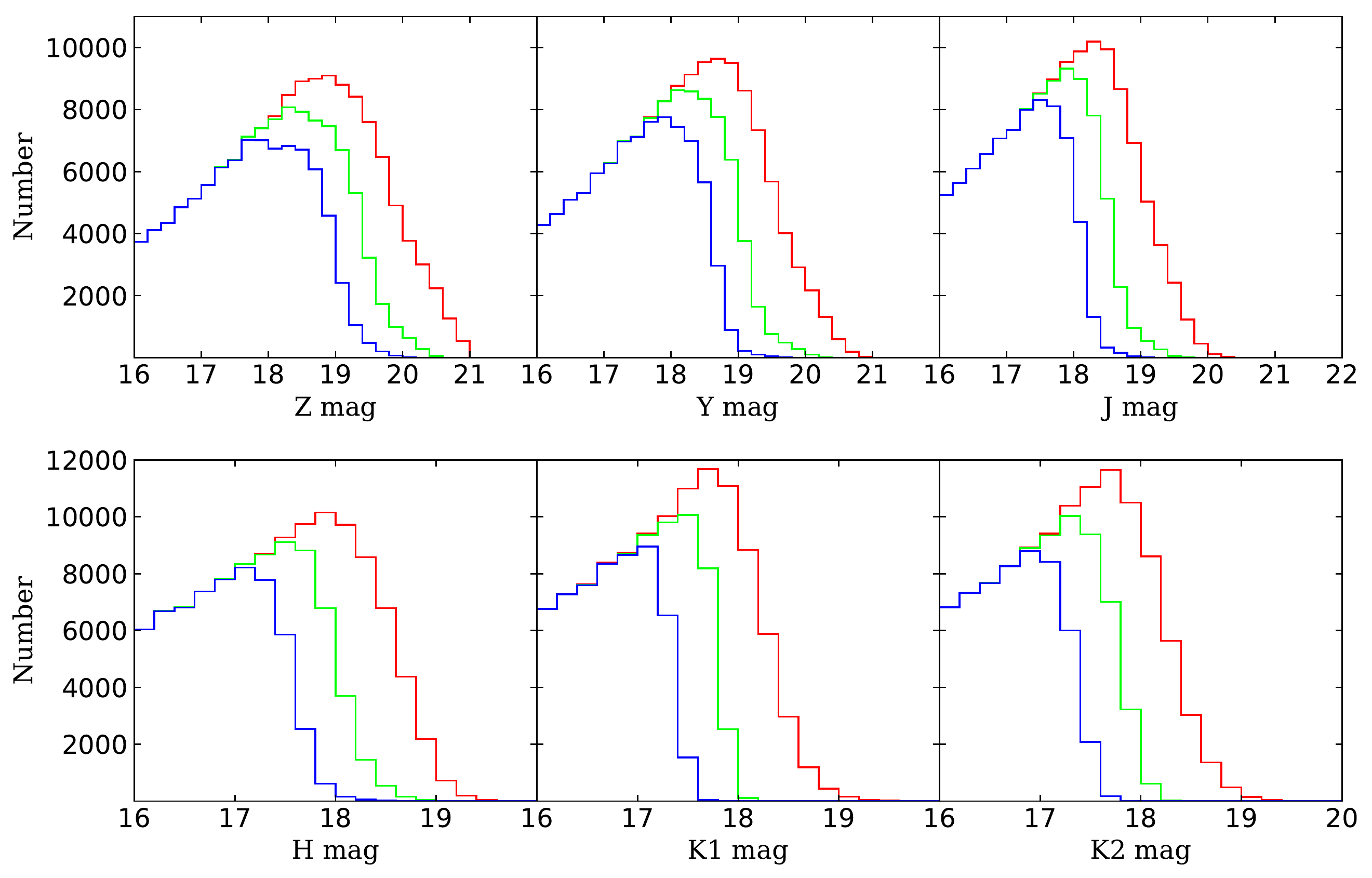}
  \caption{Number of UKIDSS/GCS stars in various bands, for all the stars (the red line) 
	  in the cluster region shown in Fig.~\ref{fig:comaCGS}, and for those with photometric 
	  errors less than 0.15~mag (green) or 0.10~mag (blue).    
       } 
  \label{fig:sense}
\end{figure}

The UKIDSS/GCS aimed to measure the very low-mass end of the stellar mass functions in 10 star clusters.  
As for other UKIDSS surveys, the Large Area Survey (LAS) covered only the edge of Coma Ber, whereas 
the Galactic Plane Survey (GPS) did not include Coma Ber at all. 
Proper motions are available for UKIDSS/GCS starting with DR9 \citep{col12,smi14}.  
For the work reported here, we use the latest data release DR10, but its spatial coverage is 
incomplete within the surveyed sky of 78.5~deg$^{2}$ toward the cluster 
(see Figure~\ref{fig:comaCGS}), missing a sky area about 7~deg$^2$ in the $Z$ and $Y$ bands, 
2.5~deg$^2$ in the $J$ band, and 3~deg$^2$ in the $H$ band, due to poor data quality \citep{bou12}.
The $K$-band observations were taken at 2 epochs to enable proper motion estimates. 
The typical proper motion error of the GCS in our data is about 5~milli-arcseconds (mas) per year.  
We have made use of the $ZYJHK$ data, with the detection limits at an error of 0.15~mag 
$Z = 20.5$, $Y = 20.3$, $J = 19.5$, $H=18.8$, $K1 = 18.0$, and $K2 = 18.1$~mag, respectively, and 
with the saturation limits $Z=11.3$, $Y=11.5$, $J=11.0$, $H =11.3$, and $K1 = 9.9$~mag \citep{lod12}.
The photometric sensitivity of each band is depicted in Figure~\ref{fig:sense}.  
Our investigation is limited to UKIDSS sources with a probability 
greater than 70\% of being a star, using the UKIDSS database flag to distinguish 
a star from a galaxy, with a 
photometric error less than 0.15~mag and being fainter than $J = 12$~mag.
The sky area of our study, limited by the UKIDSS/GCS coverage of 78.5~deg$^{2}$, or about 
a $5\degr$ radius toward Coma Ber, is chosen as the ``cluster region''.  In addition, 
a patch of sky of a $3\degr$ radius roughly $11\degr$ to the east from the cluster center 
is used in experimental design as the ``control field''.

\section{Members of the Coma Ber Star Cluster}\label{sec:members}

The early work by \citet*{tru38} led to the identification of 37 members brighter than photographic 
magnitude 10.5 within a $7\degr$ diameter on the basis of proper motions, color-magnitude 
relation, and radial velocities.  An additional 7 candidates with no radial-velocity measurements 
were also proposed.  While the bright members in Coma Ber display a density structure similar to 
those of Praesepe and the Pleiades \citep{art66}, there is a paucity of faint members, often attributed to 
stellar evaporation \citep{arg69}.  Candidates reported recently by \citet[][105 members]{cas06}, 
\citet[][149 members]{kra07} and \citet[][31 members]{mer08} are mostly bright.  
The later work by \citet[][82 stars]{mel12} expanded the member list to include 
late-M spectral types, i.e., into the brown dwarf regime.  The selection by 
\citet{mer08} included radial velocities for some candidates, but neither \citet{mer08} nor 
\citet{mel12} incorporated proper motion information into membership determination.  

The age of Coma Ber reported in the literature ranges from $\sim300$~Myr to 1~Gyr 
\citep[][summarized in their Table~V]{tsv89}, but usually an age between 400~Myr and 600~Myr 
is adopted \citep{ode98,kra07,cas14}.  

\subsection{Evolved Members} \label{sec:postMS} 

\begin{figure}[t!]
	\centering
	\includegraphics[width=0.9\columnwidth,angle=0]{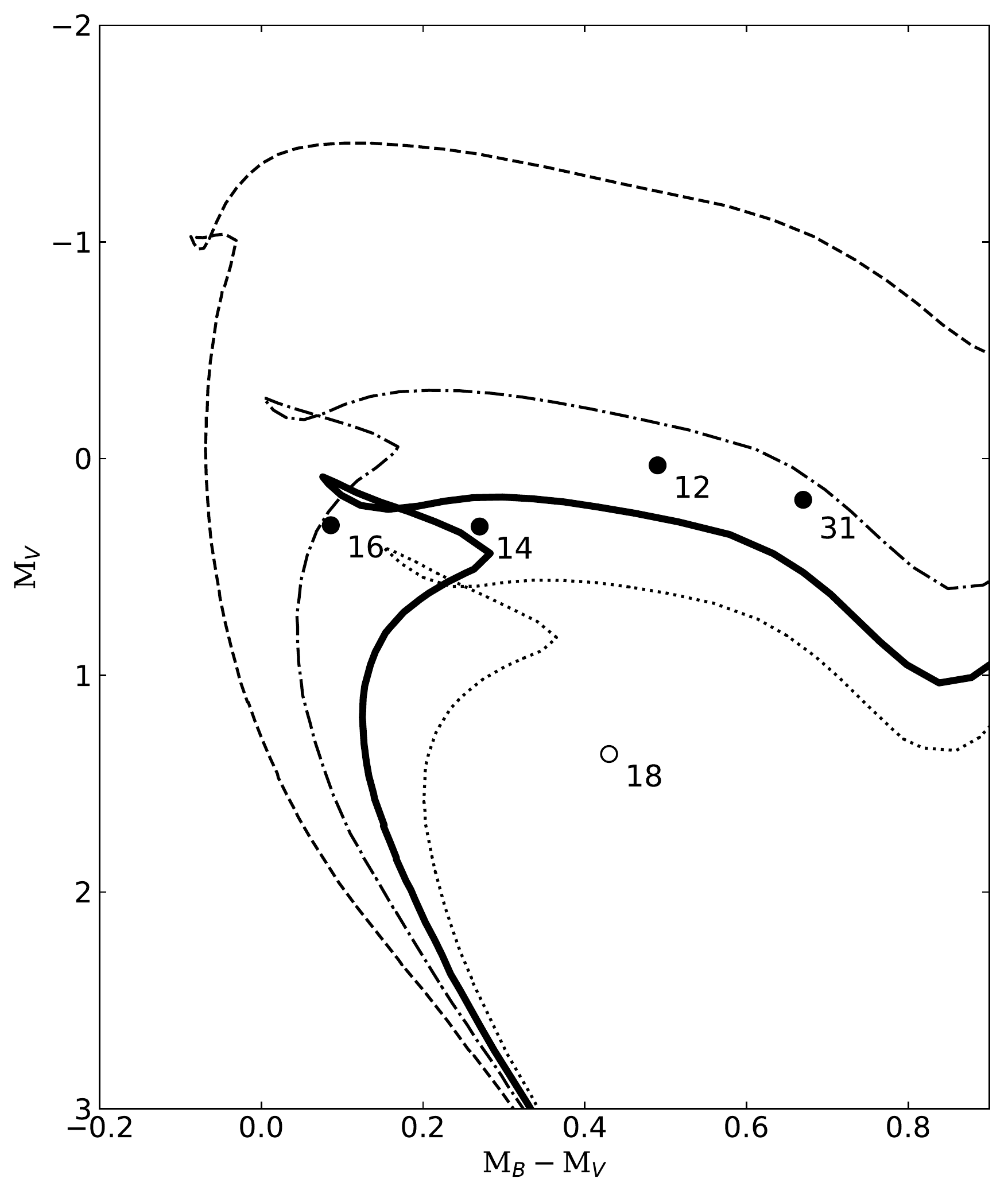}
	\caption{The optical absolute $M_V$ versus $M_B-M_V$ diagram for the brightest stars in the cluster 
	region.  Each number labels the name of a star, e.g., ``12'' means 12\,Com in Table~\ref{tab:postMS}.
	Also plotted are the PARSEC isochrones, from top to bottom, of 300~Myr, 600~Myr, 800~Myr (thick line), 
	and 1,000~Myr.  An age of 800~Myr fits the data better than younger ages.  
	}
\label{fig:postMS}
\end{figure}

We analyze evolved members to constrain the age.  In Coma Ber, any post-main sequence members are 
too bright to render reliable 2MASS photometry, so we characterize them with optical photometry.  
Table~\ref{tab:postMS} lists the parameters of the five brightest stars in the region.  For  
18\,Com, a subdwarf F5\,IV with a {\it Gaia} distance $59.8\pm0.4$~pc and proper motions 
$(\mu_\alpha \cos\delta, \mu_\delta)= (-17.57\pm0.17, 0.68\pm0.12)$~mas~yr$^{-1}$
\footnote{Note URAT1 gives very different proper motions 
$(\mu_\alpha \cos\delta, \mu_\delta)= (-12.0, 9.3)$~mas~yr$^{-1}$ with an error 5.9~mas~yr$^{-1}$ in both 
axes.}, its deviation from theoretical isochrones (shown in Figure~\ref{fig:postMS}) suggests that 
it is not a part of the cluster.  
The other four stars have distance, photometry, and kinematics consistent with membership.  
The star 12\,Com, a known member, is a double-lined spectroscopic binary \citep{gri11} consisting of 
an A2/A3 dwarf and a mid-typed giant (\citet[][F6\,III]{abt08}, \citet[][G7\,III]{gri86}).   
The coeval age of the binary system 670~Myr \citep{gri86} and a {\it Gaia} distance $84.5\pm1.7$~pc both 
indicate membership.  

The star 31\,Com, a G0\,IIIp giant, known to have varying $v\sin i$ \citep{mas08}, suggestive of 
binarity, is located at 6.8~deg from the cluster center, i.e., outside our analysis range, but 
has photometry, astrometry, and distance consistent with membership.  It has been considered 
a member by \citet{cas06} and by \citet{mer08}, and is also included in our member list.

Using the Padova isochrone \citep{bre12}, assuming null reddening \citep[$E(B-V)=0.006$,][]{nic81} and 
solar metallicity \citep{net16,fri92}, 
an age of 800~Myr gives an overall better fit than younger ages, as evidenced in Figure~\ref{fig:postMS}, 
where for each star the absolute magnitude is computed using the {\it Gaia}/DR2 parallax and the apparent 
magnitude taken from the Bright Star Catalog, without correction for extinction or reddening.   
The fit is considered satisfactory, given the known binarity of 12\,Com and 31\,Com, and non-membership 
of 18\,Com.  We therefore conclude Coma Ber to be about 800~Myr old.

\subsection{Bright Members}\label{sec:bright}

\begin{figure*}[htb!]
	\centering
	\includegraphics[width=0.7\textwidth,angle=0]{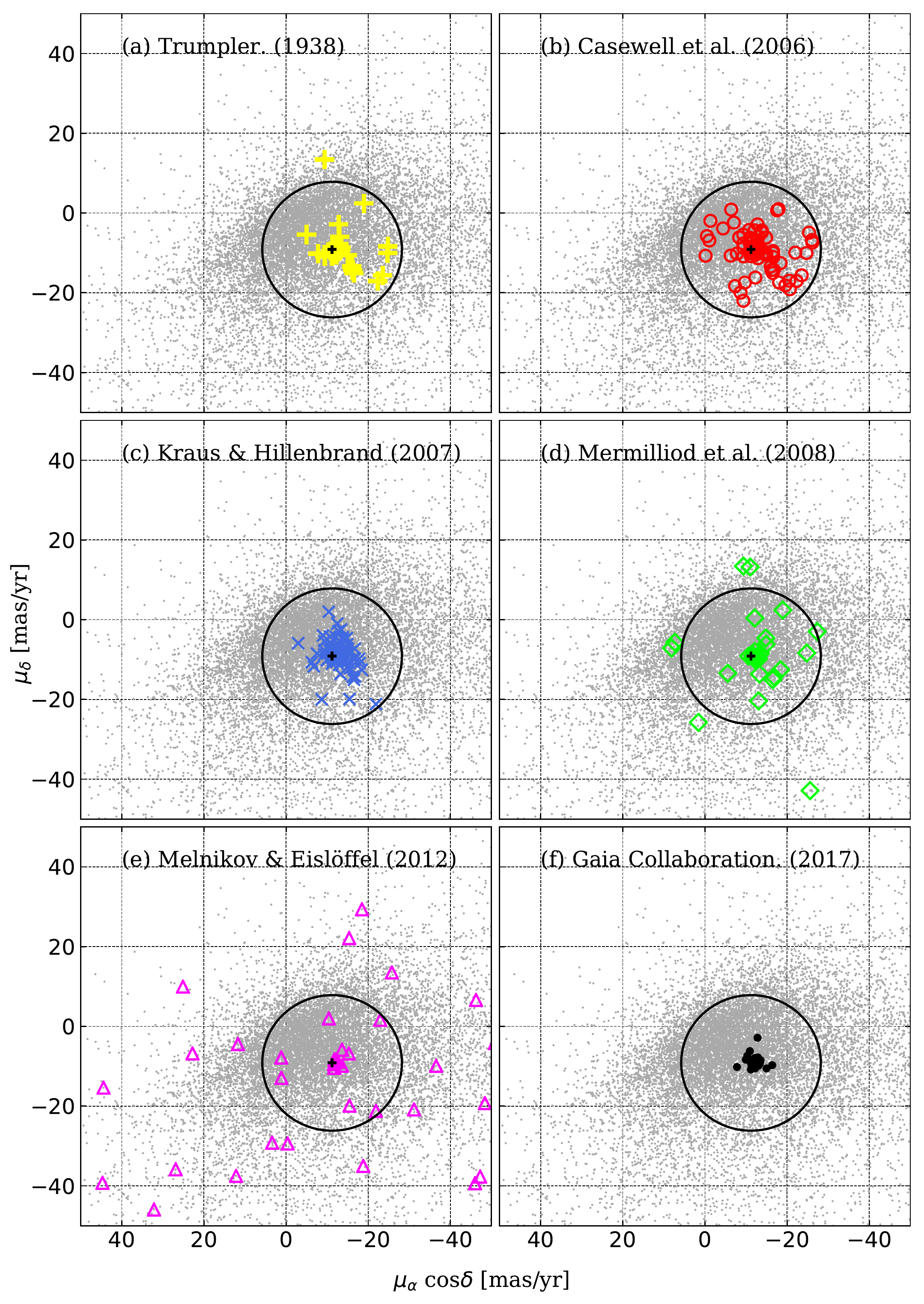}
	\caption{ {\it Gaia}/DR2 proper motion vector plots for candidates identified by 
		(a)~\citet{tru38}, (b)~\citet{cas06}, (c)~\citet{kra07}, (d)~\citet{mer08},  
		(e)~\citet{mel12}, and (f)~\citet{van17}.  In each case, the black circle 
		marks a radius 17~mas~yr$^{-1}$ centered at 
		($\mu_\alpha \cos\delta, \mu_\delta)=(-11.21, -9.16)$~mas~yr$^{-1}$,
		which indicates our selection range for bright candidates.
		For display clarify, only one in five field stars, selected randomly, is shown.
	}
  \label{fig:prevPM}
\end{figure*}

A bright candidate is selected as having proper motions, from {\it Gaia}/DR2 or 
from URAT1, within 
$17$~mas~yr$^{-1}$ from ($\mu_\alpha \cos\delta, \mu_\delta)=(-11.21, -9.16)$~mas~yr$^{-1}$, a range 
judiciously chosen to include all known proper-motion members in the literature.  
Figure~\ref{fig:prevPM} illustrates how this range encompasses the literature candidates.  
The concentration is more obvious for the samples of \citet{cas06} and \citet{kra07}, 
which included proper motions into their membership criteria, than for those of \citet{mer08} 
or \citet{mel12}, which applied no proper-motion criteria. 

Furthermore, a bright candidate is selected as being brighter than $J=14$~mag with photometric errors $< 0.15$~mag, 
and along the PARSEC isochrone \citep{bre12,che14,tan14,che15} bracketed with a color range $(-0.07, +0.3)$ 
in the $J$ versus $J-K_s$ color-magnitude diagram (CMD), 
$(-0.25, +0.15)$ in $J$ versus $J-H$, and $(-0.07, +0.11)$ in $H$ versus $H-K_s$.  
 With these criteria, binary systems would still be selected. 
After excluding 23 candidates, all fainter than about $J\sim12$~mag, considered as galaxies by SDSS (class=3), 
a total of 450 sources satisfy the initial proper-motion and CMD scrutiny.  
Of these, 393 have {\it Gaia}/DR2 counterparts.  

\begin{figure}[t!]
	\centering
	\includegraphics[width=0.9\columnwidth,angle=0]{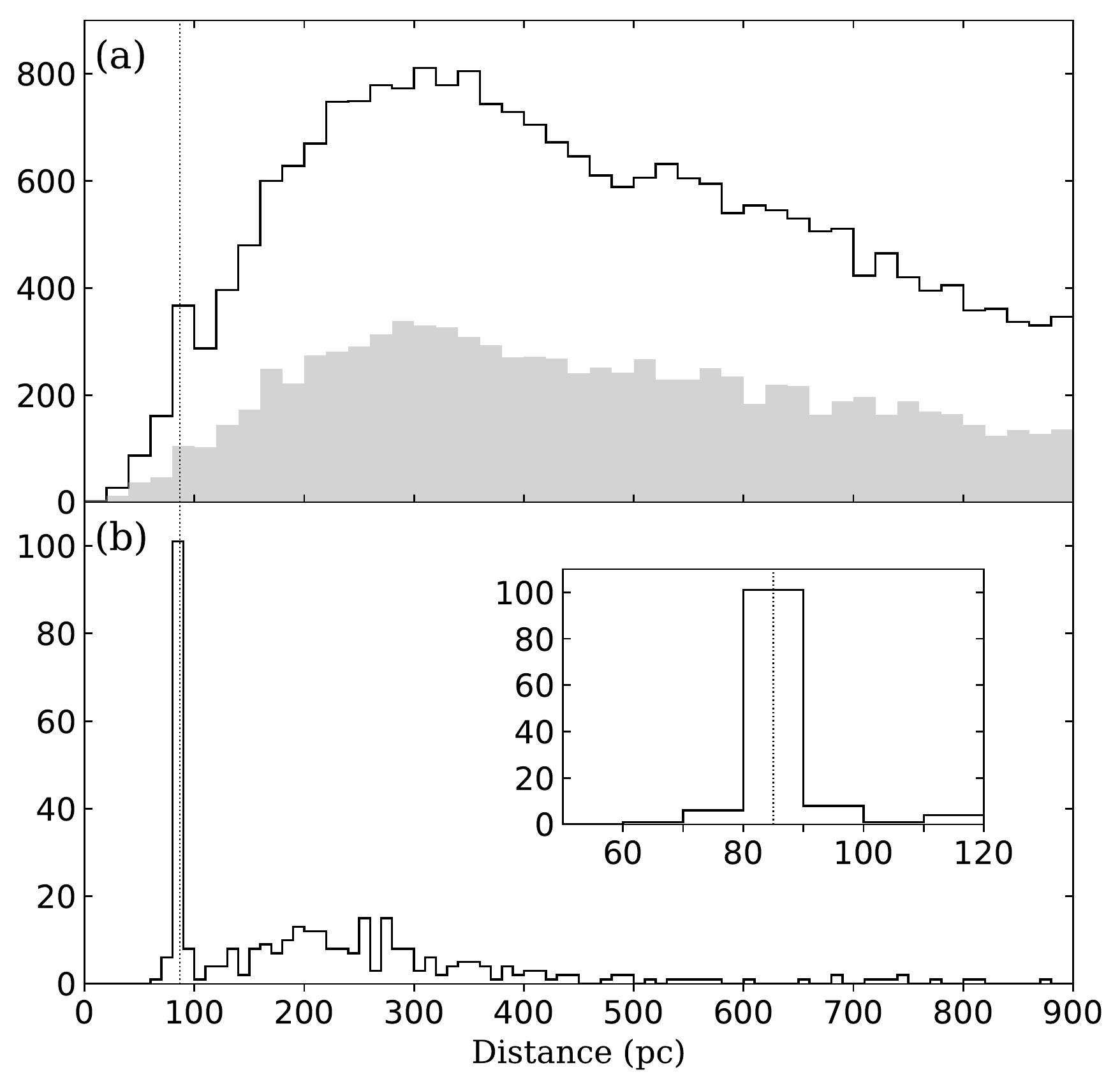}
	\caption{The {\it Gaia}/DR2 parallax measurements of (a)~all the stars toward the cluster 
	region, and all the stars toward the field field, which has a sky area 9/25 or about one-third  
	of the cluster region, and (b)~our 393 preliminary candidates that passed the proper-motion and 
	isochrone selection.  The inset expands to show the distance range 50--120~pc, in which most cluster 
	members are distributed.
	}
\label{fig:gaia}
\end{figure}

Figure~\ref{fig:gaia} plots the distance distributions of (a)~all {\it Gaia} stars in the cluster region 
(within $5\degr$ radius), and all {\it Gaia} stars in the control field ($3\degr$ radius), with 
a sky area 9/25 of the cluster region, 
and (b)~the 393 preliminary candidates with {\it Gaia} measurements available.  
The clustering around 85~pc stands out clearly, particularly in (b).  The fact that in (b) away from the 
peak the number does not increase much with distance, hence the space volume, in contrast to the case 
in (a), indicates an effective winnowing by proper motions and CMD.  

\citet{van17} analyzed a radius $10\fdg4$ around Coma Ber, and reported 50 members based on {\it Gaia}/DR1 
data.  All their members have been confirmed by our selection (40 within and 9 outside the $5\degr$ 
cluster-centric radius), except BD$+$27\,2139 which should have been in their list but is not, 
perhaps because of a glitch in editing 
\citep[][their Table~D.2 containing only 49 entries though there should have been 50]{van17}.

\begin{figure}[t!]
	\centering
  \includegraphics[width=0.9\columnwidth,angle=0]{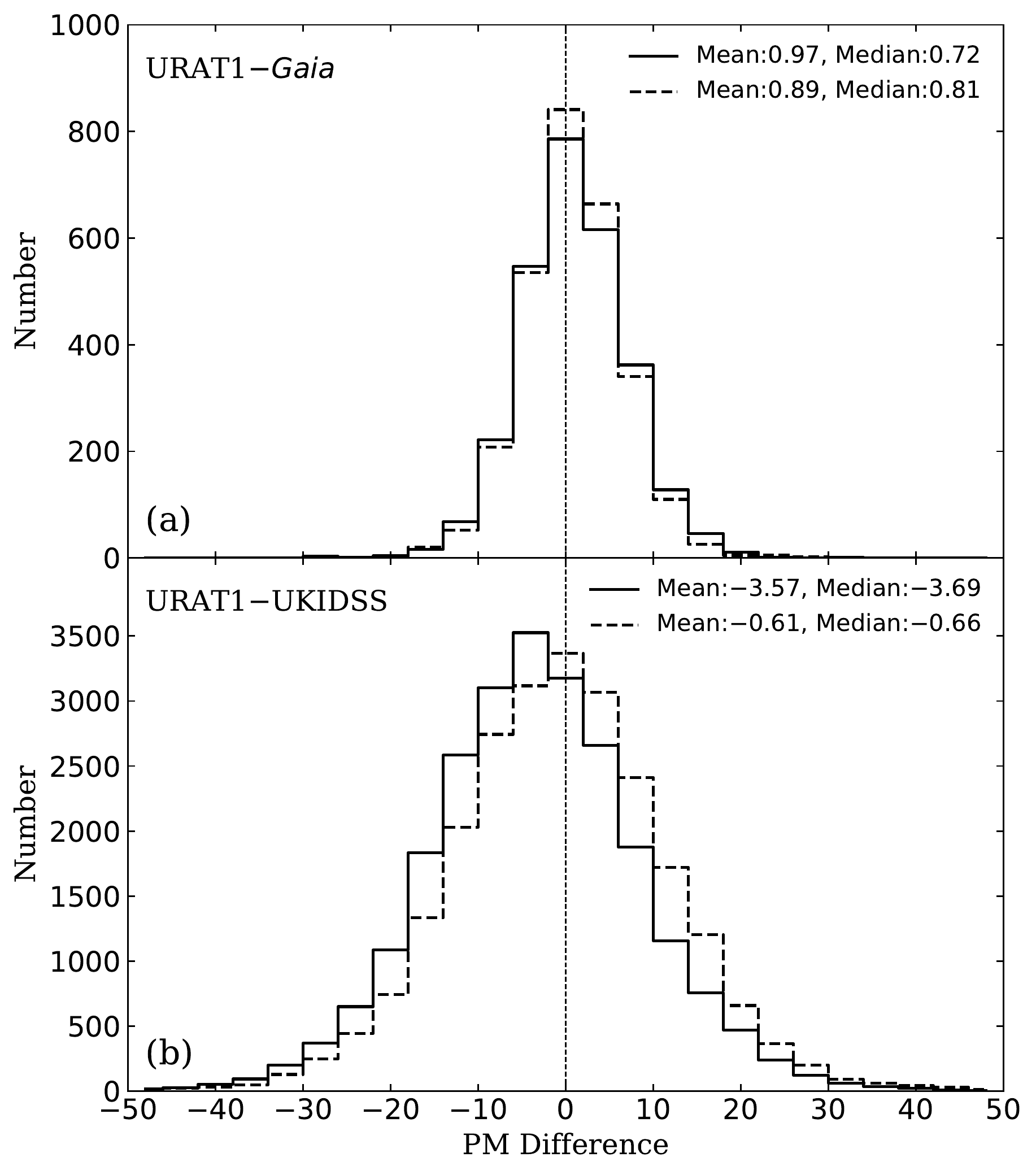}
	\caption{ (a) Comparison of URAT1 and {\it Gaia}/DR2 proper motion values, in Right Ascension 
	(solid line) and in Declination (dashed line).  (b)~same as in (a), but of URAT1 and UKIDSS/GCS 
       } 
  \label{fig:pmComp}
\end{figure}

To bootstrap the three proper-motion datasets used in this study, we compare 
the {\it Gaia}/DR2, URAT1, and UKIDSS/GCS measurements in the cluster region, 
shown in Figure~\ref{fig:pmComp}. The {\it Gaia}/DR2 and URAT1 measurements 
are consistent with each other, so are used to supplement each other for bright candidates.  
There is, however, a systematic offset of UKIDSS/GCS measurements relative to those of URAT1, 
computed for all stars with $J=12$--15~mag, i.e., common in both datasets, 
$(\Delta\mu_\alpha \cos\delta, \Delta\mu_\delta)=(-3.57, -0.61)$.  After the offset 
is applied, the UKIDSS/GCS proper-motion vector center $(-7.64, -8.55)$~mas~yr$^{-1}$ has been 
used to select faint members.

\begin{figure}[t!]
	\includegraphics[width=\columnwidth,angle=0]{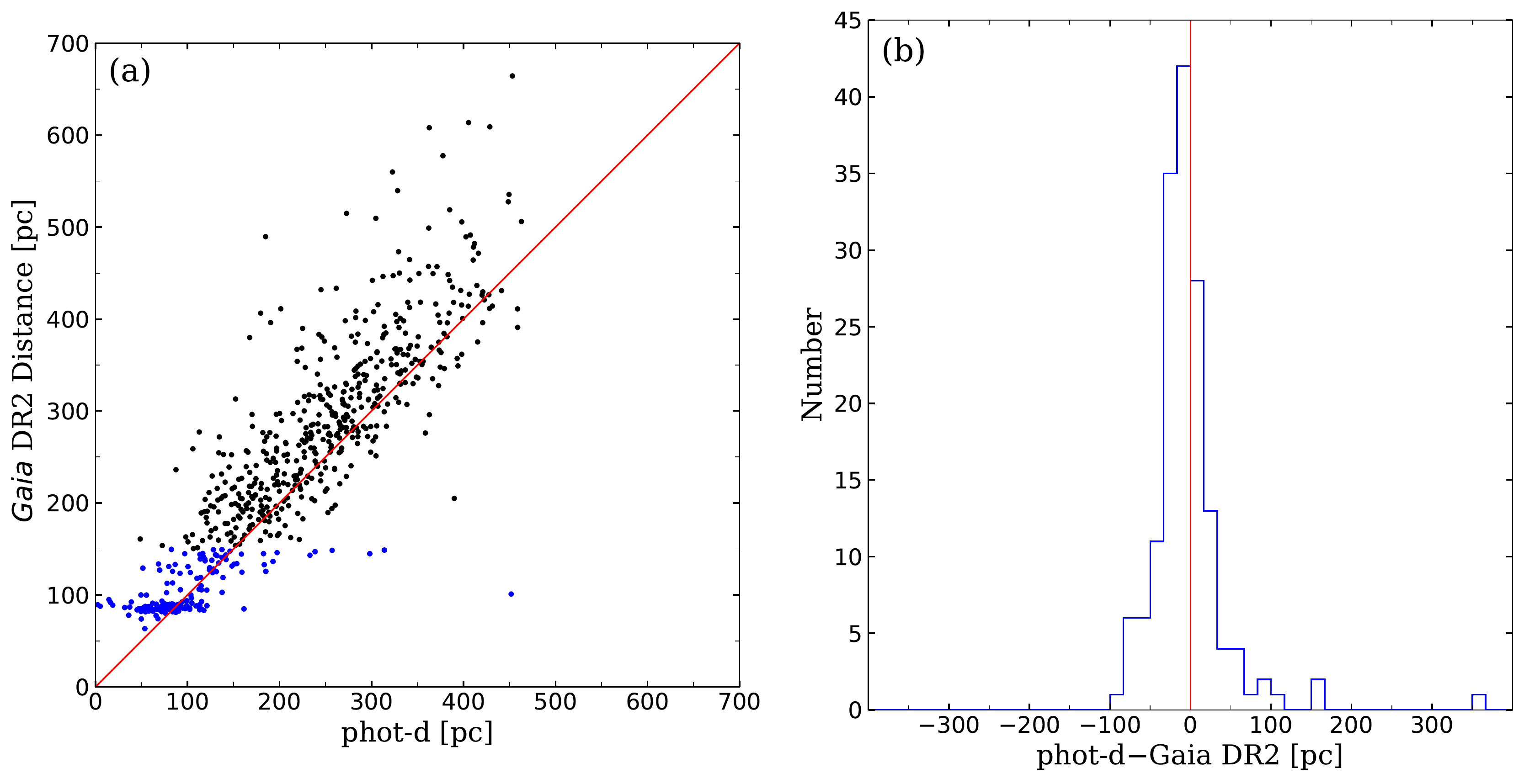}
	\caption{(a)~{\it phot-d} determined distance versus {\it Gaia} measured distance for a sample of 
	preliminary candidates toward Coma Ber satisfying proper motions and CMD criteria.  The red line 
	with a unity slope shows equality.  Black filled circles 
	mark the relatively bright stars $J\sim12$--14~mag, and the blue symbols represent the nearby 
	stars with {\it Gaia} distances closer than 150~pc.
	(b)~The difference between {\it phot-d} and {\it Gaia} distances for the nearby sample in (a).
	}
\label{fig:sptyDiff}
\end{figure}

Distance is a critical parameter in membership identification.  Although the apparent magnitude 
of a star in isochrone fitting, and the proper motion both incorporate implicitly 
the distance criterion, ambiguity exists.  
In addition to direct parallax measurements, we have developed a distance estimator using the 
photometry data taken from the PS1, SDSS, $YJHK$ from UKIDSS, plus $W1$, and $W2$ from {\it WISE}.  
Our algorithm first estimates the spectral type of a star.  This part is adapted from 
the {\it photo-type} method developed by \citet{skr15}, in which a combination of photometric 
colors of a target is compared against a database of templates of different spectral types, from which 
a match is opted, in a least-squares sense, as the most probable spectral type.  The work by \citet{skr15} 
was devised for late-M, L, and T type dwarfs only, and we expand the templates to include 
earlier spectral types (See Appendix).  Once the spectral type is determined, the distance is then derived by comparison 
of the apparent magnitude and absolute magnitude in each band, rendering a median distance when 
all bands are considered.  
Our experiments using different stellar datasets with spectral types known to be K or M types indicate an accuracy 
within 1--2 subtypes in most cases, with the majority of interlopers being extragalactic 
or post-main sequence objects, both expectedly rare in our field.  Earlier than K type, the estimator still 
works reasonably fine, albeit with larger scattering, see Figure~\ref{fig:sptyDiff}.  Currently no interstellar 
reddening is taken into account, and the algorithm is validated only for main sequence stars.  
This distance estimator, which we call {\it phot-d}, offers an effective filtering by distance of stars 
which would have contaminated our member sample by chance inclusion in proper motion and CMD selection.  
More details on {\it phot-d} are given in Appendix.

Figure~\ref{fig:bright} illustrates the $J$ versus $J-K_s$ CMD toward Coma Ber.  Member candidates 
chosen on the basis of proper motions, distances, and isochrone discussed above, are marked, together with 
those satisfying proper motion and isochrone but having inconsistent distances, which would have 
disguised as contaminants if no distance information were available.  Also shown is the false positive sample 
of the control field processed following the same selection procedure as for the cluster region.   
In this analysis, the distance range has been taken as 50--120~pc to account for possible uncertainty of 
the {\it phot-d} distance.  It is encouraging that the false positive rate is low, particularly for the 
bright candidates ($\gtrsim 1~M_\sun$)  Within this distance range, there are 131 bright {\it Gaia} members, 
averaging 87.0~pc with a standard deviation (dispersion) 8.1~pc.

\begin{figure*}[t!]
	\includegraphics[width=\textwidth,angle=0]{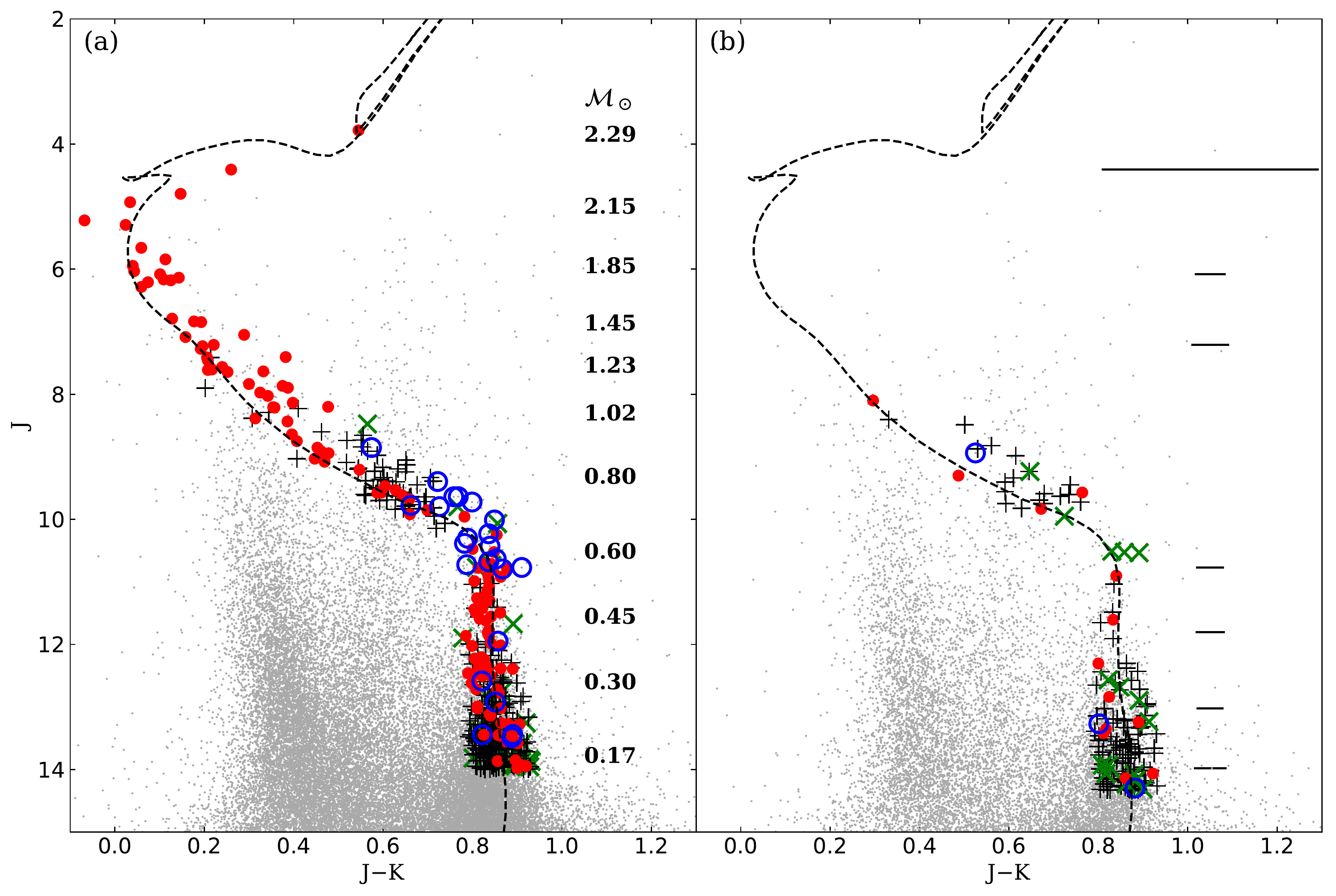}
	\caption{(a)~2MASS $J$ versus $J-Ks$ for the bright sample in the cluster region.  
	The gray dots represent all 2MASS sources.  
	The circles mark the candidates satisfying the proper motion and isochrone criteria, 
	and further constrained, respectively, by parallax distances 
	(in red filled circles) and those by {\it phot-d} distances (in blue open circles).  
	Those satisfying both proper motion and isochrone selection but otherwise 
	rejected by distances are represented by black pluses (with {\it Gaia} distances) 
	or by green crosses (with {\it phot-d} distances. 
	Stellar masses, as per the PARSEC isochrones, are indicated.
	(b)~The same but for the control sample, with the same symbols as in (a).   
        The stars that satisfied all criteria of proper motions, isochrone, and distance 
	here are false positives. 
	Typical errors in 2MASS $J-K_s$ colors are presented as horizontal bars to the right.
	}
\label{fig:bright}
\end{figure*}

\subsection{Faint Members}\label{sec:faint}
 
\begin{figure*}[t!]
	\centering
	\includegraphics[width=\textwidth,angle=0]{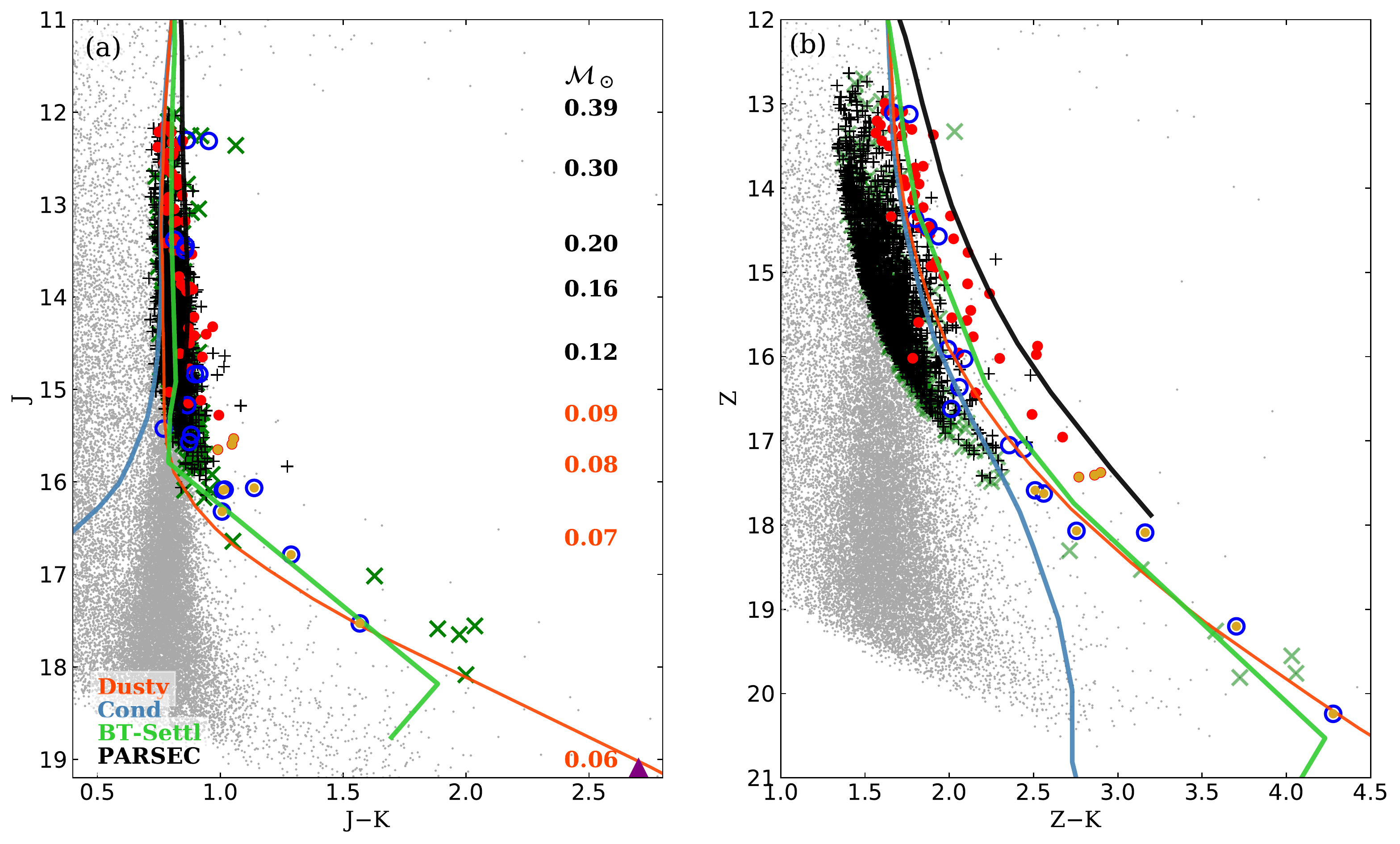}
	\caption{(a)~The $J$ versus $J-K$ CMD as in Fig.~\ref{fig:bright} with the same symbols, 
	but for the faint sample using UKIDSS/GCS data.  Additional golden dots mark the brown dwarf 
	candidates identified in this work. The stellar masses as per the PARSEC (the black line) or 
	the DUSTY model (the orange line) are indicated.  
	Additional isochrones, those of Cond (the blue line) and BT-Settl (the green line), are also shown.
	(b)~The $Z$ versus $Z-K$ CMD using the same symbols as in (a).  One source marked as a purple 
	triangle in (a) does not appear here because of no detection at $Z$.  
        For display clarity, only one in every five field stars (as gray dots), 
	selected randomly, is shown.
	}
  \label{fig:faint}
\end{figure*}

For faint stars, we have adopted the proper-motion vector center 
$(\mu_\alpha \cos\delta, \mu_\delta)=(-7.64, -8.55)$~mas~yr$^{-1}$ for UKIDSS/GCS, 
also within a radius of 17~mas~yr$^{-1}$ for membership selection.  In addition, a candidate 
is selected to be fainter than $J=12$~mag, with
photometric errors $<0.15$~mag, and along the DUSTY isochrone bracketed with a color range $(-0.35, +1.2)$ 
in the $Z$ versus $Z-K$ CMD, and $(-0.1, +0.5)$ in $J$ versus $J-K$, as shown in 
Figure~\ref{fig:faint}. The ranges of colors are chosen deliberately wide to allow for 
uncertainties in photometry/color and also in isochrones. The distance range, as for the bright sample, 
is chosen to be 50--120~pc for either the {\it Gaia} or the {\it phot-d} distances.

\begin{figure*}[t!]
	\centering
        \includegraphics[width=\textwidth,angle=0]{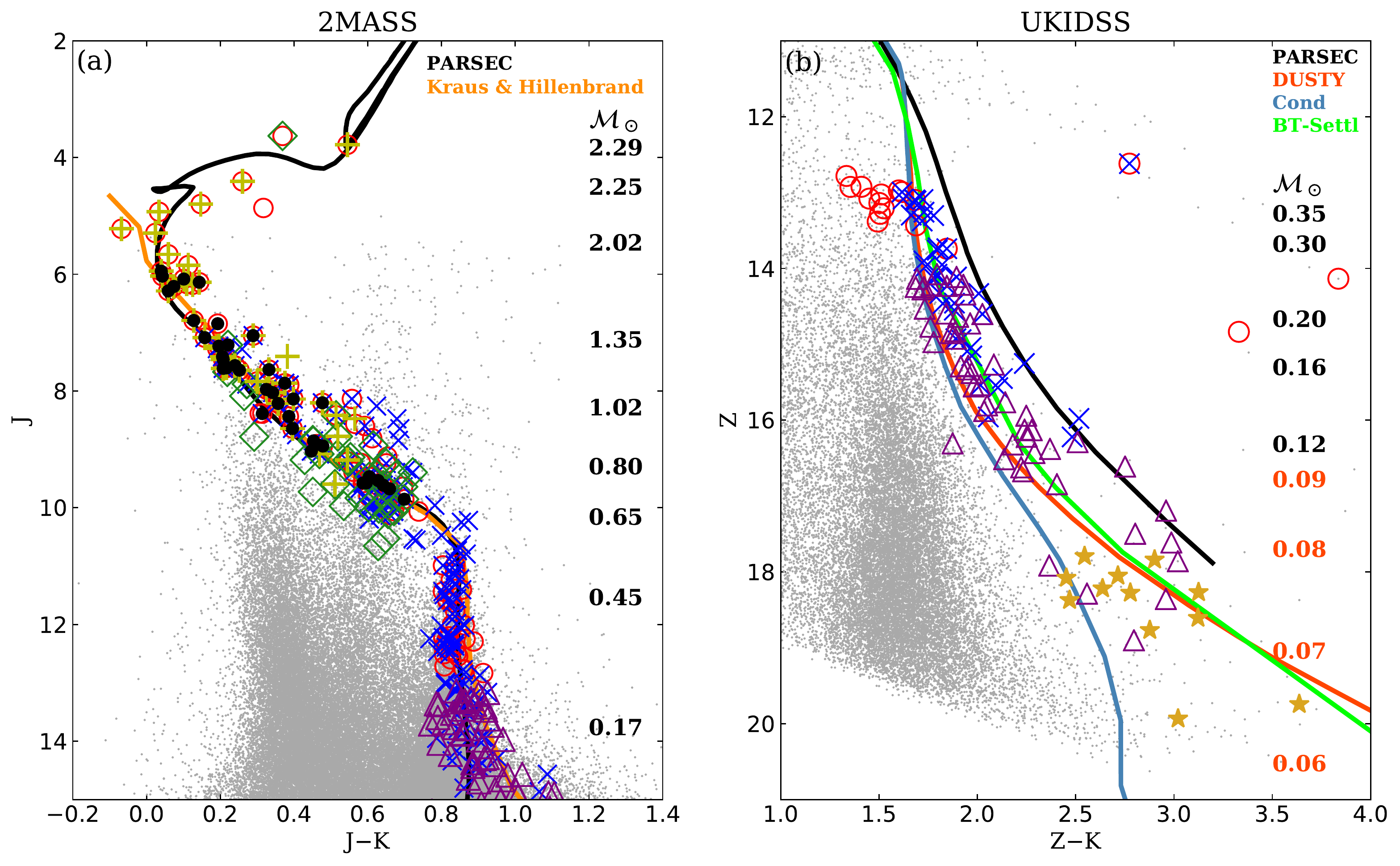}
	\caption{Color-magnitude diagrams for literature candidates in (a)~2MASS $J$ versus $J-K_s$ and 
		(b)~UKIDSS/GCS $Z$ versus $Z-K$.  Data include those from 
		\citet[][in plus symbols]{tru38}, \citet[][open circles]{cas06}
		\citet[][in cross]{kra07}, \citet[][open diamonds]{mer08}, 
		\citet[][in open triangel]{mel12}, \citet[][dots]{van17}, and
		 \citet[][their BD candidates, star symbols]{cas05}.
		Also plotted are lines depicting evolutionary models of PARSEC (in black), 
		Dusty (in orange), Cond (in blue), BT-Settl (in green), and an empirical dwarf sequence
		from \citet[][in orange]{kra07}.  The stellar masses according to PARSEC or Dusty 
		are labeled.  For display clarity, only one in every five UKIDSS field stars 
		(as gray dots), selected randomly, is shown.
	}
  \label{fig:prevCMD}
\end{figure*}

Note that in $J$ versus $J-K$, often used to identify substellar objects, the isochrone 
passes through populated regions, so such a CMD is not as discriminating as those involving shorter 
wavelengths such as $Z$ versus $Z-K$ for low-mass objects.  Our investigation hence relies primarily on $Z$ versus 
$Z-K$, though for very cool objects a marked flux suppression sometimes renders detection at neither $Y$ nor $Z$ in 
UKIDSS/GCS.  For these, analysis with $J$ versus $J-K$ would be applied.  
The lesson is that there are no preferred colors to identify cool objects, and a combination of CMDs must 
be iterated.  As a comparison, Figure~\ref{fig:prevCMD} presents how literature candidates behave in our diagnostic 
2MASS $J$ versus $J-K_s$ and UKIDSS/GCS $Z$ versus $Z-K$ CMDs, overlaid with several theoretical 
800~Myr isochrones adopting the average distance 85~pc.  We note that for a nearby cluster such as Coma Ber assuming a
single distance in a CMD would introduce intrinsic scattering unless absolute magnitudes are plotted 
(see for example Figure~\ref{sec:postMS}).  While the bright literature candidates by and large follow the model 
isochrones, the faint ones are discrepant.  

Combining the bright (\S\ref{sec:bright}) and faint (\S\ref{sec:faint}) candidates together results 
in 194 candidates within the cluster region.  This sample is spatially complete, notwithstanding the 
UKIDSS voids, and forms the basis of our characterization of the cluster. 
The candidates are listed in Table~\ref{tab:members}.  The first column gives the running number.  
Columns 2 and 3 are the coordinates, followed by columns 4 to 9 
the $J$~mag, its error, $H$~mag, its error, $K$~mag and its error.  Columns 10 to 12 give the proper motions 
and the error.  Column 13 lists the distance, followed by column 14, the references if the star is a known 
literature candidate.  The last column indicates the data source, where ``1'' stands for 2MASS photometry ($JHK_s$) 
plus {\it Gaia}/DR2 proper motions, ``2'' stands for 2MASS photometry ($JHK_s$) 
plus URAT-1 proper motions, {\bf``3'' stands for UKIDSS/GCS ($JHK$) plus {\it Gaia}/DR2 proper motions,
and ``4'' means that $JHK$ photometry and proper motion are both from UKIDSS/GCS,} 
with no transformation between 2MASS and UKIDSS photometric measurements.  

Candidates in Table~\ref{tab:members} are categorized as (1)~those with parallax distances (No.~1--154); 
this is the most reliable member list to our knowledge to date of the Coma Ber cluster; and (2)~the other 
38 with distances estimated by {\it phot-d} (No.~155--192).  For sources with parallax measurements
but $\varpi / \Delta\varpi < 10$, their {\it photo-d} distances are adopted.  In each category the entries 
are in ascending RA order.  

The criterion of the ratio $\varpi / \Delta\varpi > 10$ for {\it Gaia}/DR2 data is biased against a distant source 
for which $\varpi$ would be small (so is the ratio) even though $\Delta\varpi$ is already relatively small.  These sources 
tend to be distributed in the vertical segment of the $J$ versus $J-Ks$ CMD.  They are hence likely background giants 
and as such {\it phot-d}, valid for dwarfs only, would underestimate the distances.  At the moment we do not have 
an effective method to remove these individual contaminants from the member list, except by statistical 
subtraction by the control sample.  


There are individual objects considered as member candidates in the literature but outside the $5\degr$ radius.  
A total of 15 have been reaffirmed by our analysis, including for example 31\,Com presented above as a post-main 
sequence member (\S\ref{sec:postMS}), and two reported by \citet{kra07}:~HD\,111878 with a {\it Gaia} distance 85.1~pc, 
and HD\,109390 at 120.1~pc (close to the limit of our distance range).  This selected sample is spatially 
incomplete so not included in further analysis, but is listed in Table~\ref{tab:beyond5} for reference, 
with the same table format as in Table~\ref{tab:members}, except with no last column because all data 
are from 2MASS and {\it Gaia}/DR2.


For potential usefulness, we also summarize in Table~\ref{tab:rej} the literature candidates rejected by our selection.
The table is in a two-column format arranging in ascending R.A.\ order.  For each star, the coordinates, reference for candidacy, 
and an offending code in our analysis are given:~1=rejection by proper motions, 2=rejection by CMD, 4=rejection 
by distance.  The code is additive, so for example a literature candidate having consistent proper motions, but 
inconsistent with being a member in CMD position and in distance has a code=6.

\subsection{Brown Dwarf Members }\label{sec:bd}

\begin{figure*}[t!]
	\centering
	\includegraphics[width=0.6\textwidth,angle=0]{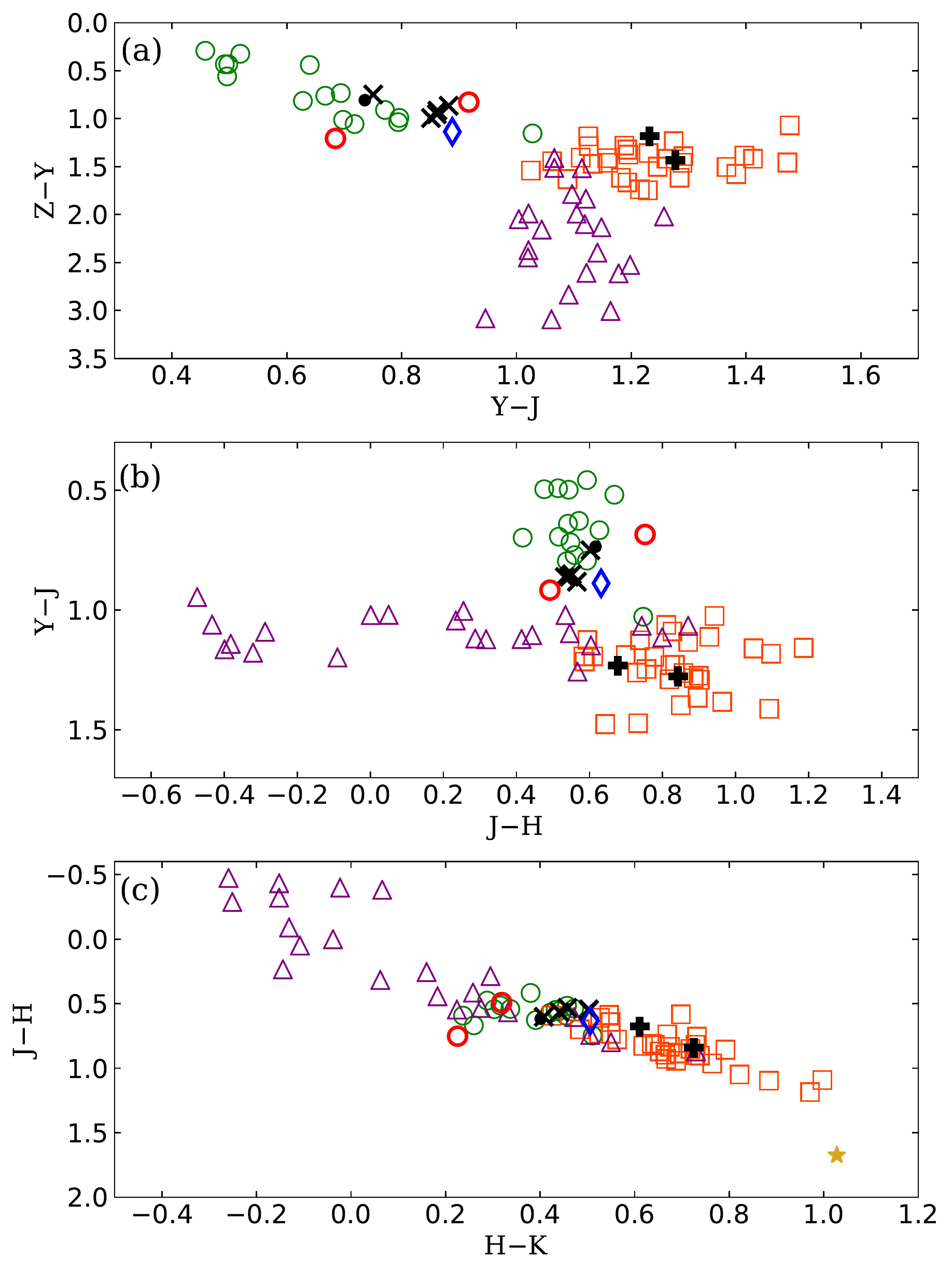}
	\caption{UKIDSS two-color diagrams of the M (green open circles), 
		L (orange open squares), and T (purple open triangles) 
		dwarfs in \citet{hew06} in (a)~($Z-Y$) versus ($Y-J$), 
		(b)~($Y-J$) versus ($J-H$), 
	and (c)~($J-H$) versus ($H-K$) diagrams. The known M dwarfs from \citet{wes11} 
	(i.e., our Nos.~6, 55, 130, 155, and 157)  
		are represented by black crosses; the M9 reported by \citet{cas14} (our No.~176) is marked 
		as a blue diamond.  The M brown dwarf and two L dwarfs found by this work are 
		in black filled circle and in black pluses.  Objects A and B are shown as thick 
		red open circles, whereas object~C, having no $Z$ or $Y$ detection, is marked with a golden 
		star symbol.  
	}
  \label{fig:tcd}
\end{figure*}

Any member fainter than $Z=17.3$~mag has a mass less than 0.08~$M_\sun$, and is  
therefore a brown dwarf.  
There are several lines of evidence to further substantiate its brown dwarf nature.  First, 
late-M, L, and T dwarfs are known to have UKIDSS colors different from those of field stars 
\citep{hew06}, and all our brown dwarf candidates indeed have colors, shown in Figure~\ref{fig:tcd}, 
consistent with being M- or L-type objects.  Secondly, all these candidates have 
spectral types estimated by {\it phot-d} as being brown dwarfs.  

Moreover, while very low-mass objects have distinctly red $W1-W2$ colors 
\citep{kir11} owing to the lack of methane absorption at $W2$ (4.6~\micron) relative to the 
flux at $W1$ (3.4~\micron), their $W3$ (12~\micron) and $W4$ (22~\micron) fluxes often fall 
below the sensitivity limits of {\it WISE} unless they are located in the solar vicinity 
\citep[see for example][] {sch11}.  

Among the efforts to identify brown dwarfs in Coma Ber, the spectroscopic study  
by \citet{cas14} led to confirmation of an M9 (their cbd34, RA=12:23:57.37, Dec.=+24:53:29.0, J2000, 
$J=15.94$~mag), an L1 (their cbd67, RA=12:18:32.71, Dec.=+27:37:31.3, J2000, $J=17.68$~mag), 
and an L2 (their cbd40, RA=12:16:59.89, Dec.=+27:20:05.5, J2000, $J=16.30$~mag) objects, among which 
cbd40 was disregarded of its membership on the basis of its brightness and colors.  
Our candidate list includes cbd34, but not cbd67 which satisfies neither the 
proper motion nor the photometric criterion, and is therefore a field brown dwarf.  
\citet{wes11} compiled a catalog of spectroscopic M dwarfs on the basis 
of the Sloan Digital Sky Survey DR7.  Among the M dwarfs in our cluster region, 
10 satisfy our selection criteria of proper motions, CMD, and distance, and indeed have been 
included in our list.  

Table~\ref{tab:substellar} summarizes the properties of the substellar objects, a subset of 
the member candidates in Table~\ref{tab:members}.  Following the same identification number as in 
Table~\ref{tab:members} in the first column, the
next columns list, respectively, the coordinates, UKIDSS $Z$, $Y$, $J$, $H$, and $K1$ magnitudes, and then the 
UKIDSS proper motions. The last two columns compare the spectral type determined with 
spectroscopy, either reported in the literature or observed by us, and the spectral type estimated with {\it phot-d}.
In general the spectral typing with {\it phot-d} is in agreement within 1--2 subtypes with observations. 
This gives us confidence in our {\it phot-d} method.  
The first six sources in Table~\ref{tab:substellar} are all of late-M types known in the literature 
\citep{wes11,cas14}, and we have confirmed their membership.  The M9 objects, namely Nos.~176, 
55, and 130, were the coolest known members in Coma Ber before our work. 

There are a few miscellaneous objects, A, B, and C listed in Table~\ref{tab:substellar} 
that are not classified as member candidates but worth clarification. Stars A and B have similar 
infrared colors (from $Z$ to {\it WISE} $W2$) as those of known brown dwarfs, as seen in the 
two-color diagrams shown in Fig.~\ref{fig:tcd}.  Yet, at shorter wavelengths object A has 
PS1 measurements $g_{P1}=19.97$~mag, $r_{P1}=18.73$~mag, and $i_{P1}=17.61$~mag, and object B has 
$g_{P1}=20.55$~mag, $r_{P1}=19.30$~mag, and $i_{P1}=18.03$~mag.  
Compared with the mean values and standard deviation of the M-type brown dwarfs in Table~\ref{tab:substellar}, 
$g_{P1}=21.51\pm0.56$~mag, $r_{P1}=21.94\pm0.19$~mag, and $i_{P1}=19.54\pm0.32$~mag, the two stars 
stand out significantly brighter. The {\it phot-d} analysis suggests both to be of early-M types. 
While binarity of a hot plus a cold component may explain the brightness inconsistency, we do not have 
evidence at the moment as to the nature of either star.

\begin{figure*}[ht!]
	\centering
	\includegraphics[width=0.8\textwidth]{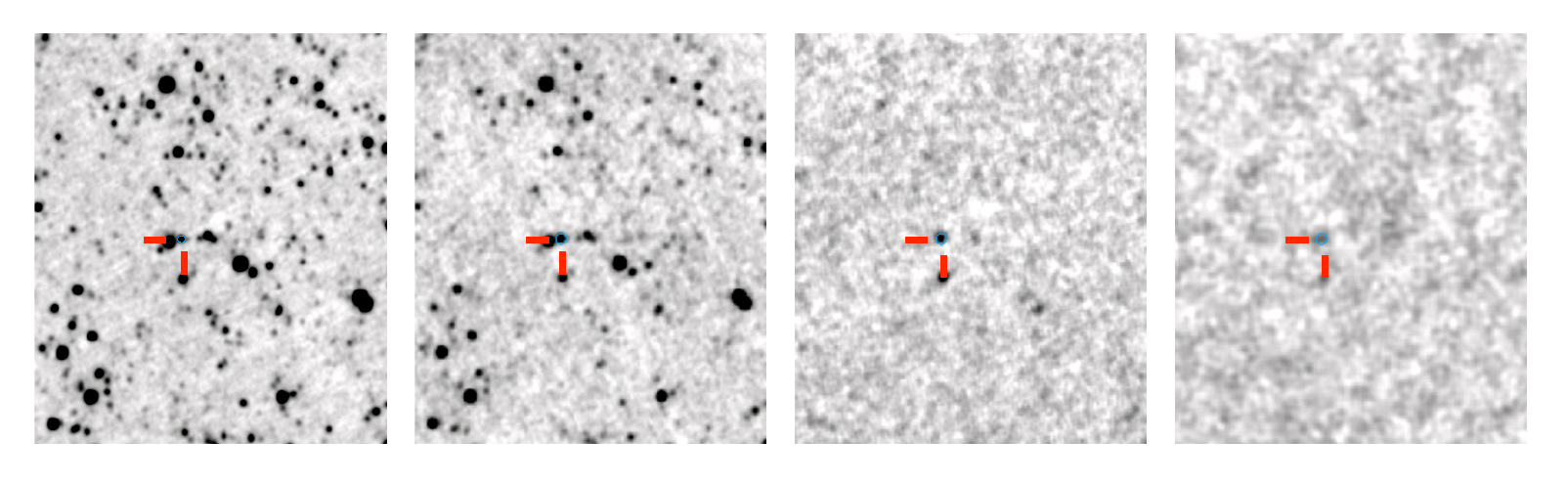}
	\caption{The {\it WISE} $W1$ to $W4$ images, from left to right, of object~C.
		}
  \label{fig:wise}
\end{figure*}

Object~C has $JHK$ magnitudes and proper motions consistent with being substellar, but has $Z$ and $Y$ magnitudes 
below the UKIDSS/GCS limits.  It is the only source in the table that has been clearly detected 
not only in $W1$ and $W2$, but also in $W3$ and $W4$ (see Figure~\ref{fig:wise}.)  It is likely a galaxy.


\subsubsection{Follow-up Spectroscopy}\label{sec:sp}

Confirmation spectra of the brown dwarf candidates were acquired with Palomar/TripleSpec or 
with Gemini/GNIRS. Three targets, No.~159, A, and B were observed using TripleSpec \citep{her08} on 
14 December 2016, with a slit width 1\arcsec.  The sky was clear and the observations 
were executed at an airmass about 1.20.  The standard A-B-B-A nodding sequence was followed 
along the slit to record target and sky spectra. The exposure time per pointing was 300~s, 
with a total integration of 1200~s for each target. Flat-fields and argon lamp spectra 
were taken after every set of target observations. A nearby A0\,V star was observed for each 
target for telluric correction as well as for flux calibration. 

SPEXTOOL package version 4.1 \citep{vac03,cus04} has been used for processing of the data taken 
with TripleSpec, which includes the pre-processing, aperture extraction and wavelength calibrations. 
The individual extracted and wavelength-calibrated spectra from a given sequence of observations, 
each with their own A0 standard star, were then scaled to a common median flux and combined using 
XCOMBSPEC in SPEXTOOL. The combined spectra were corrected for telluric absorption and 
flux-calibrated using the respective telluric standards with XTELLCOR. All calibrated sets of 
observations of a given target were then median-combined to produce the final spectrum.


The spectra of candidates Nos.~191 and 160 were acquired by Gemini Fast Turnaround 
GN-2017B-FT-18, on 7 and on 11 December 2017, respectively, both under good sky conditions, 
using the cross-dispersed mode of GNIRS \citep{eli06a,eli06b}, 
covering 0.9--2.5$\mu$m simultaneously with a resolving power $R\sim 1200$. 
The short blue camera with 32 lines/mm grating was selected with a slit width $0\farcs45$ for
No.~191 and $0\farcs675$ for No.~160.  The individual exposure time under the A-B-B-A 
nodding sequence was 150~s for No.~191, and 60~s for No.~160.  Three sets of nodding sequence were observed for 
No.~191, and one set was taken for No.~160.   

The GNIRS raw data were reduced by PyRAF using the Gemini and GRNIRS packages.
We first cleaned the pattern noise, radiation events, flat-field, sky-subtraction, then did the 
wavelength calibration by spectra of arc lamps. Each spectrum was extracted
from the combined ABBA exposure files. Telluric absorption line was removed with two 
A2\,V standard stars (HIP\,58297 and HIP\,63006) observed before or after the observing run.

\begin{figure}[t!]
	\centering
	\includegraphics[width=\columnwidth,angle=0]{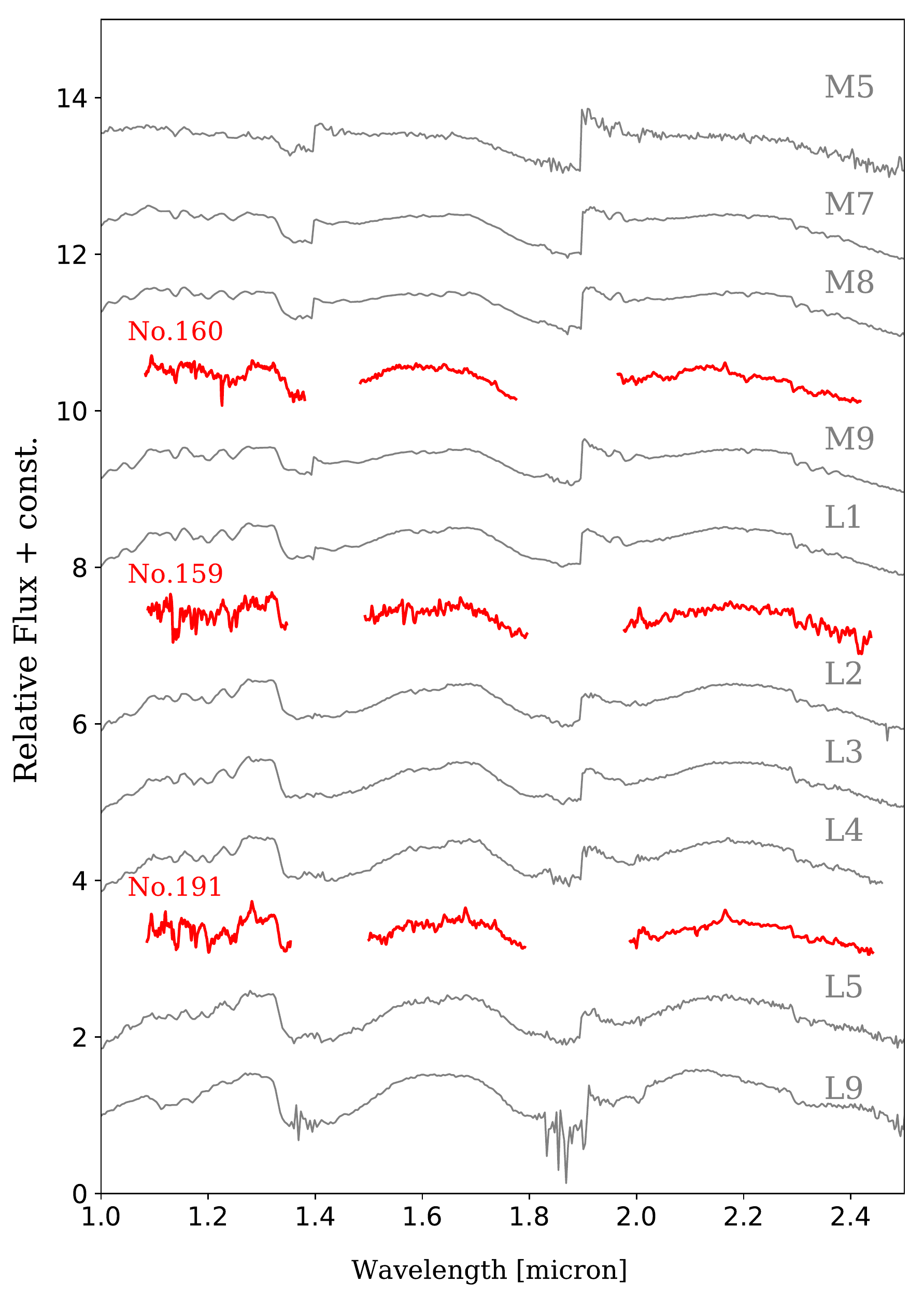} 
	\caption{Spectra of candidates Nos.~159, 160, and 191.  Also shown are brown dwarf template 
	spectra:
	M5 \citep[2MASS J01405263+0453302,][]{kir10}; 
	M7 \citep[VB\,8,][]{bur08}; 
	M8 \citep[VB\,10,][]{bur04}; 
	M9 \citep[LHS2924,][]{bur06a}; 
	L1 \citep[2MASSW~J2130446$-$084520,][]{kir10}; 
	L2 \citep[Kelu-1,][]{bur07a}; 
	L3 \citep[2MASSW~J1506544$+$132106,][]{bur07b}; 
	L4 \citep[2MASS~J21580457$-$1550098,][]{kir10}; 
	L5 \citep[SDSS~J083506.16$+$195304.4,][]{chi06}; 
	L9 \citep[DENIS-P~J0255$-$4700,][]{bur06b}  
	}
  \label{fig:bdSpec}
\end{figure}

Each reduced Palomar or Gemini one-dimensional spectrum was then compared with the low-dispersion 
template spectra of brown dwarfs from SpeX \citep{ray03}, from which the ``best'' match, judged by 
eye examination, was determined.  Three candidates turned out to be bona fide substellar objects, 
with one of late-M (No.~160), one early-L (No.~159), and one mid-L type (No.~191), shown in 
Figure~\ref{fig:bdSpec}.  The classification was warranted by the characteristic absorption 
features due to methane/water near 1.4~\micron.

\begin{figure}[thb!]
	\centering
	\includegraphics[width=\columnwidth,angle=0]{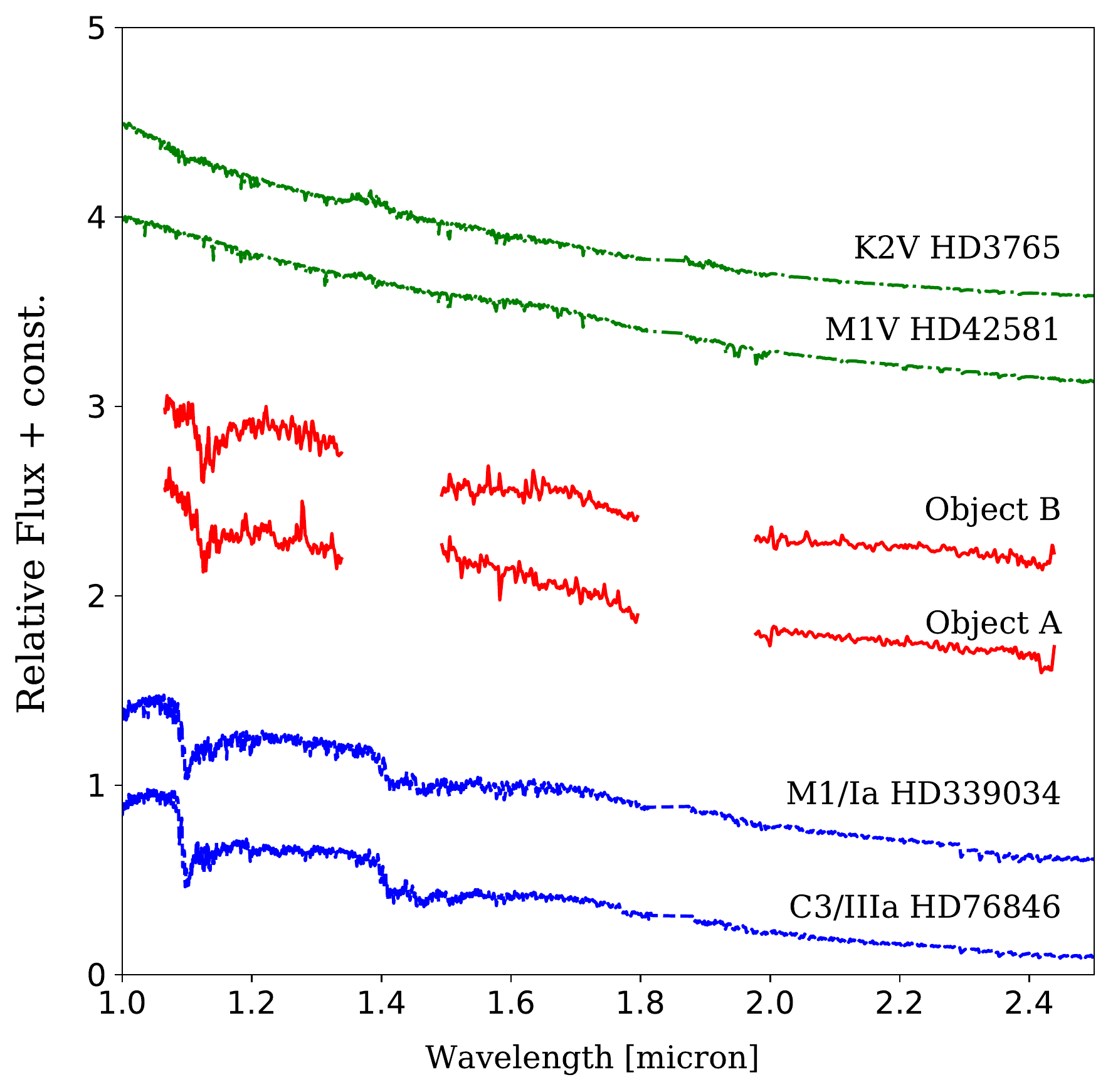} 
	\caption{Spectra of objects A and B, and template spectra of cool stars \citep{ray09}.   
	Both show the CN feature near 1.1~\micron\ seen also in the giant and supergiant spectra. }
	\label{fig:giantSpec}
\end{figure}

The Palomar/TripleSpec spectra for objects A and B are exhibited in Fig.~\ref{fig:giantSpec}, together 
with template spectra from SpeX.  Both spectra show the CN band feature near 1.1~\micron\ characteristic of 
giants and supergiants \citep{loi01,lan07}.  The {\it Gaia} measurements 
however yield $\varpi=2.5 \pm 0.2$~mas for star A, 
and $\varpi=2.6\pm 0.3$~mas for star B, respectively, placing each at $\sim400$~pc.  Their optical brightness 
($g_{P1}\sim20$~mag) is hence consistent more with M dwarfs ($M_V=9$--10~mag) than with giants ($M_V\sim-0.3$) 
\citep{cox00}.  Further spectroscopic observations are required to provide information on the nature of these two 
stars, for example whether they belong to the dwarf carbon population, 
but in any case neither of them is associated with the cluster.

\section{Discussion}\label{sec:dis}

\subsection{Age of Coma Ber}

The age we derive for Coma Ber, $\sim800$~Myr, is based on consistency of evolved members with 
model isochrones (see \S\ref{sec:postMS}), for which no rotation is taken into account.  Rotation 
affects the equations of stellar structure, and with elevated centrifugal force, acts as 
reducing the stellar mass, thereby prolonging the main sequence lifetime and core helium buring 
phase \citep{kip70}.  This age is older than the 400--600~Myr often quoted in the literature.  

Lithium abundances have been measured for solar-type members in Coma Ber \citep{jef99,for01}, 
particularly in the context of convective mixing in stellar interior in terms 
of metallicity versus age, e.g., in comparison with stars of similar spectral types in the Hyades
and Praesepe, or in the much younger Pleiades \citep[see for example][]{for01}.  
With an older age, it is not clear if the lithium depletion boundary is still applicable 
as a chronometer \citep{mar18}.  In any case, our candidate list contains quite a number of 
solar-type or cooler members to shed light on the subject.

\subsection{The Shape and Size of the Cluster} 

\begin{figure*}[thb!]
	\centering
	\includegraphics[width=\textwidth,angle=0]{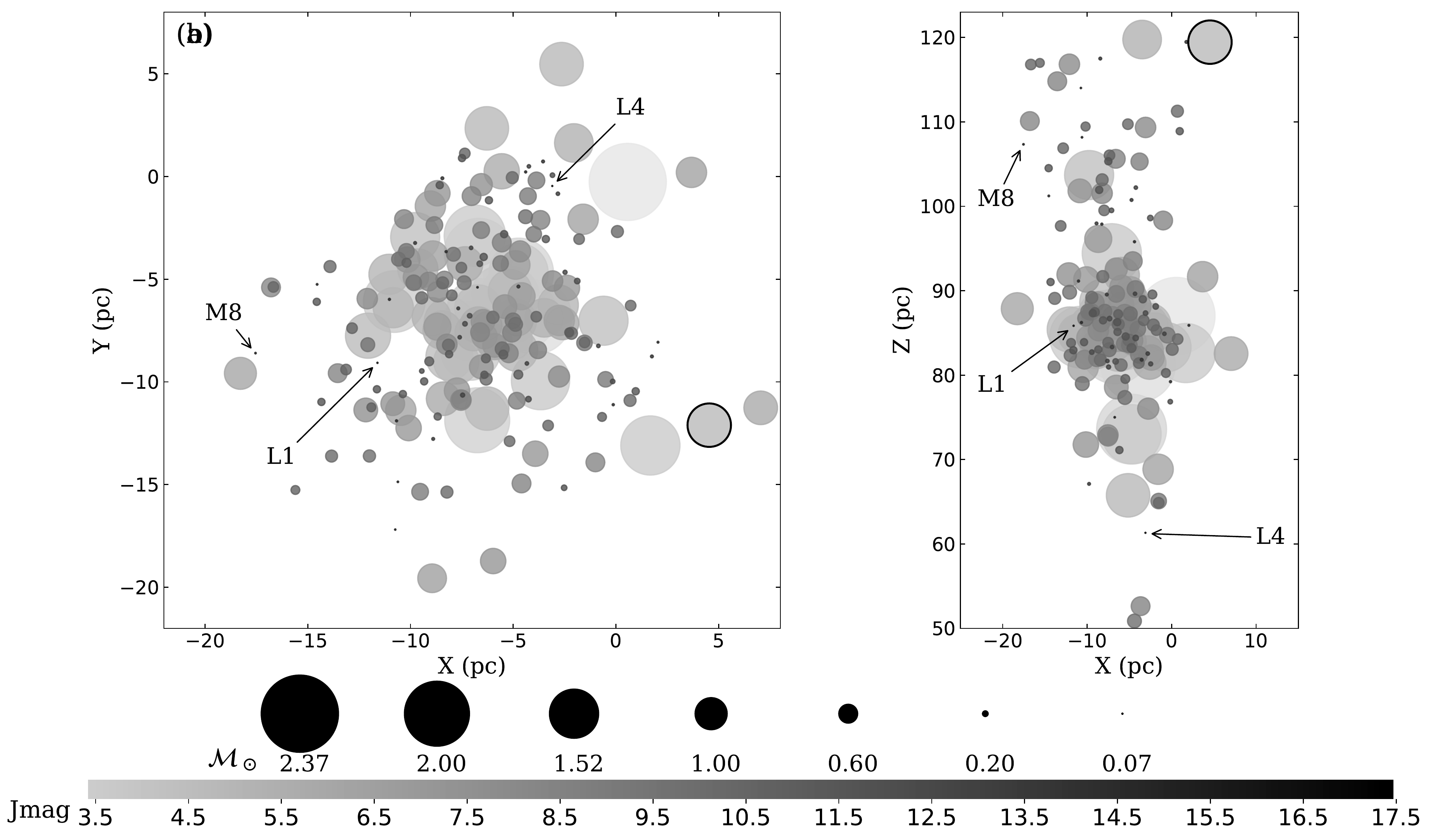}
	\caption{The spatial distribution of member candidates in the Galactic 
	(a)~$X$-$Y$ coordinates, and (b)~$X$-$Z$ coordinates. The size and 
	shade of the circle symbol represent the mass and the $J$ mag, respectively.  
	The three spectroscopically confirmed brown dwarfs reported here are labeled.  
	The one encircled with a dark boundary marks HD\,109390, a member proposed 
	by \citet{kra07} but with a distance marginally outside our selection range (see 
	Table~\ref{tab:beyond5}.)
	}
  \label{fig:3d}
\end{figure*}

Adopting a distance 85~pc, the angular radius $5\degr$ of the cluster corresponds to a linear 
radius $\sim7$~pc, to be compared with the tidal radius of the cluster 6--7~pc \citep{cas06,kra07}.
The Galactic distribution of our candidates, including those in Table~\ref{tab:members} 
and Table~\ref{tab:beyond5}, is illustrated in Figure~\ref{fig:3d}.  
Given the high Galactic latitude ($\ell\approx+84\degr$) position of the cluster, 
the distribution in the $X$-$Y$ plane, i.e., the top view on the Galactic plane, resembles that 
appears in the celestial sphere, e.g., in right ascension 
and declination coordinates, for which members are concentrated within a linear extent about 15~pc 
n roughly circular shape.  The effect of mass segregation is clearly manifest, namely, 
with more massive members (with larger-sized and lighter-shadowed symbols) 
concentrating more toward the center.  The number of members is markedly reduced beyond $\sim10$~pc 
radius, indicative that the cluster is not more extended as projected in the sky than  
the sky coverage in our analysis.      

However, the situation in the $X$-$Z$ plane, namely, the side view of the plane, is different.  
The apparent cone shape from the bottom upwards is the consequence of our pencil-beam view, and the longer 
extent in the $Z$ direction results from our analysis volume as we started out with a larger distance range 
(50--120~pc) to search for candidates than in the angular extent in the sky 
(within $~5\degr$ radius of the UKIDSS/GCS coverage).  That there are more distant sources 
beyond $\sim100$~pc than nearby ones closer than $\sim60$~pc is the consequence of the space 
volume effect, and hence many must be false positives.  
The grouping between $Z\sim65$--95~pc represents the cluster, with a linear size 
twice as extended as in the $X$-$Y$ plane, stretching toward the Galactic plane.  
\citet{ode98} reported a heliocentric space motion for the cluster 
$(U,V,W)=(-2.3,-5.5,-0.7)$~km~s$^{-1}$ with an 
error of 0.2~km~s$^{-1}$ in each component.  As such, the motion of the cluster is primarily 
along $V$, i.e., in the Galactic rotation, and the cluster is almost at its highest location 
above the plane.  The prolate spheroidal shape is likely the consequence of the 
tidal pull by the disk.  

\subsection{The Statistical Sample of Members}

Analysis of the control field (see \S\ref{sec:bright}) with the same selection criteria 
led to 14 false positives.  Considering the sky area of the control field relative to 
the cluster region, and assuming the same field distribution toward the cluster region as in 
the control field, this 
means out of the 192 candidates, there are roughly 44 field stars that coincidentally share the same 
ranges of proper motions, distances, and color-magnitude relation as true members, but are 
not physically associated with the cluster.  
Additional scrutiny of members against field stars in the same volume would have to rely 
on metallicity or chemical abundances, e.g., \citet[][by lithium abundances]{for01}.  For the faint sample, no 
control field is available because of the limit of the UKIDSS/GCS spatial coverage.  
After field subtraction for the bright sample, plus the entire faint candidate sample, there are 
148 members.  We  emphasize that this is a statistical sample in the sense that out of the 196 candidates, 
there are 148 true members, but we do not know for sure individually which ones are true members. 
This is the sample we use to derive member statistics such as the luminosity function, mass function, 
total stellar mass, etc.

In studies of cluster membership, some researchers would assign a membership probability 
for each star, considering its location relative to the cluster center, proper motions, radial velocity, 
etc., and decide on a threshold probability for membership.  As such, the probabilistic nature 
has to be taken into account when the total mass and the luminosity function are derived, e.g., a candidate 
with an 80\% probability should have its mass weighted 
by 0.8 and be counted as 0.8 stars. The uncertainty arises then if a 0.8 star is really worth twice 
as a 0.4 star in derivation of cluster parameters.  In contrast, our analysis exploits 
a control field to remove sample contaminations.  The likelihood of a star as a member 
in the cluster region depends on the number of false positives in the control field.  
As seen in Figure~\ref{fig:bright}, our member list is highly reliable for almost the entire bright \
sample, i.e. for candidates more massive than $\sim0.1$--$0.2~M_\sun$ ($J\lesssim13$~mag), with only a few 
false positives.  

\subsubsection{Luminosity Function and Mass Function}\label{sec:lfmf}

\begin{figure}[htb!]
	\centering
	\includegraphics[width=0.9\columnwidth,angle=0]{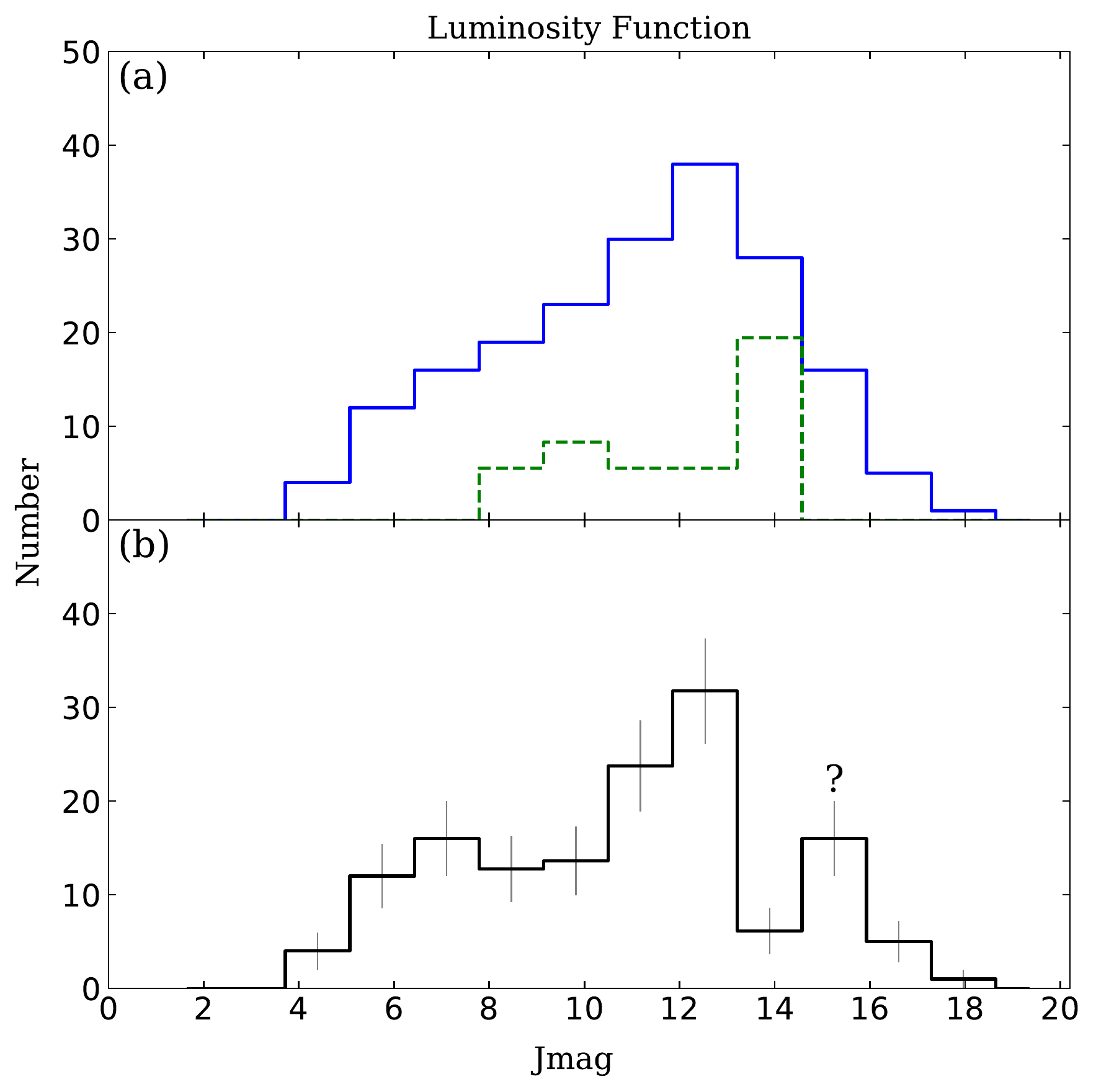}
	\caption{(a)~The $J$-band magnitude distributions toward the cluster region (histogram in 
	solid lines) and toward the control field (in dashed lines).  No control field is 
	available for the faint sample.  
	(b)~The cluster's luminosity function.  It is derived after field subtraction for the bright sample.  
	For $J$=14--16~mag, it is uncertain, thus annotated by a question mark, because of no field subtraction.  
	Fainter than $\sim J=16$~mag, field confusion is low, so the luminosity function is expected to be 
	reliable.    
	}
  \label{fig:jlumfunc}
\end{figure}

The cluster's $J$-band luminosity function is depicted in Figure~\ref{fig:jlumfunc}.  For  
the bright members, the luminosity function has been derived 
by subtraction of the $J$-magnitude distribution toward the 
cluster region by that toward the control field.  The luminosity function at the faint end, 
lacking a control field, is given as is.  This does not affect our result significantly 
because each of our substellar candidates has been examined spectroscopically so contamination 
is expected to be low. Moreover, the unique colors at very low masses, e.g., in $Z-K$ in 
Figure~\ref{fig:faint}, would 
result in little field confusion, hence creditable candidacy. The luminosity function derived 
in this work, therefore, is reliable except for $J=$14--16~mag which has not been corrected for 
field subtraction. 

The $J$-band luminosity function of the cluster increases with magnitude up to $J~\sim12$~mag, 
and falls off rapidly toward fainter magnitudes.  In the $K$-band luminosity of Coma Ber derived
by \citet{cas06}, they found a paucity around $K\sim8$--12~mag, but otherwise no difference fainter 
than $K\sim12$~mag between the cluster and a controlled sample chosen by proper motions.  
Our results show no such shortfall.  

\begin{figure}[htb!]
	\centering
	\includegraphics[width=0.9\columnwidth,angle=0]{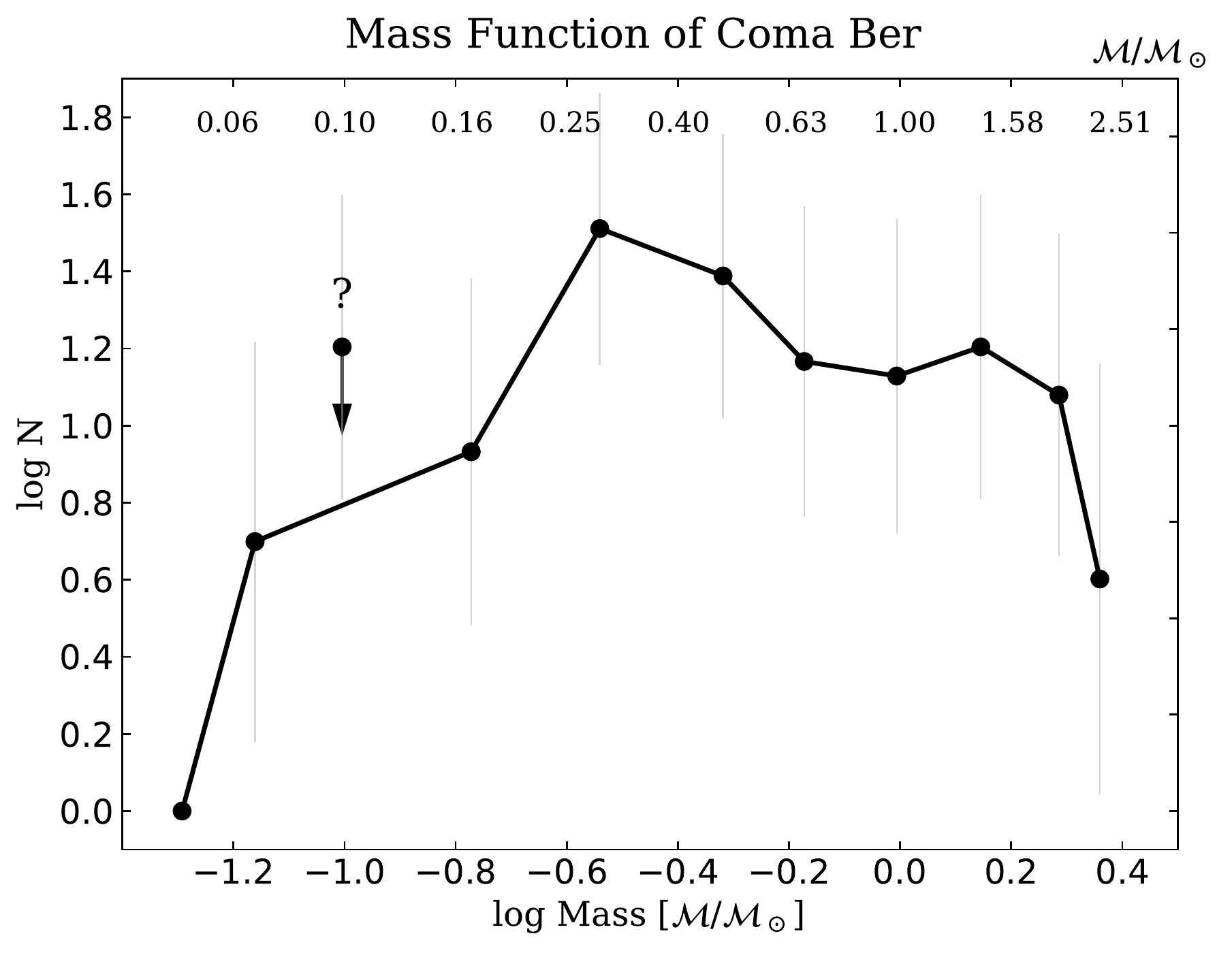}
	\caption{The mass function derived by the cluster's $J$-band luminosity function exhibited  
	in Fig.~\ref{fig:jlumfunc}b with mass estimated by the PARSEC or Dusty models. The data point 
	around 0.1~$M_\sun$, considered unreliable because of the uncertainty in the corresponding 
	luminosity function, is labelled by a question mark.
	}
  \label{fig:mf}
\end{figure}

The present-day mass function of Coma Ber is exhibited in Figure~\ref{fig:mf}.  We convert 
from the $J$ magnitude to mass according to PARSEC or, toward the fainter magnitudes, 
the AMES-Dusty model, based on the cluster luminosity function shown in Figure~\ref{fig:jlumfunc}b.  
The number of members increases in general from high toward low masses until about 
0.3~$M_\sun$, a phenomenon commonly seen in star clusters or associations \citep{bas10}.  
A linear least-squares fit gives a slope, in the sense of $dN/dm = m^{-\alpha}$, 
$\alpha\approx 0.49\pm0.03$, 
if the two most massive bins for post-main sequence objects with uncertain masses are excluded.  
This is close to the value reported by 
\citet{kra07} $\alpha=0.6\pm 0.3$ for 0.1--1~$M_\sun$, but much shallower than either 
the nominal Salpeter $\alpha=2.35$ initial mass function in the solar neighborhood for 1--10~$M_\sun$, 
the present-day mass function of field M dwarfs $\alpha\approx1.3$ for 0.1--0.7~$M_\sun$ 
\citep{rei02}, or in nearby young star clusters or associations $\alpha\approx1.3$~$M_\sun$ 
\citep{hil04}.  

Currently the statistics of the substellar population in star clusters are poorly constrained; 
less than a handful stellar systems have been surveyed comprehensively, and even these 
may be subjected to contamination.  Furthermore, the age dependence 
(hence model dependence) of the spectral type with mass hampers derivation 
of a reliable substellar mass function. \citet{mel12} derived a mass function 
with $\alpha\approx0.6$ from 0.2~$M_\sun$ to 0.14~$M_\sun$, and $\alpha\approx0$ toward
lower-masses from 0.14~$M_\sun$ to 0.06~$M_\sun$.  The sky area of their analysis, $\sim2\fdg7$ in radius, 
however, covered only part of the core of the cluster. Moreover, lacking a control sample, 
their member list could be considerably polluted.  Among the five photometric brown dwarf candidates these 
authors proposed (their Table~2), F9$-$1134 is not detected by UKIDSS/GCS; 
C3$-$250 satisfies the proper motion criterion, but has an inconsistent $Z-K$ color; 
both D3$-$1251, with $(\mu_\alpha \cos\delta \approx \mu_\delta) \approx -130$~mas~yr$^{-1}$,  
and G1$-$3083, with $(\mu_\alpha \cos\delta, \mu_\delta) \approx (-100, -79)$~mas~yr$^{-1}$, 
have too large proper motions as members.  Only E3$-$5219 turns out to be an M9 member \citep{cas05,cas14}, 
recovered also by us (No.~176).  

For our sample, the mass function below 
0.3~$M_\sun$ declines monotonically with decreasing mass, and continues into 
the substellar regime, with a slope $\alpha=-1.69\pm0.14$, if the anomaly near 0.1~$M_\sun$ 
corresponding to the unreliable bin in the luminosity function, is excluded.  
This contrasts with the flat slope that \citet{kra07} found in Coma Ber, or \citet{gol13} derived
in Hyades, which is also an intermediate-age cluster \citep[650~Myr][]{per98}.  The slopes at the low masses 
have been found even to be positive in some very young star clusters (a few Mys), e.g., 
$\alpha=0.3$--0.8 for $\lambda$ Orionis \citep{bay11}, or $\alpha\sim0.5$ for 
$\rho$ Ophiuchi \citep{luh99,bas10}. However, as we learn in our study, recognition of true members 
is susceptible to field confusion.  The situation must be escalated in star-forming regions where 
dust extinction, often variable, becomes excessive.    

\subsubsection{Dynamical Status}\label{sec:dyn}

In terms of the number of members, Coma Ber is definitely underpopulated. 
As presented above, of the 192 candidates we identified, 44 are likely field contaminations, 
leaving 148 members.  On the one hand, there may be faint members in the UKIDSS/GCS void regions,  
so not found by us.  On the other, the number of members must decrease if a control sample is 
available for the faint sample.  \citet{art66} estimated a core radius 2.6~pc and a halo radius 7~pc, 
similar to those of the Pleiades and Praesepe, both of similar ages, but Coma Ber contains half 
the number of members in the core with a relatively 
enriched halo population.  Adopting the isochrone masses for the post-main sequence members, 
the total mass of the 148 members in Coma Ber is $\sim102~M_\sun$.  
This is comparable to that \citet{kra07} reported of the 145 members earlier than M6 amounting to 
a total stellar mass $\sim 112~M_\sun$, and to that of the $102~M_\sun$ estimated by \citet{cas06}.  
Given an effective radius 10~pc within which the majority of the members 
are located, the stellar mass density is 0.024~$M_\sun$~pc$^{-3}$, one order lower
than the threshold of 0.1~$M_\sun$~pc$^{-3}$ necessary to remain dynamically stable against 
tidal disruption in the solar neighborhood \citep{bok34,lad84}.  With a high Galactic 
latitude, Coma Ber has an almost null radial velocity (RV), with an RV dispersion 
$\approx0.5$~km~s$^{-1}$ \citep{ode98}.  The proper motion data are not accurate 
enough to estimate the tangential velocity dispersion, but if we assume a space velocity dispersion 
twice as that of the RV, the cluster would have a kinetic energy comparable to the gravitational 
energy, again suggestive of a marginally bound state.  Coma Ber therefore must be disintegrating, 
evidenced also by its overall shape stretching along the Galactic motion of the cluster, 
and with a distribution of comoving, escaped stars on average fainter than members in the 
core and halo \citep{ode98}.

\section{Summary}\label{sec:summary}

We have identified and characterized the stellar and substellar member 
candidates of the Coma star cluster on the basis of photometry, colors, and 
proper motions by 2MASS, UKIDSS, and URAT1, plus distance information from {\it Gaia}/DR2.  
Out of the 192 candidates found, after field contamination is considered, 148 true members 
are expected.  The candidate 
list is largely complete within a $5\degr$ radius of the cluster, and to our knowledge is the 
most reliable to date in the literature.  We have determined the age of Coma Ber to 
be about 800~Myr, older than previously adopted.  The cluster has a shallower main-sequence 
mass function than in the field.  The mass function peaks around 0.3~$M_\sun$, and 
decreases rapidly toward the brown dwarf masses.  The cluster is mass segregated, and 
has a shape elongated toward the Galactic plane, in the process of disintegration.  
There are nine substellar members, 
six known to be of late-M types and we have confirmed their membership.  
In addition, three brown dwarf members have been spectroscopically confirmed to 
be an M8, an L1 and an L4, extending from the previously known late-M type 
for the first time to the mid-L spectral type in this elusive star cluster. 

\acknowledgments

SYT, WPC, and PSC acknowledge the financial support of the grants 
MOST 106-2112-M-008-005-MY3 and MOST 105-2119-M-008-028-MY3.
This research uses data obtained through the Telescope Access Program (TAP), 
which has been funded by the National Astronomical Observatories of China, the 
Chinese Academy of Sciences, and the Special Fund for Astronomy from the Ministry of Finance.
This study was supported by Sonderforschungsbereich SFB 881 ``The Milky Way System'' 
(subproject B7) of the German Research Foundation (DFG). 
This work has made use of data from the European Space Agency (ESA)
mission {\it Gaia} (\url{https://www.cosmos.esa.int/gaia}), processed by
the {\it Gaia} Data Processing and Analysis Consortium (DPAC,
\url{https://www.cosmos.esa.int/web/gaia/dpac/consortium}). 
Funding for the DPAC has been provided by national institutions, in particular
the institutions participating in the {\it Gaia} Multilateral Agreement.
This research has made use of the 2MASS, SDSS, UKIDSS, {\it WISE}, PS1 data, 
as well as the VizieR catalogue access tool and the SpeX Prism Spectral 
Libraries. 
We thank the anonymous referee whose constructive suggestions greatly improved the quality
of the paper.

{}


\newpage
\begin{longrotatetable}
\begin{deluxetable*}{cccc ccccc  RRR c rrc}
	\tablecaption{Post-Main Sequence Stars
		\label{tab:postMS}
		}
	\tabletypesize{\scriptsize}
 	\tablehead{
 	\colhead{Name} & \colhead{R.A. (2000)} & \colhead{Decl. (2000)} & \colhead{J} & \colhead{eJ} & \colhead{H} & \colhead{eH} & \colhead{K} & \colhead{eK} &
  	\colhead{$\mu_\alpha \cos\delta$} & \colhead{$\mu\delta$ } &\colhead{$\Delta\mu$} & \colhead{SpTy} & 
  	\colhead{B} & \colhead{V} & \colhead{Comments} \\
	\colhead{ } & \colhead{ (deg) } & \colhead{ (deg) } & \colhead{ (mag) } & \colhead{ (mag) } & \colhead{ (mag) } & 
	\colhead{ (mag) } & \colhead{ (mag) } & \colhead{ (mag) } & 	\colhead{(mas~yr$^{-1}$)} &\colhead{(mas~yr$^{-1}$)} &
	\colhead{(mas~yr$^{-1}$)}  & \colhead{} & 
	\colhead{ (mag) } & \colhead{ (mag) } & \colhead{ } 
	  }
	\startdata
12 Com & 185.62626 & +25.84614 & 3.781 & 0.254 & 3.401 & 0.216 & 3.236 & 0.244 &  -10.9 & -9.6 & 0.5 & F6\,III+A3V & 5.30  & 4.81 &  \\
14 Com & 186.60021 & +27.26820 & 4.409 & 0.240 & 4.235 & 0.194 & 4.149 & 0.036 & -16.0 & -13.4 & 0.4 & F0p       & 5.22 & 4.95  & \\
16 Com & 186.74703 & +26.82568 & 4.796 & 0.192 & 4.727 & 0.020 & 4.649 & 0.024 & -11.5 & -9.2 & 0.4 & A4\,V       & 5.05 & 4.96 & \\
31 Com & 192.92465 & +27.54068 & 3.629 & 0.292 & 3.367 & 0.218 & 3.260 & 0.286 & -11.0 & -8.3 & 0.3 & G0\,IIIp   & 4.39 & 4.94  & Outside 5 degs\\
\hline
18 Com & 187.36261 & +24.10894 & 4.864 & 0.194 & 4.572 & 0.036 & 4.574 & 0.288 & -17.6 & 0.7 & 0.2 & F5\,IV   & 5.90 & 5.47  & not a member\\
	\enddata
\end{deluxetable*} 
\end{longrotatetable}

\newpage

\begin{longrotatetable}
\begin{deluxetable}{rcc cccccc RRR rrc}
\tablecaption{Candidate Members of Coma Ber \label{tab:members}}

\tabletypesize{\scriptsize}
 \tablehead{
	 \colhead{No.} & \colhead{R.A. (2000)} & \colhead{Decl. (2000)} & \colhead{J} & \colhead{Jerr} & \colhead{H}  & 
	 \colhead{Herr} & \colhead{Ks} & \colhead{Kserr} &
	\colhead{$\mu_\alpha \cos\delta$} & \colhead{$\mu\delta$ }	&\colhead{$\Delta\mu$} & 
	\colhead{Dist} & \colhead{Ref.$^\dagger$} & \colhead{Data$^\ddagger$}  \\ 
 	 \colhead{ }  & \colhead{(deg)} & \colhead{ (deg)} & \colhead{ (mag)} & \colhead{(mag)} & \colhead{(mag)}  & 
	 \colhead{(mag)} & \colhead{(mag)} & \colhead{(mag)} &
	\colhead{mas~yr$^{-1}$} & \colhead{mas~yr$^{-1}$} & \colhead{ mas~yr$^{-1}$} & \colhead{ (pc)} & 
	\colhead{ } & \colhead{ }  \\
  \colhead{(1)}  & \colhead{(2)} & \colhead{(3)} & \colhead{(4)} & \colhead{(5)} & \colhead{(6)} & 
 \colhead{(7)} &  \colhead{(8)} & \colhead{(9)} & \colhead{(10)} & \colhead{(11)} & \colhead{(12)} & \colhead{(13) } 
	& \colhead{(14) } & \colhead{(15) } 
	   } 
\startdata
\multicolumn{15}{c}{ Distance by Parallax}  \\ 
\hline
1 & 181.09398 & 24.02144 & 12.787 & 0.002 & 12.297 & 0.001 & 11.962 & 0.001 & -12.3 & -9.4 & 0.2 & 85.4 & c & 3 \\
2 & 181.09694 & 24.82066 & 8.750 & 0.023 & 8.414 & 0.018 & 8.343 & 0.019 & -12.5 & -9.9 & 0.1 & 82.4 & cf & 1 \\
3 & 181.36050 & 26.33820 & 10.524 & 0.031 & 9.871 & 0.031 & 9.676 & 0.035 & -3.8 & -17.8 & 0.1 & 74.1 &  & 1 \\
4 & 181.36131 & 26.33937 & 10.783 & 0.022 & 10.152 & 0.018 & 9.937 & 0.019 & -4.9 & -20.6 & 0.1 & 73.9 &  & 1 \\
5 & 181.38475 & 22.90031 & 13.488 & 0.002 & 12.966 & 0.002 & 12.656 & 0.002 & -7.8 & -12.5 & 0.2 & 92.8 &  & 3 \\
6 & 181.39774 & 25.02538 & 15.531 & 0.007 & 14.980 & 0.007 & 14.477 & 0.008 & -12.9 & -9.7 & 1.1 & 87.7 &  & 3 \\
7 & 181.53316 & 24.35440 & 12.719 & 0.022 & 12.118 & 0.021 & 11.864 & 0.020 & -11.3 & -17.4 & 0.2 & 119.0 &  & 1 \\
8 & 181.63802 & 23.86454 & 13.482 & 0.027 & 12.958 & 0.032 & 12.592 & 0.025 & -12.4 & -9.6 & 0.2 & 84.4 & c & 1 \\
9 & 181.79992 & 26.06119 & 13.116 & 0.024 & 12.572 & 0.023 & 12.278 & 0.022 & -12.2 & -9.2 & 0.2 & 84.3 & c & 1 \\
10 & 181.92396 & 24.21646 & 9.959 & 0.022 & 9.332 & 0.022 & 9.177 & 0.017 & -9.3 & -8.7 & 0.1 & 93.4 & c & 1 \\
11 & 181.99044 & 25.58648 & 9.534 & 0.022 & 9.028 & 0.019 & 8.906 & 0.017 & -9.9 & -8.3 & 0.2 & 92.7 & cf & 1 \\
12 & 182.03835 & 24.72508 & 13.279 & 0.024 & 12.711 & 0.023 & 12.374 & 0.023 & -11.7 & -8.0 & 0.3 & 85.2 & c & 1 \\
13 & 182.37284 & 29.32707 & 10.790 & 0.019 & 10.126 & 0.021 & 9.913 & 0.017 & -17.4 & -15.4 & 0.3 & 94.8 &  & 1 \\
14 & 182.67053 & 26.42563 & 14.321 & 0.004 & 13.802 & 0.003 & 13.352 & 0.003 & -11.3 & -8.3 & 0.6 & 82.0 &  & 3 \\
15 & 182.69202 & 27.28147 & 5.659 & 0.029 & 5.668 & 0.031 & 5.600 & 0.033 & -12.4 & -9.5 & 0.2 & 86.9 & ab & 1 \\
16 & 182.78072 & 25.99015 & 8.387 & 0.023 & 8.115 & 0.017 & 8.073 & 0.033 & -11.9 & -9.3 & 0.1 & 86.8 & bcf & 1 \\
17 & 182.89646 & 29.37899 & 9.575 & 0.022 & 9.053 & 0.027 & 8.979 & 0.024 & -11.8 & -8.8 & 0.1 & 89.3 & cf & 1 \\
18 & 182.94819 & 24.87139 & 13.938 & 0.003 & \nodata & \nodata & 13.075 & 0.003 & -12.3 & -10.3 & 0.2 & 83.8 & c & 3 \\
19 & 183.10368 & 27.38006 & 7.274 & 0.024 & 7.130 & 0.021 & 7.082 & 0.021 & -12.2 & -9.4 & 0.1 & 85.7 & abc & 1 \\
20 & 183.17821 & 25.22843 & 13.406 & 0.022 & 12.835 & 0.026 & 12.518 & 0.025 & -11.8 & -7.9 & 0.2 & 88.4 & c & 1 \\
21 & 183.22177 & 26.25037 & 9.577 & 0.018 & 9.106 & 0.016 & 8.990 & 0.018 & -12.1 & -9.5 & 0.1 & 86.3 & bcf & 1 \\
22 & 183.33962 & 23.37229 & 12.619 & 0.023 & 12.053 & 0.023 & 11.821 & 0.022 & 4.3 & -10.3 & 0.1 & 99.0 &  & 1 \\
23 & 183.43287 & 22.88796 & 7.211 & 0.027 & 7.052 & 0.017 & 6.990 & 0.033 & -13.5 & -8.4 & 0.2 & 86.6 & cdf & 1 \\
24 & 183.44299 & 30.34077 & 13.582 & 0.027 & 13.012 & 0.033 & 12.681 & 0.026 & -11.2 & -19.3 & 0.7 & 99.8 &  & 1 \\
25 & 183.61048 & 23.23399 & 11.806 & 0.024 & 11.244 & 0.031 & 10.972 & 0.022 & -11.3 & -8.9 & 0.1 & 91.0 &  & 1 \\
26 & 183.81815 & 29.35072 & 12.506 & 0.021 & 11.952 & 0.021 & 11.684 & 0.018 & -11.8 & -9.3 & 0.1 & 87.4 & c & 1 \\
27 & 183.88110 & 25.06697 & 13.022 & 0.021 & 12.423 & 0.020 & 12.159 & 0.023 & -12.1 & -9.1 & 0.2 & 88.4 & c & 1 \\
28 & 184.00348 & 28.09662 & 11.077 & 0.024 & 10.525 & 0.033 & 10.240 & 0.018 & -10.9 & -7.7 & 0.6 & 86.8 & bc & 1 \\
29 & 184.03485 & 25.76034 & 7.232 & 0.026 & 7.117 & 0.047 & 7.036 & 0.017 & -12.2 & -10.5 & 0.1 & 85.3 & abcf & 1 \\
30 & 184.09773 & 28.41948 & 13.866 & 0.026 & 13.263 & 0.027 & 13.010 & 0.027 & -10.4 & 2.0 & 0.3 & 85.2 & ce & 1 \\
31 & 184.15536 & 26.89944 & 12.225 & 0.022 & 11.680 & 0.024 & 11.420 & 0.018 & -12.3 & -9.4 & 0.1 & 86.8 & bc & 1 \\
32 & 184.17173 & 26.77747 & 13.141 & 0.020 & 12.563 & 0.024 & 12.301 & 0.023 & -11.7 & -10.4 & 0.2 & 85.9 & ce & 1 \\
33 & 184.33914 & 25.48123 & 13.274 & 0.023 & 12.671 & 0.023 & 12.394 & 0.021 & -11.9 & -8.3 & 0.2 & 87.3 & ce & 1 \\
34 & 184.42752 & 30.98329 & 12.900 & 0.002 & 12.389 & 0.001 & 12.054 & 0.001 & -21.8 & -13.6 & 0.1 & 81.1 &  & 3 \\
35 & 184.46208 & 25.57132 & 7.086 & 0.018 & 6.982 & 0.024 & 6.928 & 0.017 & -11.3 & -9.6 & 0.1 & 89.4 & abcf & 1 \\
36 & 184.55316 & 26.82093 & 12.016 & 0.022 & 11.459 & 0.021 & 11.153 & 0.020 & -11.0 & -6.3 & 0.7 & 78.0 & bc & 1 \\
37 & 184.59100 & 25.40361 & 14.498 & 0.004 & \nodata & \nodata & 13.623 & 0.004 & -13.6 & -9.9 & 0.4 & 82.4 & e & 3 \\
38 & 184.59132 & 27.74508 & 12.992 & 0.024 & 12.392 & 0.030 & 12.181 & 0.018 & -12.2 & -9.8 & 0.2 & 83.9 & c & 1 \\
39 & 184.65070 & 23.12004 & 7.635 & 0.019 & 7.386 & 0.023 & 7.303 & 0.020 & -15.0 & -10.5 & 0.6 & 85.7 & abcf & 1 \\
40 & 184.75610 & 24.84615 & 7.837 & 0.056 & 7.555 & 0.036 & 7.537 & 0.018 & -16.2 & -14.3 & 0.2 & 85.0 & abc & 1 \\
41 & 184.75839 & 26.00832 & 6.082 & 0.026 & 6.004 & 0.049 & 5.981 & 0.023 & -12.5 & -8.5 & 0.3 & 86.4 & abf & 1 \\
42 & 184.81311 & 28.28071 & 12.385 & 0.029 & 11.769 & 0.030 & 11.522 & 0.022 & -3.0 & -15.6 & 0.1 & 110.6 &  & 1 \\
43 & 184.82992 & 23.03464 & 5.948 & 0.021 & 5.969 & 0.021 & 5.908 & 0.020 & -12.6 & -9.6 & 0.1 & 85.4 & abf & 1 \\
44 & 184.86810 & 24.28422 & 7.867 & 0.019 & 7.557 & 0.044 & 7.492 & 0.027 & -12.8 & -2.9 & 0.3 & 83.8 & abcf & 1 \\
45 & 184.90816 & 26.57905 & 12.776 & 0.029 & 12.239 & 0.033 & 11.917 & 0.023 & -16.6 & -9.6 & 0.7 & 87.9 & bc & 1 \\
46 & 184.96089 & 28.46432 & 6.209 & 0.019 & 6.192 & 0.034 & 6.135 & 0.024 & -12.3 & -9.0 & 0.1 & 85.6 & abf & 1 \\
47 & 184.96843 & 31.16615 & 15.278 & 0.006 & 14.718 & 0.006 & 14.284 & 0.006 & -11.9 & -10.6 & 0.8 & 80.0 &  & 3 \\
48 & 185.06034 & 25.43537 & 12.214 & 0.001 & \nodata & \nodata & 11.464 & 0.001 & -11.8 & -8.7 & 0.1 & 92.4 & c & 3 \\
49 & 185.16139 & 29.65041 & 12.392 & 0.024 & 11.790 & 0.028 & 11.502 & 0.025 & -7.7 & -23.4 & 0.2 & 65.4 &  & 1 \\
50 & 185.16438 & 29.64768 & 11.495 & 0.022 & 10.871 & 0.028 & 10.633 & 0.025 & -7.8 & -23.9 & 0.1 & 65.6 &  & 1 \\
51 & 185.18983 & 25.76584 & 7.974 & 0.021 & 7.740 & 0.033 & 7.649 & 0.026 & -12.2 & -8.3 & 0.1 & 84.4 & abcf & 1 \\
52 & 185.31501 & 26.15388 & 9.614 & 0.019 & 9.087 & 0.024 & 8.972 & 0.020 & -11.8 & -9.4 & 0.1 & 84.9 & bcdf & 1 \\
53 & 185.36137 & 24.99700 & 6.792 & 0.020 & 6.742 & 0.026 & 6.664 & 0.017 & -12.0 & -9.5 & 0.1 & 87.7 & abf & 1 \\
54 & 185.41950 & 27.13081 & 10.712 & 0.023 & 10.125 & 0.025 & 9.887 & 0.019 & -11.8 & -8.9 & 0.1 & 87.3 &  & 1 \\
55 & 185.42258 & 21.68386 & 15.651 & 0.008 & 15.119 & 0.009 & 14.661 & 0.011 & -10.6 & -8.2 & 1.3 & 102.4 &  & 3 \\
56 & 185.45422 & 26.54907 & 8.214 & 0.026 & 7.863 & 0.027 & 7.857 & 0.027 & -12.5 & -8.1 & 0.1 & 86.5 & abcf & 1 \\
57 & 185.48395 & 27.30948 & 7.565 & 0.024 & 7.399 & 0.042 & 7.325 & 0.020 & -13.4 & -9.0 & 0.1 & 85.3 & abcf & 1 \\
58 & 185.55134 & 30.85934 & 13.940 & 0.028 & 13.344 & 0.026 & 13.030 & 0.028 & -22.5 & -14.0 & 0.3 & 77.5 &  & 1 \\
59 & 185.56031 & 25.44897 & 13.980 & 0.028 & 13.397 & 0.032 & 13.080 & 0.024 & -11.6 & -9.5 & 0.3 & 87.3 & ce & 1 \\
60 & 185.60311 & 22.46410 & 7.604 & 0.019 & 7.394 & 0.018 & 7.387 & 0.020 & -11.8 & -9.9 & 0.1 & 85.7 & abcf & 1 \\
61 & 185.62626 & 25.84614 & 3.781 & 0.254 & 3.401 & 0.216 & 3.236 & 0.244 & -10.8 & -9.5 & 0.5 & 84.5 & ab & 1 \\
62 & 185.63074 & 25.82848 & 7.406 & 0.029 & 7.150 & 0.026 & 7.024 & 0.018 & -5.0 & -5.4 & 1.2 & 66.2 & a & 1 \\
63 & 185.66225 & 27.77868 & 14.418 & 0.004 & 13.904 & 0.003 & 13.524 & 0.004 & -11.8 & -8.3 & 0.4 & 90.2 & ce & 3 \\
64 & 185.75796 & 21.56379 & 12.469 & 0.024 & 11.874 & 0.023 & 11.661 & 0.020 & -23.2 & -0.9 & 0.1 & 118.1 &  & 1 \\
65 & 185.76285 & 22.13122 & 11.863 & 0.022 & 11.294 & 0.022 & 11.078 & 0.020 & -1.4 & -22.6 & 0.1 & 79.8 &  & 1 \\
66 & 185.78496 & 25.85135 & 8.027 & 0.020 & 7.762 & 0.021 & 7.685 & 0.023 & -12.3 & -8.7 & 0.3 & 87.7 & abcf & 1 \\
67 & 185.80006 & 23.93748 & 12.204 & 0.022 & 11.614 & 0.022 & 11.384 & 0.018 & -11.7 & -9.1 & 0.2 & 89.9 & bc & 1 \\
68 & 185.80644 & 26.03844 & 12.025 & 0.020 & 11.519 & 0.016 & 11.226 & 0.019 & -14.9 & -12.0 & 0.5 & 81.7 & c & 1 \\
69 & 185.86747 & 25.89440 & 9.920 & 0.022 & 9.354 & 0.022 & 9.260 & 0.020 & -12.0 & -9.8 & 0.1 & 86.2 & bcd & 1 \\
70 & 185.92082 & 26.97991 & 7.461 & 0.021 & 7.334 & 0.080 & 7.253 & 0.021 & -12.2 & -7.9 & 0.1 & 87.5 & abcf & 1 \\
71 & 185.92423 & 26.60147 & 8.137 & 0.023 & 7.791 & 0.033 & 7.739 & 0.020 & -13.0 & -9.5 & 0.2 & 92.4 & abcf & 1 \\
72 & 185.94668 & 23.24565 & 9.677 & 0.021 & 9.129 & 0.022 & 9.018 & 0.018 & -12.6 & -10.0 & 0.1 & 85.1 & bf & 1 \\
73 & 185.95401 & 24.13215 & 10.474 & 0.024 & 9.880 & 0.030 & 9.674 & 0.019 & -17.9 & -10.9 & 0.4 & 85.9 & c & 1 \\
74 & 185.98129 & 23.41443 & 11.591 & 0.020 & 10.995 & 0.023 & 10.775 & 0.017 & -11.3 & -9.3 & 0.1 & 88.2 & bc & 1 \\
75 & 186.01438 & 25.85121 & 6.179 & 0.024 & 6.075 & 0.047 & 6.054 & 0.018 & -13.0 & -5.9 & 0.5 & 95.0 & ab & 1 \\
76 & 186.02385 & 26.12857 & 9.080 & 0.027 & 8.762 & 0.065 & 8.611 & 0.021 & -11.7 & -9.0 & 0.1 & 88.9 & acd & 1 \\
77 & 186.04529 & 23.99336 & 12.266 & 0.021 & 11.664 & 0.020 & 11.452 & 0.018 & -12.8 & -9.0 & 0.1 & 82.4 & bc & 1 \\
78 & 186.04673 & 26.88793 & 10.921 & 0.031 & 10.265 & 0.030 & 10.058 & 0.018 & -11.3 & -9.4 & 0.1 & 86.3 &  & 1 \\
79 & 186.07718 & 26.09857 & 4.930 & 0.037 & 4.943 & 0.063 & 4.896 & 0.023 & -24.7 & -10.0 & 0.5 & 86.7 & ab & 1 \\
80 & 186.11157 & 25.58248 & 5.844 & 0.019 & 5.778 & 0.031 & 5.731 & 0.016 & -9.4 & -10.9 & 0.1 & 85.9 & ab & 1 \\
81 & 186.12976 & 25.08834 & 14.777 & 0.005 & 14.242 & 0.004 & 13.902 & 0.005 & -21.9 & -21.2 & 0.4 & 90.1 & ce & 3 \\
82 & 186.18144 & 30.29726 & 11.423 & 0.023 & 10.801 & 0.028 & 10.601 & 0.022 & -11.7 & -10.3 & 0.1 & 85.3 & c & 1 \\
83 & 186.25937 & 25.56064 & 7.051 & 0.018 & 6.849 & 0.016 & 6.762 & 0.031 & -12.6 & -8.2 & 0.1 & 87.1 & abcf & 1 \\
84 & 186.26095 & 26.71060 & 11.621 & 0.019 & 11.028 & 0.016 & 10.791 & 0.020 & -11.3 & -8.8 & 0.1 & 84.0 & bc & 1 \\
85 & 186.34369 & 23.22904 & 7.644 & 0.024 & 7.480 & 0.027 & 7.392 & 0.018 & -11.2 & -10.8 & 0.1 & 86.5 & abf & 1 \\
86 & 186.35434 & 23.84796 & 10.715 & 0.020 & 10.072 & 0.020 & 9.873 & 0.017 & -11.9 & -7.6 & 0.1 & 89.2 & c & 1 \\
87 & 186.46641 & 26.77665 & 7.411 & 0.024 & 7.303 & 0.059 & 7.205 & 0.026 & -13.4 & -8.6 & 0.1 & 86.3 & abcf & 1 \\
88 & 186.47588 & 26.86073 & 11.984 & 0.022 & 11.389 & 0.028 & 11.143 & 0.020 & -11.5 & -7.8 & 0.1 & 87.7 & c & 1 \\
89 & 186.50104 & 24.15579 & 10.979 & 0.021 & 10.356 & 0.029 & 10.142 & 0.025 & -11.6 & -6.6 & 0.1 & 86.7 & bc & 1 \\
90 & 186.53527 & 24.65868 & 11.863 & 0.021 & 11.279 & 0.030 & 11.026 & 0.025 & -11.8 & -9.3 & 0.1 & 83.5 & c & 1 \\
91 & 186.60021 & 27.26820 & 4.409 & 0.240 & 4.235 & 0.194 & 4.149 & 0.036 & -16.0 & -13.4 & 0.5 & 81.6 & ab & 1 \\
92 & 186.66775 & 27.31204 & 12.462 & 0.022 & 11.896 & 0.029 & 11.672 & 0.025 & -13.5 & -9.2 & 0.2 & 81.8 & c & 1 \\
93 & 186.71256 & 26.26715 & 9.855 & 0.022 & 9.275 & 0.026 & 9.156 & 0.020 & -12.1 & -8.0 & 0.1 & 86.0 & bcdf & 1 \\
94 & 186.73599 & 22.67396 & 11.556 & 0.020 & 10.951 & 0.022 & 10.715 & 0.017 & -11.0 & -8.3 & 0.1 & 87.4 & c & 1 \\
95 & 186.74702 & 26.82568 & 4.796 & 0.192 & 4.727 & 0.020 & 4.649 & 0.024 & -11.5 & -9.2 & 0.6 & 85.7 & ab & 1 \\
96 & 186.76784 & 25.68371 & 13.946 & 0.030 & 13.395 & 0.040 & 13.026 & 0.030 & -13.6 & -5.9 & 1.1 & 100.0 & c & 1 \\
97 & 186.77604 & 26.84567 & 8.642 & 0.037 & 8.327 & 0.026 & 8.246 & 0.036 & -12.9 & -7.8 & 0.3 & 89.6 & abcf & 1 \\
98 & 186.83616 & 23.32981 & 8.912 & 0.021 & 8.537 & 0.021 & 8.451 & 0.017 & -12.4 & -9.1 & 0.1 & 84.0 & bcf & 1 \\
99 & 186.90981 & 25.91208 & 6.285 & 0.023 & 6.222 & 0.027 & 6.226 & 0.017 & -12.2 & -9.3 & 0.1 & 84.4 & abf & 1 \\
100 & 186.95118 & 28.19438 & 8.436 & 0.023 & 8.050 & 0.046 & 8.050 & 0.023 & -13.4 & -9.4 & 0.1 & 81.8 & bcf & 1 \\
101 & 187.01885 & 24.35210 & 12.392 & 0.023 & 11.835 & 0.032 & 11.579 & 0.021 & -12.0 & -9.3 & 0.1 & 84.3 & bc & 1 \\
102 & 187.03613 & 24.96473 & 12.507 & 0.026 & 11.903 & 0.031 & 11.666 & 0.023 & -3.3 & -17.7 & 0.1 & 100.0 &  & 1 \\
103 & 187.08792 & 28.04054 & 8.943 & 0.024 & 8.472 & 0.044 & 8.465 & 0.027 & -12.7 & -9.0 & 0.1 & 83.2 & bcf & 1 \\
104 & 187.11487 & 28.56222 & 12.284 & 0.022 & 11.709 & 0.028 & 11.471 & 0.023 & -12.8 & -8.4 & 0.1 & 86.3 & c & 1 \\
105 & 187.14427 & 29.54501 & 14.337 & 0.004 & 13.835 & 0.003 & 13.467 & 0.003 & -12.1 & -8.8 & 0.3 & 85.3 & c & 3 \\
106 & 187.15893 & 26.22693 & 6.137 & 0.052 & 6.022 & 0.029 & 5.994 & 0.017 & -7.8 & -10.2 & 0.2 & 86.4 & abf & 1 \\
107 & 187.18560 & 25.89926 & 6.165 & 0.026 & 6.100 & 0.024 & 6.056 & 0.023 & -22.3 & -17.1 & 0.2 & 73.3 & ab & 1 \\
108 & 187.22784 & 25.91280 & 5.221 & 0.020 & 5.297 & 0.034 & 5.289 & 0.017 & -23.5 & -15.6 & 0.4 & 73.9 & ab & 1 \\
109 & 187.23505 & 26.54925 & 9.208 & 0.026 & 8.768 & 0.031 & 8.661 & 0.023 & -12.9 & -9.2 & 0.1 & 84.3 & bcd & 1 \\
110 & 187.24021 & 27.78009 & 10.989 & 0.023 & 10.349 & 0.028 & 10.185 & 0.022 & -13.8 & -4.9 & 0.1 & 105.7 & bc & 1 \\
111 & 187.33392 & 24.74286 & 13.553 & 0.026 & 12.984 & 0.032 & 12.679 & 0.021 & -12.3 & -8.4 & 0.2 & 82.0 & c & 1 \\
112 & 187.33491 & 28.43433 & 13.025 & 0.023 & 12.426 & 0.033 & 12.214 & 0.020 & -12.0 & -8.4 & 0.2 & 89.9 & c & 1 \\
113 & 187.35376 & 21.78045 & 10.714 & 0.023 & 10.048 & 0.028 & 9.881 & 0.021 & -12.1 & -9.8 & 0.1 & 82.5 &  & 1 \\
114 & 187.42048 & 24.52071 & 8.202 & 0.019 & 7.841 & 0.047 & 7.725 & 0.029 & -11.2 & -8.6 & 0.1 & 87.9 & abcf & 1 \\
115 & 187.52024 & 24.04274 & 11.774 & 0.022 & 11.182 & 0.016 & 10.937 & 0.019 & -12.0 & -8.7 & 0.1 & 88.0 & bc & 1 \\
116 & 187.69229 & 23.76363 & 12.718 & 0.024 & 12.167 & 0.030 & 11.910 & 0.022 & -14.7 & -9.8 & 0.1 & 110.0 & b & 1 \\
117 & 187.73906 & 22.77080 & 11.245 & 0.021 & 10.645 & 0.029 & 10.420 & 0.021 & -12.4 & -9.3 & 0.1 & 82.5 & bc & 1 \\
118 & 187.75230 & 24.56715 & 5.293 & 0.037 & 5.308 & 0.027 & 5.269 & 0.017 & -12.4 & -9.7 & 0.4 & 83.4 & ab & 1 \\
119 & 187.76286 & 27.73031 & 7.612 & 0.019 & 7.463 & 0.055 & 7.404 & 0.018 & -12.4 & -8.6 & 0.1 & 86.2 & abcf & 1 \\
120 & 187.78398 & 24.27641 & 14.402 & 0.004 & 13.869 & 0.003 & 13.460 & 0.003 & -14.0 & -8.4 & 0.4 & 83.7 & c & 3 \\
121 & 187.86551 & 25.39438 & 11.438 & 0.023 & 10.842 & 0.030 & 10.634 & 0.020 & -12.0 & -8.6 & 0.1 & 84.0 & bc & 1 \\
122 & 187.90212 & 24.87421 & 13.452 & 0.026 & 12.878 & 0.028 & 12.594 & 0.025 & -11.8 & -9.0 & 0.2 & 86.6 & ce & 1 \\
123 & 187.96062 & 29.31413 & 6.847 & 0.027 & 6.745 & 0.057 & 6.654 & 0.020 & -10.9 & -6.2 & 0.1 & 83.6 & bf & 1 \\
124 & 188.18704 & 23.41905 & 6.836 & 0.029 & 6.720 & 0.044 & 6.659 & 0.020 & 3.7 & -3.1 & 0.1 & 104.2 &  & 1 \\
125 & 188.25256 & 27.71240 & 9.470 & 0.030 & 8.940 & 0.030 & 8.866 & 0.018 & -13.0 & -9.9 & 0.1 & 82.3 & bcdf & 1 \\
126 & 188.33335 & 22.40649 & 8.855 & 0.019 & 8.470 & 0.023 & 8.402 & 0.020 & -10.3 & -7.4 & 0.1 & 83.9 & cf & 1 \\
127 & 188.36980 & 26.44913 & 15.029 & 0.006 & 14.500 & 0.005 & 14.236 & 0.007 & -20.6 & -6.0 & 1.1 & 101.0 &  & 3 \\
128 & 188.39255 & 24.28296 & 6.032 & 0.030 & 5.988 & 0.031 & 5.989 & 0.026 & -11.9 & -9.0 & 0.1 & 86.7 & abf & 1 \\
129 & 188.42545 & 25.94274 & 9.031 & 0.029 & 8.601 & 0.036 & 8.584 & 0.020 & -16.4 & -9.7 & 0.1 & 90.2 & cf & 1 \\
130 & 188.46503 & 31.11809 & 15.593 & 0.007 & 15.048 & 0.007 & 14.545 & 0.007 & -11.3 & -8.6 & 0.9 & 86.3 &  & 3 \\
131 & 188.63078 & 25.75006 & 10.249 & 0.019 & 9.577 & 0.027 & 9.395 & 0.020 & -13.2 & -9.3 & 0.4 & 84.3 & c & 1 \\
132 & 188.68958 & 27.38686 & 14.217 & 0.003 & 13.703 & 0.002 & 13.324 & 0.004 & -12.9 & -9.4 & 0.2 & 81.5 & c & 3 \\
133 & 188.71789 & 25.15673 & 10.893 & 0.023 & 10.289 & 0.028 & 10.058 & 0.020 & -16.7 & -7.3 & 0.3 & 87.5 & c & 1 \\
134 & 188.72619 & 27.45559 & 7.897 & 0.029 & 7.583 & 0.040 & 7.510 & 0.020 & -16.6 & -10.0 & 0.6 & 120.0 & bc & 1 \\
135 & 188.82267 & 24.46504 & 11.142 & 0.023 & 10.508 & 0.028 & 10.309 & 0.022 & -11.7 & -7.8 & 0.1 & 89.9 & c & 1 \\
136 & 188.89199 & 25.01716 & 13.446 & 0.027 & 12.882 & 0.032 & 12.621 & 0.026 & -12.3 & -8.4 & 0.2 & 84.8 & c & 1 \\
137 & 188.96796 & 27.84477 & 13.851 & 0.027 & 13.228 & 0.032 & 12.955 & 0.028 & -11.8 & -8.5 & 0.2 & 88.3 &  & 1 \\
138 & 189.03665 & 29.80297 & 12.371 & 0.022 & 11.759 & 0.029 & 11.535 & 0.021 & -11.4 & -10.3 & 0.2 & 84.5 & c & 1 \\
139 & 189.31846 & 30.06635 & 15.147 & 0.006 & 14.630 & 0.006 & 14.276 & 0.006 & -2.7 & 5.3 & 0.4 & 119.8 &  & 3 \\
140 & 189.48458 & 25.86257 & 11.491 & 0.024 & 10.893 & 0.033 & 10.684 & 0.021 & -12.7 & -9.0 & 0.1 & 85.6 & c & 1 \\
141 & 189.54769 & 23.55611 & 10.776 & 0.022 & 10.163 & 0.023 & 9.963 & 0.020 & -12.0 & -10.3 & 0.1 & 87.3 &  & 1 \\
142 & 189.69180 & 26.31618 & 13.257 & 0.026 & 12.625 & 0.032 & 12.393 & 0.024 & -12.2 & -8.6 & 0.2 & 89.1 & c & 1 \\
143 & 190.19120 & 27.20596 & 12.407 & 0.029 & 11.849 & 0.032 & 11.586 & 0.024 & -12.8 & -8.5 & 0.1 & 85.4 & c & 1 \\
144 & 190.66186 & 25.16037 & 11.298 & 0.020 & 10.689 & 0.021 & 10.460 & 0.018 & -11.3 & -7.5 & 0.1 & 90.3 & c & 1 \\
145 & 190.77738 & 24.25476 & 12.240 & 0.020 & 11.653 & 0.021 & 11.414 & 0.018 & -12.2 & -8.6 & 0.1 & 84.3 & c & 1 \\
146 & 190.86251 & 23.99538 & 8.207 & 0.021 & 7.895 & 0.021 & 7.853 & 0.021 & 0.7 & -4.2 & 0.1 & 88.3 &  & 1 \\
147 & 190.89413 & 23.43523 & 13.495 & 0.003 & 12.957 & 0.002 & 12.684 & 0.002 & -3.4 & 1.0 & 0.1 & 105.6 &  & 3 \\
148 & 191.06186 & 23.45737 & 12.623 & 0.002 & 12.186 & 0.001 & 11.840 & 0.001 & -25.8 & -7.4 & 0.1 & 106.3 &  & 3 \\
149 & 191.12490 & 24.93406 & 11.262 & 0.022 & 10.696 & 0.030 & 10.452 & 0.021 & -12.2 & -8.8 & 0.1 & 82.5 &  & 1 \\
150 & 191.12880 & 28.16368 & 12.242 & 0.024 & 11.655 & 0.030 & 11.417 & 0.024 & -12.7 & -8.9 & 0.1 & 83.1 & c & 1 \\
151 & 191.13160 & 25.78914 & 14.649 & 0.004 & 14.143 & 0.004 & 13.722 & 0.004 & -12.2 & -7.8 & 0.4 & 82.6 &  & 3 \\
152 & 191.67721 & 25.40008 & 13.938 & 0.027 & 13.351 & 0.031 & 13.027 & 0.037 & -11.9 & -7.2 & 0.2 & 87.4 & c & 1 \\
153 & 191.69063 & 24.98089 & 14.612 & 0.004 & 14.116 & 0.003 & 13.777 & 0.004 & -7.4 & -3.3 & 0.9 & 102.3 &  & 3 \\
154 & 191.84451 & 24.76349 & 15.117 & 0.006 & 14.628 & 0.004 & 14.196 & 0.006 & -11.8 & -9.3 & 0.6 & 81.9 &  & 3 \\
\hline
\multicolumn{15}{c}{ Distance by {\it phot-d}  }  \\ 
\hline
155 & 181.54707 & 26.83880 & 16.320 & 0.013 & 15.755 & 0.011 & 15.313 & 0.015 & -14.8 & -6.4 & 3.2 & 115.8 &  & 4 \\
156 & 182.06965 & 27.51282 & 12.312 & 0.001 & 11.709 & 0.001 & 11.359 & 0.001 & 0.3 & -11.4 & 2.8 & 104.6 &  & 4 \\
157 & 182.16738 & 26.29052 & 16.088 & 0.011 & 15.486 & 0.008 & 15.078 & 0.013 & -13.7 & -23.5 & 3.2 & 109.7$^\star$ &  & 4 \\
158 & 182.76838 & 26.46638 & 14.833 & 0.005 & 14.242 & 0.003 & 13.918 & 0.004 & -25.2 & -8.1 & 2.9 & 99.2$^\star$  &  & 4 \\
159 & 182.81209 & 23.59442 & 16.787 & 0.020 & 16.109 & 0.013 & 15.499 & 0.014 & -2.2 & -9.1 & 6.9 & 87.1$^\star$  &  & 4 \\
160 & 183.35818 & 21.50937 & 16.082 & 0.009 & 15.466 & 0.010 & 15.064 & 0.011 & -17.5 & -3.3 & 3.5 & 109.1$^\star$  &  & 4 \\
161 & 183.48757 & 25.08627 & 9.797 & 0.020 & 9.130 & 0.014 & 9.071 & 0.013 & -8.8 & 1.4 & 5.1 & 103.0$^\star$  &  & 2 \\
162 & 183.67470 & 26.49705 & 9.724 & 0.020 & 9.068 & 0.015 & 8.925 & 0.016 & 0.7 & 1.1 & 5.1 & 79.4$^\star$  &  & 2 \\
163 & 184.51479 & 23.83123 & 10.725 & 0.022 & 10.054 & 0.020 & 9.938 & 0.018 & -4.1 & -6.7 & 5.2 & 116.0$^\star$  &  & 2 \\
164 & 184.55135 & 29.10878 & 10.232 & 0.022 & 9.540 & 0.022 & 9.396 & 0.020 & -5.0 & -10.7 & 5.5 & 76.7$^\star$  &  & 2 \\
165 & 184.60960 & 26.96763 & 12.921 & 0.029 & 12.329 & 0.036 & 12.070 & 0.022 & -23.0 & -17.5 & 5.9 & 80.2 &  & 2 \\
166 & 184.61984 & 30.78003 & 10.680 & 0.026 & 9.980 & 0.030 & 9.844 & 0.022 & -7.4 & -12.8 & 6.0 & 99.3$^\star$  &  & 2 \\
167 & 185.20981 & 22.08020 & 13.441 & 0.046 & 12.778 & 0.044 & 12.551 & 0.041 & -24.5 & -11.8 & 5.2 & 105.7 &  & 2 \\
168 & 185.21537 & 23.32075 & 12.586 & 0.021 & 12.016 & 0.023 & 11.765 & 0.020 & -1.8 & -4.7 & 5.3 & 107.9 &  & 2 \\
169 & 185.25005 & 21.91837 & 9.638 & 0.022 & 9.040 & 0.024 & 8.870 & 0.017 & -1.2 & -6.8 & 5.2 & 72.6$^\star$  &  & 2 \\
170 & 185.31003 & 21.17550 & 10.633 & 0.021 & 9.910 & 0.022 & 9.780 & 0.018 & -8.5 & -1.7 & 5.2 & 111.5$^\star$  & c & 2 \\
171 & 185.37505 & 23.18350 & 15.490 & 0.007 & 14.930 & 0.005 & 14.608 & 0.007 & -5.5 & -2.4 & 3.4 & 92.0$^\star$  &  & 4 \\
172 & 185.55564 & 21.48469 & 14.834 & 0.005 & 14.265 & 0.004 & 13.937 & 0.005 & -4.1 & -15.9 & 3.4 & 67.9$^\star$  &  & 4 \\
173 & 185.71817 & 26.64010 & 9.777 & 0.027 & 9.263 & 0.032 & 9.115 & 0.021 & -9.4 & 2.7 & 5.8 & 91.0 & bc & 2 \\
174 & 185.74752 & 24.98287 & 9.396 & 0.027 & 8.811 & 0.044 & 8.674 & 0.019 & -13.8 & -7.9 & 5.8 & 96.8$^\star$  & cd & 2 \\
175 & 185.79713 & 25.56805 & 10.430 & 0.018 & 9.735 & 0.016 & 9.591 & 0.019 & -0.9 & -6.9 & 5.1 & 102.2$^\star$  &  & 2 \\
176 & 185.98912 & 24.89141 & 16.065 & 0.012 & 15.433 & 0.009 & 14.927 & 0.012 & -17.1 & -6.0 & 3.3 & 75.5$^\star$  & e & 4 \\
177 & 186.69448 & 23.61378 & 10.303 & 0.022 & 9.645 & 0.020 & 9.514 & 0.017 & -14.5 & -17.9 & 5.2 & 117.6$^\star$  &  & 2 \\
178 & 186.85393 & 26.22864 & 10.770 & 0.024 & 9.999 & 0.029 & 9.860 & 0.020 & -8.1 & -5.7 & 5.9 & 106.1$^\star$  &  & 2 \\
179 & 187.63748 & 30.36935 & 12.303 & 0.001 & 11.663 & 0.001 & 11.439 & 0.001 & -17.6 & -12.1 & 2.5 & 111.8 &  & 4 \\
180 & 187.83843 & 30.44276 & 13.445 & 0.023 & 12.906 & 0.030 & 12.623 & 0.021 & -11.9 & -9.4 & 5.5 & 109.4 &  & 2 \\
181 & 187.89421 & 24.09135 & 11.948 & 0.021 & 11.318 & 0.028 & 11.091 & 0.021 & -21.1 & -19.2 & 6.0 & 51.1 &  & 2 \\
182 & 188.02089 & 24.15844 & 15.424 & 0.006 & 14.866 & 0.006 & 14.653 & 0.010 & 1.9 & -14.2 & 3.2 & 98.3$^\star$  &  & 4 \\
183 & 188.33619 & 24.95474 & 10.794 & 0.030 & 10.123 & 0.032 & 9.927 & 0.020 & -16.5 & -16.6 & 5.9 & 52.8 & c & 2 \\
184 & 189.06597 & 23.41478 & 13.492 & 0.002 & 12.995 & 0.002 & 12.636 & 0.002 & -8.3 & 1.9 & 3.0 & 71.4 &  & 4 \\
185 & 189.07309 & 22.65018 & 9.640 & 0.020 & 9.069 & 0.031 & 8.882 & 0.014 & -11.2 & -7.4 & 5.5 & 88.8$^\star$  &  & 2 \\
186 & 189.35725 & 27.17839 & 10.385 & 0.022 & 9.708 & 0.028 & 9.603 & 0.021 & -6.1 & -4.1 & 5.9 & 109.5$^\star$  &  & 2 \\
187 & 189.80086 & 23.18844 & 13.490 & 0.023 & 12.866 & 0.022 & 12.602 & 0.024 & -25.0 & -14.2 & 4.9 & 102.3$^\star$  &  & 2 \\
188 & 190.22918 & 23.83889 & 10.011 & 0.021 & 9.316 & 0.019 & 9.162 & 0.017 & -5.4 & -16.1 & 4.9 & 92.8$^\star$  &  & 2 \\
189 & 190.37988 & 23.67833 & 15.170 & 0.006 & 14.673 & 0.004 & 14.304 & 0.006 & -4.9 & -3.2 & 3.4 & 117.8 &  & 4 \\
190 & 190.41708 & 26.93449 & 8.852 & 0.018 & 8.346 & 0.038 & 8.278 & 0.020 & -14.0 & 7.4 & 5.8 & 68.9 &  & 2 \\
191 & 190.72371 & 24.91863 & 17.530 & 0.035 & 16.688 & 0.025 & 15.963 & 0.028 & -16.6 & -1.6 & 3.7 & 61.4 &  & 4 \\
192 & 191.40753 & 24.84447 & 15.569 & 0.008 & 15.070 & 0.006 & 14.696 & 0.009 & 2.0 & -11.9 & 3.2 & 95.9$^\star$  &  & 4 \\
\enddata
\tablecomments{
$^\dagger$ a:~\citet{tru38}, b:~\citet{cas06}, c:~\citet{kra07},
	d:~\citet{mer08}, e:~\citet{mel12}, f:~\citet{van17},  g:~\citet{cas05} \\
$\ddagger$ 1:~2MASS photometry and {\it Gaia}/DR2 proper motions, 2:~2MASS photometry and
	URAT-1 proper motions, 3:~UKIDSS/GCS photometry and {\it Gaia}/DR2 proper motions,
	4:~both photometry and proper motions are from UKIDSS/GCS \\
$\star$:~{\it Gaia}/DR2 parallax measurements with $\varpi / \Delta\varpi < 10$}
\end{deluxetable}
\end{longrotatetable}

\newpage

\begin{longrotatetable}
\begin{deluxetable}{rcc cccccc RRR rrc}
\tablecaption{Selected Literature Members Beyond 5 degree Radius Confirmed by This Work \label{tab:beyond5}}

\tabletypesize{\scriptsize}
 \tablehead{
	 \colhead{No.} & \colhead{R.A. (2000)} & \colhead{Decl. (2000)} & \colhead{J} & \colhead{Jerr} & \colhead{H}  & 
	 \colhead{Herr} & \colhead{Ks} & \colhead{Kserr} &
	\colhead{$\mu_\alpha \cos\delta$} & \colhead{$\mu\delta$ }	&\colhead{$\Delta\mu$} & 
	\colhead{Dist} & \colhead{Ref.$^\dagger$}  \\ 
 	 \colhead{ }  & \colhead{(deg)} & \colhead{ (deg)} & \colhead{ (mag)} & \colhead{(mag)} & \colhead{(mag)}  & 
	 \colhead{(mag)} & \colhead{(mag)} & \colhead{(mag)} &
	\colhead{mas~yr$^{-1}$} & \colhead{mas~yr$^{-1}$} & \colhead{ mas~yr$^{-1}$} & \colhead{ (pc)} & 
	\colhead{ }  \\
  \colhead{(1)}  & \colhead{(2)} & \colhead{(3)} & \colhead{(4)} & \colhead{(5)} & \colhead{(6)} & 
 \colhead{(7)} &  \colhead{(8)} & \colhead{(9)} & \colhead{(10)} & \colhead{(11)} & \colhead{(12)} & \colhead{(13) } 
	& \colhead{(14) } 
	   } 
\startdata
1 & 177.15700 & 28.27510 & 9.02 & 0.026 & 8.63 & 0.029 & 8.59 & 0.02 & -12.06 & -9.36 & 0.09 & 89.29 & f  \\
2 & 178.88890 & 29.72820 & 9.69 & 0.022 & 9.19 & 0.022 & 9.06 & 0.017 & -15.71 & -9.67 & 0.11 & 90.45 & f \\
3 & 179.31637 & 24.65142 & 12.05 & 0.024 & 11.44 & 0.024 & 11.2 & 0.023 & -12.32 & -10.85 & 0.14 & 84.31 & c \\
4 & 179.72550 & 23.85242 & 12.21 & 0.02 & 11.65 & 0.024 & 11.38 & 0.018 & -11.75 & -8.79 & 0.17 & 91.18 & c \\
5 & 179.77171 & 26.74294 & 11.22 & 0.019 & 10.58 & 0.019 & 10.37 & 0.019 & -12.02 & -8.79 & 0.06 & 89.33 & c \\
6 & 180.61090 & 20.12300 & 8.65 & 0.018 & 8.39 & 0.034 & 8.31 & 0.02 & -12.14 & -7.95 & 0.08 & 90.26 & f \\
7 & 184.47371 & 20.37667 & 12.33 & 0.021 & 11.72 & 0.022 & 11.5 & 0.018 & -13.98 & -8.02 & 0.30 & 82.26 & c \\
8 & 184.62110 & 32.74890 & 6.13 & 0.019 & 6.05 & 0.024 & 6.02 & 0.016 & -13.01 & -9.28 & 0.13 & 83.66 & f \\
9 & 188.12946 & 35.33119 & 8.41 & 0.019 & 8.13 & 0.023 & 8.09 & 0.018 & -12.22 & -10.34 & 0.07 & 83.61 & df \\
10 & 188.52692 & 32.02686 & 7.28 & 0.02 & 7.06 & 0.02 & 7.02 & 0.023 & -10.17 & -11.71 & 0.08 & 120.14 & c \\
11 & 191.77800 & 22.61680 & 7.32 & 0.032 & 7.08 & 0.038 & 7.03 & 0.017 & -12.59 & -8.92 & 0.10 & 83.30 & f \\
12 & 192.92465 & 27.54068 & 3.629 & 0.292 & 3.367 & 0.218 & 3.26 & 0.286 & -10.99 & -8.31 & 0.34 & 87.01 & bd \\
13 & 193.04837 & 25.37350 & 7.88 & 0.021 & 7.65 & 0.021 & 7.61 & 0.015 & -11.47 & -8.55 & 0.08 & 85.13 & cdf \\
14 & 194.40350 & 28.97910 & 8.9 & 0.026 & 8.54 & 0.046 & 8.47 & 0.02 & -11.57 & -5.35 & 0.10 & 91.74 & f \\
15 & 195.14658 & 23.65175 & 7.38 & 0.021 & 7.22 & 0.018 & 7.18 & 0.02 & -11.97 & -8.88 & 0.11 & 86.15 & df \\
\enddata
\tablecomments{
$^\dagger$ a:~\citet{tru38}, b:~\citet{cas06}, c:~\citet{kra07}, d:~\citet{mer08}, e:~\citet{mel12}, f:~\citet{van17}}
\end{deluxetable}
\end{longrotatetable}



\begin{deluxetable*}{rcc ccccc RR cc}
\tablecolumns{12}
\tablewidth{0pt}
\tabletypesize{\scriptsize}
\tablecaption{
    Brown Dwarf Members of Coma Ber and Miscellaneous Objects  \label{tab:substellar}
	     }
\tablehead{ \colhead{No.} &  \colhead{R.A.} & \colhead{Decl.} &  
	     \colhead{Z } & \colhead{Y } & \colhead{J } &  \colhead{H } &  \colhead{K1 } &  
	      \colhead{ $\mu_\alpha \cos\delta$ } &  \colhead{$\mu_\delta$} & 
	       \colhead{Spectroscopy} & \colhead{\it photo-type}  \\
   \colhead{}   & \colhead{ (deg) } &   \colhead{ (deg) } &  
	     \colhead{(mag)} & \colhead{(mag)}  & \colhead{(mag)} & \colhead{(mag)} & \colhead{(mag)} &
	     \colhead{(mas~yr$^{-1}$) } & \colhead{(mas~yr$^{-1}$) } & 
	      \colhead{(SpT)} & \colhead{(SpT)} 
	   } 
\startdata 
\multicolumn{12}{c}{Brown Dwarf Members} \\ \hline 
6     & 181.39774 & +25.02538 & 17.3756 & 16.3825 & 15.5315 & 14.9799 & 14.4771 & -12.9  & -9.9 & M8$^a$ & M7 \\ 
155 & 181.54707 & +26.83880 & 18.0680 & 17.2021 & 16.3201 & 15.7545 & 15.3133 & -11.3 & -5.8 & M8$^a$ & M7 \\ 
157 & 182.16738 & +26.29052 & 17.5872 & 16.8390 & 16.0885 & 15.4859 & 15.0781 & -10.2 & -22.9 & M8$^a$ &  M7 \\ 
176 & 185.98912 & +24.89141 & 18.0889 & 16.9529 & 16.0646 & 15.4329 & 14.9268 & -13.5 & -5.3 & M9$^b$ & M9 \\ 
55   & 185.42258 & +21.68386 & 17.4297 & 16.5134 & 15.6511 & 15.1188 & 14.6610 & -10.6  & -8.2 & M9$^a$ & M7 \\ 
130 & 188.46503 & +31.11809 & 17.4072 & 16.4539 & 15.5926 & 15.0481 & 14.5452 & -11.3 & -8.6 & M9$^a$ & M9 \\
159 & 182.81209 & +23.59442 & 19.2019 & 18.0187 & 16.7868 & 16.1094 & 15.4987 & 1.4   & -8.5  & L2$^c$ & L2 \\ 
160 & 183.35818 & +21.50937 & 17.6250 & 16.8178 & 16.0820 & 15.4655 & 15.0643 & -13.9 & -2.7 & M8$^c$  & M7 \\  
191 & 190.72371 & +24.91863 & 20.2402 & 18.8068 & 17.5300 & 16.6876 & 15.9625 & -13.0 & -1.0 & L4$^c$ & L5 \\ 
 \hline
\multicolumn{12}{c}{Miscellaneous Objects } \\ 
 \hline
 A  & 183.19468 & +25.50436 & 17.7934 & 16.5876 & 15.9029 & 15.1508 & 14.9253 & -13.0 & 2.1 &  MIII?  & M1.5\\
 B  & 188.90996 & +23.19031 & 17.7026 & 16.8755 & 15.9585 & 15.4668 & 15.1479 & -3.0  & -7.7 &  MIII?  & M3\\ 
 C  & 186.48545 & +22.78955 & \nodata & \nodata & 19.1199 & 17.4459 & 16.4182 & -7.3  & -4.6  & \nodata  &  \nodata\\
\enddata
\tablenotetext{a}{\citet{wes11}}
\tablenotetext{b}{\citet{cas14}}
\tablenotetext{c}{This work}
\end{deluxetable*}


\newpage
\startlongtable
\begin{deluxetable*}{cCcc c cCcc}
\tablecaption{ Literature Candidates Rejected by Our Analysis \label{tab:rej}
             }
\tablehead{
	        \colhead{R.A.} & \colhead{Decl.} &
		\colhead{Ref.$^\dagger$} & \colhead{ Rej.$^\ddagger$} & \colhead{} &
	        \colhead{R.A.} & \colhead{Decl.} &
		\colhead{Ref.$^\dagger$} & \colhead{ Rej.$^\ddagger$}  \\
	        	\colhead{ (deg) } &   \colhead{ (deg) } &  
		\colhead{} & \colhead{ } & \colhead{}& 
		\colhead{ (deg) } &   \colhead{ (deg) } &  
		\colhead{} & \colhead{ }
	   } 
\startdata 
181.094000 & 24.021440 & c & 1 &  &   186.260417 & 26.717778 & e & 1 \\
181.558040 & 26.780640 & c & 6 &  &   186.268333 & 23.699444 & e & 1 \\
181.974620 & 25.929310 & c & 4 &  &   186.270417 & 25.312500 & e & 5 \\
182.081080 & 21.082720 & c & 1 &  &   186.292250 & 27.662440 & bc & 4 \\
182.301833 & 26.660806 & bc & 4 &  &  186.388750 & 24.356667 & e & 7 \\
183.146080 & 27.494080 & c & 5 &  &   186.391667 & 25.241111 & e & 1 \\
183.389280 & 27.472287 & g & 7 &  &   186.482333 & 29.127306 & b & 4 \\
183.433042 & 27.311437 & g & 5 &  &   186.517083 & 24.314444 & e & 1 \\
183.533920 & 22.840920 & c & 4 &  &   186.522833 & 26.743972 & bc & 6 \\
183.582417 & 25.179611 & b & 7 &  &   186.600040 & 25.261940 & c & 6 \\
183.599875 & 28.354611 & b & 4 &  &   186.655500 & 22.581500 & b & 4 \\
183.820542 & 28.747222 & b & 4 &  &   186.722917 & 25.731944 & e & 1 \\
183.824362 & 27.919314 & g & 5 &  &   186.753375 & 29.610528 & b & 2 \\
183.891708 & 26.261917 & bc & 2 &  &  186.762917 & 28.546667 & e & 1 \\
183.955417 & 27.326944 & e & 5 &  &   186.767917 & 25.682778 & e & 5 \\
183.980833 & 26.949167 & e & 4 &  &   186.785667 & 27.023028 & b & 6 \\
184.010417 & 28.048667 & b & 4 &  &   186.821250 & 26.355833 & e & 4 \\
184.079540 & 26.927080 & c & 6 &  &   186.836565 & 27.169444 & a & 4 \\
184.086667 & 25.306667 & e & 1 &  &   186.859500 & 24.782500 & d & 5 \\
184.095167 & 24.316972 & b & 4 &  &   186.955184 & 27.991243 & g & 1 \\
184.121708 & 23.542472 & b & 4 &  &   186.969417 & 25.095750 & ad & 2 \\
184.206000 & 24.855861 & d & 5 &  &   187.027500 & 24.293333 & e & 5 \\
184.233333 & 25.947778 & e & 1 &  &   187.042083 & 28.581111 & e & 1 \\
184.249439 & 27.334819 & g & 1 &  &   187.067083 & 26.178889 & e & 5 \\
184.309167 & 25.515278 & e & 3 &  &   187.106500 & 29.898444 & d & 4 \\
184.335833 & 25.081667 & e & 1 &  &   187.114625 & 29.878417 & d & 7 \\
184.356000 & 27.242310 & c & 2 &  &   187.142875 & 23.541833 & b & 4 \\
184.417917 & 24.364444 & e & 5 &  &   187.157917 & 28.096389 & e & 1 \\
184.434583 & 25.827778 & e & 1 &  &   187.161250 & 25.986944 & b & 6 \\
184.439167 & 27.295556 & e & 1 &  &   187.181667 & 24.431111 & e & 4 \\
184.449167 & 28.119722 & e & 5 &  &   187.188333 & 27.189167 & e & 1 \\
184.461667 & 23.825278 & e & 5 &  &   187.208667 & 27.294917 & b & 6 \\
184.496250 & 26.535000 & e & 3 &  &   187.270000 & 25.070000 & e & 5 \\
184.561710 & 21.533000 & c & 4 &  &   187.305417 & 23.794167 & e & 1 \\
184.574042 & 23.642444 & b & 4 &  &   187.362667 & 24.108917 & b & 7 \\
184.611250 & 25.883560 & c & 1 &  &  187.375083 & 29.512722 & b & 4 \\
184.636255 & 27.625217 & g & 7 &  &   187.417500 & 26.332222 & e & 1 \\
184.677083 & 24.413333 & e & 1 &  &   187.425670 & 28.620750 & c & 6 \\
184.712625 & 26.323444 & d & 1 &  &   187.436083 & 25.543194 & d & 1 \\
184.720833 & 27.217222 & e & 1 &  &   187.558750 & 25.028417 & ad & 1 \\
184.738625 & 25.886417 & bcd & 4 &  & 187.559167 & 24.635000 & e & 1 \\
184.744583 & 26.058583 & cd & 2 &  &  187.611667 & 24.893611 & e & 1 \\
184.771583 & 26.184556 & d & 3 &  &   187.674715 & 28.001458 & g & 7 \\
184.785500 & 25.053222 & d & 6 &  &   187.695833 & 26.057500 & e & 5 \\
184.811040 & 27.930640 & c & 2 &  &  187.751167 & 26.940306 & b & 4 \\
184.817750 & 25.436250 & ad & 4 &  &  187.769917 & 24.262611 & bc & 2 \\
184.889206 & 24.546867 & g & 7 &  &   187.855417 & 24.541111 & e & 1 \\
184.945000 & 27.443611 & e & 1 &  &   187.964583 & 25.042778 & e & 1 \\
185.023792 & 25.910222 & d & 4 &  &   187.989250 & 25.145139 & bc & 4 \\
185.101250 & 25.139444 & e & 5 &  &   188.033708 & 28.901806 & bc & 4 \\
185.112083 & 26.775556 & e & 5 &  &   188.049285 & 27.115717 & g & 7 \\
185.131667 & 24.603889 & e & 6 &  &   188.115000 & 23.765833 & e & 1 \\
185.150833 & 28.451944 & e & 2 &  &   188.157500 & 24.656944 & e & 5 \\
185.151250 & 25.092778 & e & 5 &  &   188.257500 & 24.654444 & e & 4 \\
185.185833 & 27.262500 & e & 1 &  &   188.375792 & 26.166694 & bcd & 4 \\
185.226250 & 25.432222 & e & 1 &  &   188.376292 & 28.215528 & b & 4 \\
185.260153 & 26.367863 & g & 1 &  &   188.380625 & 24.202528 & b & 5 \\
185.328735 & 26.708572 & g & 5 &  &   188.383750 & 24.773056 & e & 5 \\
185.337125 & 32.176444 & d & 1 &  &   188.393750 & 24.656111 & e & 1 \\
185.383750 & 23.754722 & e & 1 &  &   188.475917 & 27.134639 & cd & 2 \\
185.582500 & 25.290833 & e & 1 &  &   188.559250 & 28.378310 & c & 2 \\
185.607583 & 25.317583 & d & 7 &  &   188.695542 & 24.160472 & bcd & 2 \\
185.669375 & 25.669972 & d & 1 &  &   188.751375 & 30.192667 & d & 1 \\
185.673040 & 27.247560 & ce & 4 &  &  188.820958 & 26.056056 & b & 5 \\
185.717670 & 25.066670 & c & 2 &  &   188.929420 & 25.922970 & c & 2 \\
185.740000 & 24.809167 & e & 7 &  &   189.019330 & 27.959890 & c & 2 \\
185.761667 & 27.645278 & e & 1 &  &   189.142917 & 25.010639 & b & 5 \\
185.799958 & 29.249972 & b & 5 &  &   189.218708 & 24.024194 & d & 3 \\
185.839710 & 21.712560 & c & 2 &  &   189.450710 & 26.963110 & c & 2 \\
185.869580 & 22.848860 & bc & 3 &  &  189.519667 & 25.855139 & b & 4 \\
185.880000 & 23.610556 & e & 5 &  &   189.562250 & 26.357806 & b & 3 \\
186.020000 & 25.563889 & e & 1 &  &   189.730667 & 25.893917 & d & 1 \\
186.043208 & 29.488778 & b & 4 &  &   189.941250 & 28.299083 & d & 3 \\
186.070417 & 24.892500 & e & 1 &  &   189.968458 & 25.775833 & b & 4 \\
186.071458 & 24.324556 & abcd & 2 &  & 190.054167 & 28.222500 & d & 1 \\
186.095625 & 25.915611 & d & 6 &  &   190.103710 & 27.918310 & c & 4 \\
186.107540 & 21.604860 & c & 4 &  &   190.270708 & 25.704972 & d & 4 \\
186.111040 & 25.752140 & c & 4 &  &   190.560620 & 28.603560 & c & 6 \\
186.223333 & 23.718028 & bc & 4 &  &  190.854460 & 26.785470 & c & 4 \\
\enddata
\tablecomments{
$^\dagger$ a:~\citet{tru38}, b:~\citet{cas06}, c:~\citet{kra07}, d:~\citet{mer08}, e:~\citet{mel12}, 
  f:~\citet{van17}, g:~\citet{cas05}  \\
$\ddagger$ 1:~rejection by proper motion, 2:~rejection by CMD, 4:~rejection by distance (additive) 
}
\end{deluxetable*} 

\appendix

\section{Photometric Spectral Typing and Distance}

We have modified the method {\it photo-type} proposed by \citet{skr15,skr16} for spectral typing, which 
then in turn is used for distance estimation.  \citet{skr15} included spectral templates from types 
M5 to T9.  To identify cluster members in Coma Ber, we extend to include main-sequence templates.  
For types from B2 to M4, the median colors and absolute magnitudes are taken from \citet{pec12} 
and \citet{pec13}, converting their Johnson colors to the SDSS system by the equations in \citet{jor05}.   
For spectral types later than M7, we adopt the median colors and absolute magnitudes from \citet{bes18}, 
using the equations in \citet{hew06} to convert between 2MASS and UKIDSS systems.
For M dwarfs, we use the median colors from \citet{bes18} and the absolute magnitudes from 
\citet{pec12} and \citet{pec13}.  Table~\ref{tab:b2m4}, Table~\ref{tab:mdwarfs}, and Table~\ref{tab:bds} 
list, respectively, these median colors.   

The algorithm performs classification by finding the minimum $\chi^{2}$ among spectral templates.  
Given an object with photometry $m_{b}$ ($b=g, r, i... W2$), we define the ``reference'' magnitude, $m_{(B,t)}$,   
at the band $B$ ($B=J$ band, in this study) for each spectral type $t$ by, 
 \begin{equation}
   m_{(B,t)} = \frac{\sum\limits_{b=1}^{N_{b}}\frac{m_{b} - c_{(b,t)}} {\sigma_{b}^{2}}} {\sum\limits_{b=1}^{N_{b}}\frac{1} {\sigma_{b}^{2}}}, 
\end{equation}
where the parameters $c_{(b,t)}$ are the expected colors of each spectral type $t$ listed in Tables~\ref{tab:b2m4}, 
Table~\ref{tab:mdwarfs}, and Table~\ref{tab:bds}.  The $m_{(B,t)}$ could be regarded as the pseudo-$J$ magnitude 
for the $t$ spectral type. 

Next, the weighted $\chi^{2}$ is derived, 
\begin{equation} 
 \chi^{2} (\{m_{b} \},\{\sigma_{b}\}, m_{(B,t)}, t) = \sum\limits_{b=1}^{N_{b}} (\frac{m_{b} - m_{(B,t)} - c_{(b,t)}} {\sigma_{b}} )^{2}
\end{equation}

\begin{figure}[htbp]
\begin{center}
	\includegraphics[width=0.5\columnwidth,angle=0]{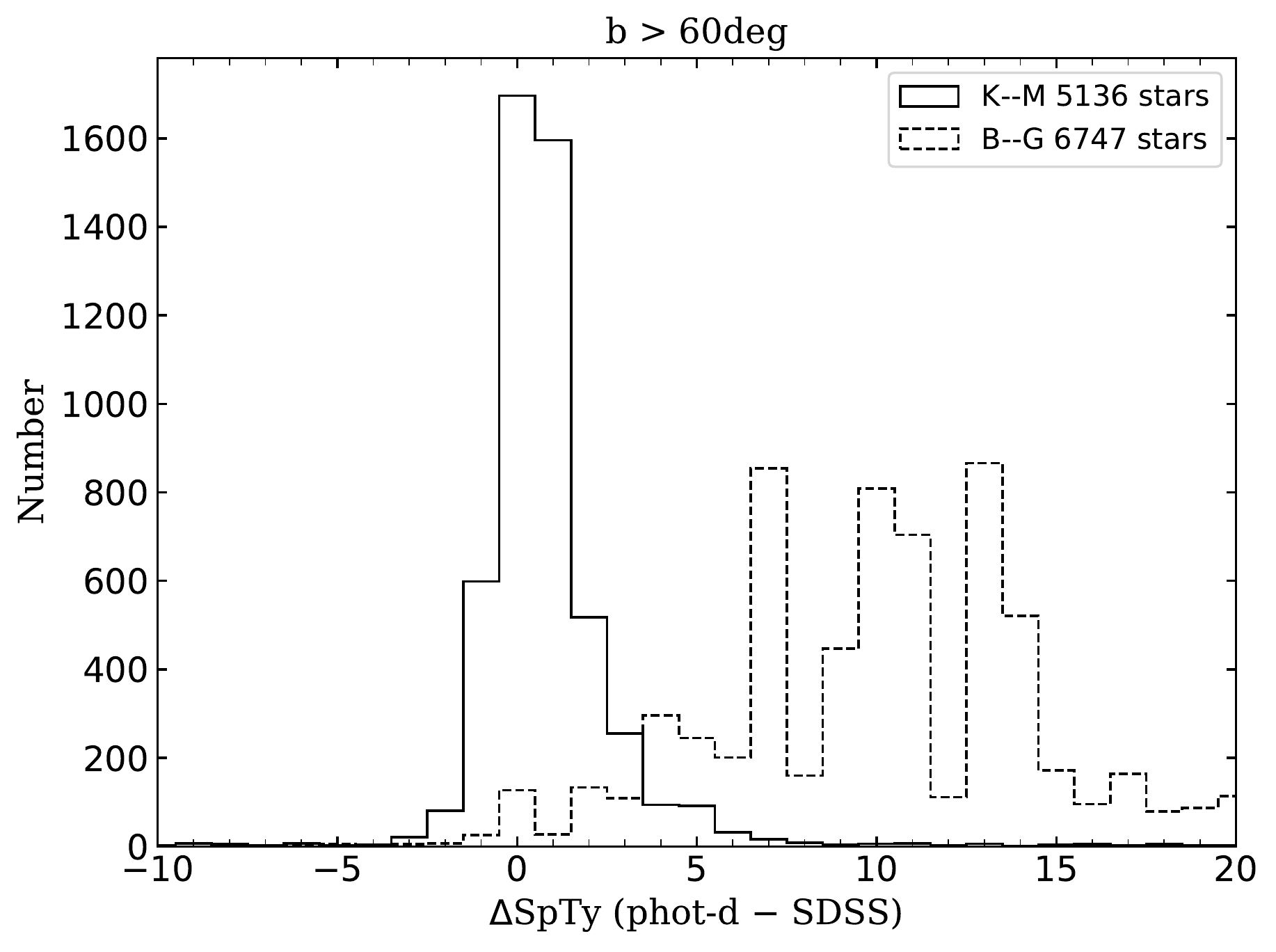}
	\caption{Differences between {\it phot-d} estimated spectral types and SDSS observed spectral types
	for K and M (solid lines) and for B--G dwarfs (dashed lines).  Only stars above the Galactic latitude 
	$60\degr$ have been selected to minimize the effect of interstellar reddening.  
	        }
\label{fig:spComp}
\end{center}
\end{figure}

For each spectral type $t$, $m_{b} - m_{(B,t)}$ is the ``observed'' value while $c_{(b,t)}$ is the expectation. 
Finally, the spectral type having the minimum $\chi^{2}$ among all templates is considered the best-fit type for 
the target.  Figure~\ref{fig:spComp} shows a general consistency with $\pm2$ subtypes between the {\it photo-type} results and 
those measured by SDSS/DR14.  Once the spectral type is determined, the absolute magnitudes listed in Table~\ref{tab:absmags} 
are used to estimate the distance for each band.  The median value of all bands is thereby adopted as the distance to the object.  



\newpage
\startlongtable


\begin{deluxetable*}{cRRR RRRR}
\tablecaption{Stellar Colors from B2 to M4 \label{tab:b2m4}
             }
 \tablehead{
	 \colhead{SpTy} & \colhead{$g-r$} & \colhead{$r-i$} & \colhead{$i-z$} & 
	 \colhead{$z-J$} & \colhead{$J-H$} & \colhead{$H-K_s$} & \colhead{$K-W1$} 
	 	  }
\startdata
B1.5 & -0.330 & -0.397 & -0.169 & 0.168 & -0.132 & -0.05 & 0.035 \\
B2 & -0.289 & -0.369 & -0.166 & 0.232 & -0.113 & -0.027 & 0.036 \\
B2.5 & -0.277 & -0.357 & -0.167 & 0.266 & -0.105 & -0.025 & 0.036 \\
B3 & -0.259 & -0.347 & -0.166 & 0.289 & -0.098 & -0.022 & 0.036 \\
B4 & -0.246 & -0.338 & -0.166 & 0.305 & -0.092 & -0.018 & 0.036 \\
B5 & -0.238 & -0.331 & -0.166 & 0.326 & -0.089 & -0.021 & 0.036 \\
B6 & -0.222 & -0.320 & -0.164 & 0.348 & -0.081 & -0.009 & 0.035 \\
B7 & -0.211 & -0.313 & -0.165 & 0.368 & -0.077 & -0.003 & 0.035 \\
B8 & -0.192 & -0.299 & -0.163 & 0.395 & -0.067 & 0.007 & 0.034 \\
B9 & -0.153 & -0.273 & -0.157 & 0.449 & -0.05 & 0.02 & 0.032 \\
B9.5 & -0.132 & -0.259 & -0.154 & 0.483 & -0.044 & 0.024 & 0.031 \\
A0 & -0.087 & -0.238 & -0.146 & 0.512 & -0.032 & 0.032 & 0.03 \\
A1 & -0.047 & -0.217 & -0.139 & 0.520 & -0.024 & 0.034 & 0.03 \\
A2 & -0.011 & -0.193 & -0.127 & 0.537 & -0.01 & 0.04 & 0.029 \\
A3 & 0.005 & -0.184 & -0.123 & 0.558 & -0.002 & 0.032 & 0.029 \\
A4 & 0.057 & -0.155 & -0.107 & 0.583 & 0.022 & 0.038 & 0.029 \\
A5 & 0.078 & -0.144 & -0.100 & 0.598 & 0.031 & 0.039 & 0.029 \\
A6 & 0.088 & -0.138 & -0.099 & 0.606 & 0.036 & 0.044 & 0.029 \\
A7 & 0.130 & -0.116 & -0.085 & 0.633 & 0.055 & 0.045 & 0.029 \\
A8 & 0.172 & -0.092 & -0.072 & 0.660 & 0.075 & 0.045 & 0.028 \\
A9 & 0.177 & -0.089 & -0.070 & 0.663 & 0.078 & 0.042 & 0.028 \\
F0 & 0.219 & -0.066 & -0.057 & 0.699 & 0.098 & 0.042 & 0.028 \\
F1 & 0.262 & -0.043 & -0.042 & 0.694 & 0.119 & 0.051 & 0.028 \\
F2 & 0.304 & -0.019 & -0.029 & 0.710 & 0.14 & 0.05 & 0.028 \\
F3 & 0.320 & -0.011 & -0.024 & 0.720 & 0.147 & 0.053 & 0.028 \\
F4 & 0.345 & 0.003 & -0.015 & 0.737 & 0.159 & 0.051 & 0.028 \\
F5 & 0.373 & 0.018 & -0.004 & 0.740 & 0.173 & 0.057 & 0.028 \\
F6 & 0.419 & 0.041 & 0.010 & 0.755 & 0.199 & 0.061 & 0.028 \\
F7 & 0.446 & 0.054 & 0.019 & 0.772 & 0.213 & 0.057 & 0.027 \\
F8 & 0.466 & 0.064 & 0.025 & 0.779 & 0.225 & 0.065 & 0.027 \\
F9 & 0.488 & 0.074 & 0.034 & 0.791 & 0.237 & 0.063 & 0.027 \\
G0 & 0.534 & 0.096 & 0.051 & 0.805 & 0.262 & 0.068 & 0.027 \\
G1 & 0.542 & 0.100 & 0.053 & 0.806 & 0.267 & 0.073 & 0.027 \\
G2 & 0.588 & 0.119 & 0.070 & 0.853 & 0.293 & 0.077 & 0.028 \\
G3 & 0.598 & 0.124 & 0.073 & 0.841 & 0.299 & 0.071 & 0.028 \\
G4 & 0.611 & 0.129 & 0.077 & 0.847 & 0.307 & 0.073 & 0.028 \\
G5 & 0.617 & 0.131 & 0.080 & 0.840 & 0.31 & 0.08 & 0.028 \\
G6 & 0.640 & 0.141 & 0.087 & 0.865 & 0.324 & 0.076 & 0.028 \\
G7 & 0.649 & 0.145 & 0.092 & 0.863 & 0.329 & 0.081 & 0.028 \\
G8 & 0.672 & 0.155 & 0.099 & 0.888 & 0.342 & 0.078 & 0.028 \\
G9 & 0.712 & 0.171 & 0.113 & 0.904 & 0.365 & 0.085 & 0.029 \\
K0 & 0.751 & 0.186 & 0.130 & 0.9189 & 0.387 & 0.093 & 0.03 \\
K1 & 0.781 & 0.199 & 0.139 & 0.835 & 0.402 & 0.098 & 0.03 \\
K2 & 0.832 & 0.222 & 0.164 & 0.966 & 0.432 & 0.098 & 0.031 \\
K3 & 0.934 & 0.266 & 0.209 & 1.024 & 0.49 & 0.11 & 0.034 \\
K4 & 1.074 & 0.344 & 0.284 & 1.051 & 0.544 & 0.126 & 0.039 \\
K5 & 1.139 & 0.384 & 0.315 & 1.067 & 0.568 & 0.132 & 0.042 \\
K6 & 1.250 & 0.459 & 0.357 & 1.105 & 0.601 & 0.149 & 0.049 \\
K7 & 1.351 & 0.539 & 0.374 & 1.174 & 0.622 & 0.168 & 0.06 \\
K8 & 1.384 & 0.578 & 0.371 & 1.211 & 0.623 & 0.177 & 0.081 \\
K9 & 1.417 & 0.617 & 0.368 & 1.237 & 0.625 & 0.185 & 0.101 \\
M0 & 1.451 & 0.657 & 0.365 & 1.284 & 0.626 & 0.194 & 0.122 \\
M0.5 & 1.497 & 0.730 & 0.348 & 1.343 & 0.62 & 0.21 & 0.13 \\
M1 & 1.520 & 0.838 & 0.295 & 1.456 & 0.613 & 0.227 & 0.137 \\
M1.5 & 1.531 & 0.891 & 0.273 & 1.522 & 0.607 & 0.233 & 0.105 \\
M2 & 1.545 & 0.954 & 0.245 & 1.610 & 0.6 & 0.23 & 0.11 \\
M2.5 & 1.568 & 1.055 & 0.208 & 1.818 & 0.589 & 0.151 & 0.117 \\
M3 & 1.591 & 1.1416 & 0.182 & 1.799 & 0.579 & 0.281 & 0.122 \\
M3.5 & 1.650 & 1.339 & 0.141 & 2.049 & 0.558 & 0.272 & 0.132 \\
M4 & 1.702 & 1.454 & 0.126 & 2.175 & 0.557 & 0.283 & 0.139 \\
\enddata
\tablecomments{
Stellar median colors from \citet{pec12} and \citet{pec13}.  
Photometric systems are in SDSS, 2MASS, and {\it WISE}.
              }
\end{deluxetable*} 


\newpage
\begin{deluxetable*}{cRRRR RRRRR}
	\tablecaption{Colors of M dwarfs \label{tab:mdwarfs} }
 \tablehead{
	 \colhead{SpTy} & \colhead{$g-r$} & \colhead{$r-i$} & \colhead{$i-z$} & 
	 \colhead{$z-y$} & \colhead{$y-J$} & \colhead{$J-H$} & \colhead{$H-K_s$} & \colhead{$K_s-W1$} & \colhead{$W1-W2$}  
	 	  }
\startdata
M0 & 1.19 & 0.67 & 0.31 & 0.17 & 1.12 & 0.66 & 0.18 & 0.1 & 0.02 \\
M1 & 1.22 & 0.85 & 0.39 & 0.2 & 1.14 & 0.64 & 0.21 & 0.12 & 0.07 \\
M2 & 1.21 & 1.02 & 0.46 & 0.23 & 1.16 & 0.62 & 0.22 & 0.12 & 0.12 \\
M3 & 1.21 & 1.22 & 0.55 & 0.27 & 1.2 & 0.6 & 0.24 & 0.14 & 0.16 \\
M4 & 1.23 & 1.46 & 0.67 & 0.32 & 1.25 & 0.59 & 0.26 & 0.16 & 0.18 \\
M5 & 1.31 & 1.88 & 0.87 & 0.44 & 1.34 & 0.59 & 0.31 & 0.19 & 0.21 \\
M6 & 1.33 & 2.13 & 0.98 & 0.51 & 1.4 & 0.6 & 0.33 & 0.21 & 0.22 \\
M7 & 1.4 & 2.55 & 1.2 & 0.67 & 1.54 & 0.63 & 0.39 & 0.22 & 0.23 \\
M8 & 1.53 & 2.7 & 1.38 & 0.81 & 1.66 & 0.68 & 0.43 & 0.26 & 0.23 \\
M9 & 1.79 & 2.58 & 1.44 & 0.92 & 1.77 & 0.71 & 0.48 & 0.31 & 0.26 \\
\enddata
\tablecomments{
Stellar median colors of M dwarfs from \citet{bes18}.  
	Photometric systems are in PS1, 2MASS, and {\it WISE}.
              }
\end{deluxetable*}

\newpage
\begin{deluxetable*}{cRRR RRRR}
\tablecaption{Median Colors of Brown Dwarfs \label{tab:bds} }
 \tablehead{
	 \colhead{SpTy} & \colhead{$i-z$} & \colhead{$z-Y$} & \colhead{$Y-J$} & \colhead{$J-H$} & 
	 \colhead{$H-K$} & \colhead{$K-W1$} & \colhead{$W1-W2$}  
	 	  }
\startdata
M5 & 0.91 & 0.47 & 0.55 & 0.45 & 0.32 & 0.11 & 0.17 \\
M6 & 1.45 & 0.6 & 0.67 & 0.53 & 0.39 & 0.22 & 0.21 \\
M7 & 1.36 & 0.55 & 0.68 & 0.54 & 0.38 & 0.17 & 0.2 \\
M8 & 1.68 & 0.69 & 0.79 & 0.56 & 0.44 & 0.19 & 0.22 \\
M9 & 1.86 & 0.79 & 0.87 & 0.59 & 0.49 & 0.22 & 0.23 \\
L0 & 2.01 & 0.86 & 1.04 & 0.63 & 0.54 & 0.29 & 0.27 \\
L1 & 2.02 & 0.88 & 1.11 & 0.67 & 0.58 & 0.33 & 0.28 \\
L2 & 2.04 & 0.9 & 1.18 & 0.73 & 0.63 & 0.4 & 0.28 \\
L3 & 2.1 & 0.92 & 1.23 & 0.79 & 0.67 & 0.48 & 0.29 \\
L4 & 2.2 & 0.94 & 1.27 & 0.86 & 0.71 & 0.56 & 0.3 \\
L5 & 2.33 & 0.97 & 1.31 & 0.91 & 0.74 & 0.65 & 0.32 \\
L6 & 2.51 & 1.0 & 1.33 & 0.96 & 0.75 & 0.72 & 0.36 \\
L7 & 2.71 & 1.04 & 1.35 & 0.97 & 0.75 & 0.77 & 0.41 \\
L8 & 2.93 & 1.09 & 1.21 & 0.96 & 0.71 & 0.79 & 0.48 \\
L9 & 3.15 & 1.16 & 1.2 & 0.9 & 0.65 & 0.79 & 0.57 \\
T0 & 3.36 & 1.23 & 1.19 & 0.8 & 0.56 & 0.76 & 0.68 \\
T1 & 3.55 & 1.33 & 1.19 & 0.65 & 0.45 & 0.71 & 0.82 \\
T2 & 3.7 & 1.43 & 1.18 & 0.46 & 0.31 & 0.65 & 0.99 \\
T3 & 3.82 & 1.55 & 1.18 & 0.25 & 0.16 & 0.59 & 1.19 \\
T4 & 3.9 & 1.68 & 1.17 & 0.02 & 0.01 & 0.55 & 1.43 \\
T5 & 3.95 & 1.81 & 1.16 & -0.19 & -0.11 & 0.54 & 1.7 \\
T6 & 3.98 & 1.96 & 1.16 & -0.35 & -0.19 & 0.59 & 2.02 \\
T7 & 4.01 & 2.11 & 1.15 & -0.43 & -0.2 & 0.7 & 2.38 \\
T8 & 4.08 & 2.26 & 1.15 & -0.36 & -0.09 & 0.9 & 2.79 \\
\enddata
\tablecomments{
Stellar median colors of brown dwarfs from \citet{skr15} and \citet{skr16}
Photometric systems are in SDSS (Vega), UKIDSS, and {\it WISE}.
              }
\end{deluxetable*}

\newpage
\begin{longrotatetable}
\begin{deluxetable}{lRRRR RRRRR R}
	\tablecaption{Absolute Magnitudes \label{tab:absmags}  }
 \tablehead{
	 \colhead{SpTy} & \colhead{$g$} & \colhead{$r$} & \colhead{$i$} & \colhead{$z$} & 
	 \colhead{$y$} & \colhead{$J$} & \colhead{$H$} & \colhead{$K_s$} & \colhead{$W1$} & \colhead{$W2$} 
	 	  } 
\startdata
B1.5 & -2.959 & -2.628   & -2.231 & -2.062   & \nodata & -2.23 & -2.098 & -2.05 & -2.085 & \nodata \\
B2    & -1.832 & -1.543   & -1.175 & -1.008   & \nodata & -1.24 & -1.127 & -1.1 & -1.136 & \nodata \\
B2.5 & -1.525 & -1.248   & -0.891 & -0.724   & \nodata & -0.99 & -0.885 & -0.86 & -0.896 & \nodata \\
B3    & -1.212 & -0.953   & -0.607 & -0.441   & \nodata & -0.73 & -0.632 & -0.61 & -0.646 & \nodata \\
B4    & -1.104 & -0.858   & -0.520 & -0.355   & \nodata & -0.66 & -0.568 & -0.55 & -0.586 & \nodata \\
B5    & -0.999 & -0.761   & -0.429 & -0.264   & \nodata & -0.59 & -0.501 & -0.48 & -0.516 & -0.471 \\
B6    & -0.588 & -0.367   & -0.047 & 0.118    & \nodata & -0.23 & -0.149 & -0.14 & -0.175 & -0.13 \\
B7    & -0.481 & -0.269    & 0.043 & 0.208    & \nodata & -0.16 & -0.083 & -0.08 & -0.115 & -0.07 \\
B8    & -0.269 & -0.077    & 0.222 & 0.385    & \nodata & -0.01 & 0.057 & 0.05 & 0.016 & 0.062 \\
B9    &  0.656 &  0.809    & 1.082 & 1.239    & \nodata &0.79 & 0.84 & 0.82 & 0.788 & 0.851 \\
B9.5 &  0.769 &  0.900    & 1.160 & 1.313    & \nodata &0.83 & 0.874 & 0.85 & 0.819 & 0.863 \\
A0    &  1.11   &  1.197    & 1.436 & 1.582    & \nodata &1.07 & 1.102 & 1.07 & 1.04 & 1.081 \\
A1    &  1.367 &  1.414    & 1.631 & 1.780    & \nodata &1.25 & 1.274 & 1.24 & 1.21 & 1.246 \\
A2    &  1.527 &  1.537    & 1.730 & 1.857    & \nodata &1.32 & 1.33 & 1.29 & 1.261 & 1.295 \\
A3    &  1.607 &  1.601    & 1.785 & 1.908    & \nodata &1.35 & 1.352 & 1.32 & 1.291 & 1.324 \\
A4    &  1.848 &  1.791    & 1.946 & 2.053    & \nodata &1.47 & 1.448 & 1.41 & 1.381 & 1.412 \\
A5    &  1.941 &  1.863    & 2.007 & 2.108    & \nodata &1.51 & 1.479 & 1.44 & 1.411 & 1.441 \\
A6    &  1.997 &  1.909    & 2.047 & 2.146    & \nodata &1.54 & 1.504 & 1.46 & 1.431 & 1.461 \\
A7    &  2.202 &  2.072    & 2.188 & 2.273    & \nodata &1.64 & 1.585 & 1.54 & 1.511 & 1.541 \\
A8    &  2.448 &  2.275    & 2.368 & 2.440    & \nodata &1.78 & 1.705 & 1.66 & 1.632 & 1.66 \\
A9    &  2.461 &  2.283    & 2.372 & 2.443    & \nodata &1.78 & 1.702 & 1.66 & 1.632 & 1.66 \\
F0    &  2.695 &  2.476    & 2.543 & 2.599    & \nodata &1.90 & 1.802 & 1.76 & 1.732 & 1.758 \\
F1    &  3.000 &  2.739    & 2.782 & 2.824    & \nodata &2.13 & 2.011 & 1.96 & 1.932 & 1.958 \\
F2    &  3.226 &  2.922    & 2.941 & 2.970    & \nodata &2.26 & 2.12 & 2.07 & 2.042 & 2.069 \\
F3    &  3.325 &  3.005    & 3.016 & 3.040    & \nodata &2.32 & 2.173 & 2.12 & 2.092 & 2.12 \\
F4    &  3.490 &  3.145    & 3.142 & 3.157    & \nodata &2.42 & 2.261 & 2.21 & 2.182 & 2.211 \\
F5    &  3.676 &  3.303    & 3.286 & 3.290    & \nodata &2.55 & 2.377 & 2.32 & 2.292 & 2.322 \\
F6    &  4.005 &  3.586    & 3.544 & 3.535    & \nodata &2.78 & 2.581 & 2.52 & 2.492 & 2.525 \\
F7    &  4.191 &  3.745    & 3.691 & 3.672    & \nodata &2.90 & 2.687 & 2.63 & 2.603 & 2.639 \\
F8    &  4.344 &  3.878    & 3.814 & 3.789    & \nodata &3.01 & 2.785 & 2.72 & 2.693 & 2.732 \\
F9    &  4.498 &  4.009    & 3.935 & 3.901    & \nodata &3.11 & 2.873 & 2.81 & 2.783 & 2.824 \\
G0   &  4.825 &  4.292    & 4.196 & 4.145    & \nodata &3.34 & 3.078 & 3.01 & 2.983 & 3.026 \\
G1   &  4.881 &  4.339    & 4.239 & 4.186    & \nodata &3.38 & 3.113 & 3.04 & 3.013 & 3.057 \\
G2   &  5.200 &  4.612    & 4.492 & 4.423    & \nodata &3.57 & 3.277 & 3.2 & 3.172 & 3.222 \\
G3   &  5.276 &  4.678    & 4.554 & 4.481    & \nodata &3.64 & 3.341 & 3.27 & 3.242 & 3.292 \\
G4   &  5.365 &  4.754    & 4.625 & 4.547    & \nodata & 3.70 & 3.393 & 3.32 & 3.292 & 3.344 \\
G5   &  5.408 &  4.792    & 4.660 & 4.580    & \nodata &3.74 & 3.43 & 3.35 & 3.322 & 3.374 \\
G6   &  5.574 &  4.934    & 4.792 & 4.705    & \nodata &3.84 & 3.516 & 3.44 & 3.412 & 3.465 \\
G7   &  5.629 &  4.980    & 4.835 & 4.743    & \nodata &3.88 & 3.551 & 3.47 & 3.442 & 3.496 \\
G8   &  5.784 &  5.112    & 4.957 & 4.858    & \nodata &3.97 & 3.628 & 3.55 & 3.522 & 3.579 \\
G9   &  6.040 &  5.328    & 5.157 & 5.044    & \nodata &4.14 & 3.775 & 3.69 & 3.661 & 3.721 \\
K0   &  6.274 &  5.523    & 5.337 & 5.208    & \nodata &4.29 & 3.903 & 3.81 & 3.78 & 3.843 \\
K1   &  6.424 &  5.643    & 5.444 & 5.305    & \nodata &4.47 & 4.068 & 3.97 & 3.94 & 4.004 \\
K2   &  6.753 &  5.921    & 5.699 & 5.536    & \nodata &4.57 & 4.138 & 4.04 & 4.009 & 4.077 \\
K3   &  7.194 &  6.259    & 5.993 & 5.784    & \nodata &4.76 & 4.27 & 4.16 & 4.126 & 4.197 \\
K4   &  7.733 &  6.659    & 6.314 & 6.031    & \nodata &4.98 & 4.436 & 4.31 & 4.271 & 4.344 \\
K5   &  8.085 &  6.946    & 6.562 & 6.247    & \nodata &5.18 & 4.612 & 4.48 & 4.438 & 4.511 \\
K6   &  8.581 &  7.332    & 6.872 & 6.515    & \nodata &5.41 & 4.809 & 4.66 & 4.611 & \nodata \\
K7   &  9.088 &  7.737    & 7.198 & 6.824    & \nodata &5.65 & 5.028 & 4.86 & 4.8 & \nodata\\
K8   &  9.324 &  7.940    & 7.362 & 6.991    & \nodata &5.78 & 5.157 & 4.98 & 4.899 & \nodata \\
K9   &  9.561 &  8.143    & 7.526 & 7.157    & \nodata &5.92 & 5.295 & 5.11 & 5.009 & \nodata \\
M0   & 9.797 &  8.346    & 7.690 & 7.324    & \nodata &6.04 & 5.414 & 5.22 & 5.098 & \nodata \\
M1   & 10.619 & 9.099   & 8.261 & 7.966    & \nodata &6.51 & 5.897 & 5.67 & 5.533 & \nodata \\
M2   & 11.245 & 9.700   & 8.745 & 8.500    & \nodata &6.89 & 6.29 & 6.06 & 5.95 & \nodata \\
M3   & 12.113 & 10.522 & 9.380 & 9.199    & \nodata &7.40 & 6.821 & 6.54 & 6.418 & \nodata \\
M4   & 13.846 & 12.145 & 10.691 & 10.565 & \nodata &8.39 & 7.833 & 7.55 & 7.411 & \nodata \\
M5   & 15.481 & 13.597 & 11.804 & 11.693 & \nodata &9.25 & 8.67 & 8.36 & \nodata & \nodata \\
M6   & 17.880 & 15.802 & 13.357 & 13.099 & 10.28 & \nodata &9.675 & 9.32 & \nodata & \nodata \\
M7 & 18.11 & 16.75 & 14.18 & 12.97 & 12.31 & 10.77 & 10.14 & 9.75 & 9.52 & 9.3 \\
M8 & 19.19 & 17.73 & 14.98 & 13.59 & 12.8 & 11.14 & 10.46 & 10.02 & 9.77 & 9.54 \\
M9 & 19.95 & 18.15 & 15.63 & 14.19 & 13.25 & 11.48 & 10.77 & 10.29 & 9.96 & 9.7 \\
L0 & 20.41 & 18.42 & 16.05 & 14.58 & 13.63 & 11.81 & 11.05 & 10.57 & 10.27 & 9.99 \\
L1 & 20.86 & 18.75 & 16.42 & 14.94 & 13.97 & 12.04 & 11.24 & 10.72 & 10.37 & 10.12 \\
L2 & 21.24 & 19.02 & 16.73 & 15.31 & 14.33 & 12.32 & 11.41 & 10.83 & 10.41 & 10.14 \\
L3 & \nodata & 19.71 & 17.52 & 16.01 & 15.02 & 12.89 & 11.94 & 11.3 & 10.78 & 10.51 \\
L4 & \nodata & 20.46 & 18.25 & 16.56 & 15.56 & 13.41 & 12.35 & 11.77 & 11.07 & 10.75 \\
L5 & \nodata & 20.66 & 18.74 & 16.94 & 15.87 & 13.7 & 12.65 & 12.03 & 11.28 & 10.98 \\
L6 & \nodata & 21.2 & 19.26 & 17.34 & 16.26 & 14.17 & 13.18 & 12.54 & 11.84 & 11.48 \\
L7 & \nodata & \nodata & 20.11 & 18.2 & 17.14 & 14.95 & 13.79 & 13.08 & 12.39 & 12.0 \\
L8 & \nodata & 22.88 & 20.44 & 18.1 & 17.03 & 14.9 & 13.77 & 13.08 & 12.22 & 11.73 \\
L9 & \nodata & \nodata & 20.64 & 18.16 & 17.03 & 14.93 & 13.86 & 13.27 & 12.42 & 11.94 \\
T0 & \nodata & \nodata & 20.22 & 17.99 & 16.76 & 14.56 & 13.63 & 13.11 & 12.49 & 11.94 \\
T1 & \nodata & \nodata & 21.03 & 18.87 & 17.44 & 15.25 & 14.37 & 14.07 & 13.44 & 12.69 \\
T2 & \nodata & \nodata & 21.51 & 18.28 & 16.79 & 14.56 & 13.76 & 13.49 & 12.96 & 12.07 \\
T3 & \nodata & \nodata & \nodata & 18.0 & 16.46 & 14.19 & 13.6 & 13.37 & 12.71 & 11.64 \\
T4 & \nodata & \nodata & \nodata & 18.07 & 16.37 & 13.94 & 13.62 & 13.56 & 13.36 & 11.93 \\
T5 & \nodata & \nodata & 22.69 & 19.2 & 17.43 & 14.94 & 14.75 & 14.77 & 14.46 & 12.69 \\
T6 & \nodata & \nodata & \nodata & 19.82 & 18.07 & 15.53 & 15.48 & 15.37 & 15.08 & 13.02 \\
T7 & \nodata & \nodata & \nodata & 21.14 & 19.33 & 16.78 & 16.7 & 16.7 & 16.25 & 14.04 \\
T8 & \nodata & \nodata & \nodata & 21.52 & 19.75 & 17.18 & 17.09 & \nodata & 16.45 & 13.77 \\
T9 & \nodata & \nodata & \nodata & 21.82 & 20.37 & 17.75 & 17.51 & \nodata & 16.7 & 13.81 \\
\enddata
\tablecomments{
Absolute magnitudes of spectral types.  Types earlier than M6 are from \citet{pec12} and \citet{pec13} in 
SDSS, 2MASS, and {\it WISE} systems.  Types M7 and later are from \citet{bes18} in PS1, 2MASS, and {\it WISE}.
              }
\end{deluxetable} 
\end{longrotatetable}


\begin{thebibliography}{}


\bibitem[Abt(2008)]{abt08}
	Abt, H.\ 2008, \apjs, 176, 216

\bibitem[Alam et al.(2015)]{ala15} 
	Alam, S., Albareti, F.~D., Allende Prieto, C., et al.\ 2015, \apjs, 219, 12

\bibitem[Allard et al.(2012)]{all12} 
	Allard, F., Homeier, D., \& Freytag, B.\ 2012, Philosophical Transactions 
		of the Royal Society of London Series A, 370, 2765 

\bibitem[Argue \& Kenworthy(1969)]{arg69} 
	Argue, A.~N., \& Kenworthy, C.~M.\ 1969, \mnras, 146, 479 

\bibitem[Artyukhina \& Kholopov(1966)]{art66} 
	Artyukhina, N.~M., \& Kholopov, P.~N.\ 1966, \sovast, 10, 448

\bibitem[Bayo et al.(2011)]{bay11} 
	Bayo, A., Barrado, D., Stauffer, J., et al.\ 2011, \aap, 536, A63 
	
\bibitem[Bastian et al.(2010)]{bas10} 
	Bastian, N., Covey, K.~R., \& Meyer, M.~R.\ 2010, \araa, 48, 339 

\bibitem[Best et al.(2018)]{bes18} 
	Best, W.~M.~J., Magnier, E.~A., Liu, M.~C., et al.\ 2018, \apjs, 234, 1 
	
\bibitem[Bhattacharya et al.(2017)]{bha17} 
	Bhattacharya, S., Mishra, I., Vaidya, K., \& Chen, W.~P.\ 2017, \apj, 847, 138 

\bibitem[Binney \& Tremaine(1987)]{bin87}
	Binney, J., \& Tremaine, S.,\ 1987, Galactic Dynamics, Princeton University Press

\bibitem[Bok(1934)]{bok34} 
	Bok, B.~J.\ 1934, Harvard College Observatory Circular, 384, 1 

\bibitem[Boudreault et al.(2012)]{bou12} 
	Boudreault, S., Lodieu, N., Deacon, N.~R., \& Hambly, N.~C.\ 2012, \mnras, 426, 3419 

\bibitem[Brandner et al.(2008)]{bra08} 
	Brandner, W., Clark, J.~S., Stolte, A., et al.\ 2008, \aap, 478, 137 

\bibitem[Bressan et al.(2012)]{bre12}
	Bressan, A., Marigo, P. Girardi, L., et al.\ 2012, \mnras. 427, 127

\bibitem[Burgasser et al.(2004)]{bur04} 
	Burgasser, A.~J., McElwain, M.~W., Kirkpatrick, J.~D., et al.\ 2004, \aj, 127, 2856 

\bibitem[Burgasser \& McElwain(2006)]{bur06a} 
	Burgasser, A.~J., \& McElwain, M.~W.\ 2006, \aj, 131, 1007 

\bibitem[Burgasser et al.(2006)]{bur06b} 
	Burgasser, A.~J., Geballe, T.~R., Leggett, S.~K., Kirkpatrick, J.~D., \& Golimowski, D.~A.\ 2006, \apj, 637, 1067  

\bibitem[Burgasser et al.(2007)]{bur07a} 
	Burgasser, A.~J., Looper, D.~L., Kirkpatrick, J.~D., \& Liu, M.~C.\ 2007, \apj, 658, 557 

\bibitem[Burgasser(2007)]{bur07b}
	Burgasser, A.~J.\ 2007, \apj, 659, 655 

\bibitem[Burgasser et al.(2008)]{bur08}
	Burgasser, A.~J., Liu, M.~C., Ireland, M.~J., Cruz, K.~L., \& Dupuy, T.~J.\ 2008, \apj, 681, 579-593 

\bibitem[Carrera \& Pancino(2011)]{car11} 
	Carrera, R., \& Pancino, E.\ 2011, \aap, 535, A30 

\bibitem[Casewell et al.(2005)]{cas05} 
	Casewell, S.~L., Jameson, R.~F., \& Dobbie, P.~D.\ 2005, Astronomische Nachrichten, 326, 991 

\bibitem[Casewell et al.(2006)]{cas06} 
	Casewell, S.~L., Jameson, R.~F., \& Dobbie, P.~D.\ 2006, \mnras, 365, 447 

\bibitem[Casewell et al.(2014)]{cas14} 
	Casewell, S.~L., Littlefair, S.~P., Burleigh, M.~R., \& Roy, M.\ 2014, \mnras, 441, 2644 

\bibitem[Cayrel de Strobel(1990)]{cay90} 
	Cayrel de Strobel, G.\ 1990, \memsai, 61, 613 

\bibitem[Chambers et al.(2016)]{cha16} 
	Chambers, K.~C., Magnier, E.~A., Metcalfe, N., et al.\ 2016, arXiv:1612.05560 

\bibitem[Chappelle et al.(2005)]{cha05} 
	Chappelle, R.~J., Pinfield, D.~J., Steele, I.~A., Dobbie, P.~D., \& Magazz{\`u}, A.\ 2005, 
	 \mnras, 361, 1323

\bibitem[Chen \& Chen(2010)]{che10} 
	Chen, C.~W., \& Chen, W.~P.\ 2010, \apj, 721, 1790 

\bibitem[Chen et al.(2014)]{che14} 
	Chen, Y., Girardi, L., Bressan, A., et al.\ 2014, \mnras, 444, 2525 

\bibitem[Chen et al.(2015)]{che15} 
	Chen, Y., Bressan, A., Girardi, L., et al.\ 2015, \mnras, 452, 1068 

\bibitem[Chen et al.(2004)]{che04} 
	Chen, W.~P., Chen, C.~W., \& Shu, C.~G.\ 2004, \aj, 128, 2306 

\bibitem[Chiu et al.(2006)]{chi06} 
	Chiu, K., Fan, X., Leggett, S.~K., et al.\ 2006, \aj, 131, 2722

\bibitem[Collins \& Hambly(2012)]{col12}
	Collins, R., \& Hambly, N.\ 2012, Astronomical Data Analysis Software and Systems XXI, ASPC, 461, 525

\bibitem[Cox(2000)]{cox00} 
	Cox, A.~N.\ 2000, Allen's Astrophysical Quantities (IV, ed.; New York: AIP)
\bibitem[Cruz et al.(2004)]{cru04} 
	Cruz, K.~L., Burgasser, A.~J., Reid, I.~N., \& Liebert, J.\ 2004, \apjl, 604, L61

\bibitem[Cushing et al.(2004)]{cus04} 
	Cushing, M.~C., Vacca, W.~D., \& Rayner, J.~T.\ 2004, \pasp, 116, 362 

\bibitem[Cutri et al.(2013)]{cut13}
	Cutri, R.~M., Wright, E.~L., Conrow, T., et al.\ 2013, Explanatory 
	Supplement to the AllWISE Data Release Products

\bibitem[de Grijs \& Parmentier(2007)]{deg07} 
	de Grijs, R., \& Parmentier, G.\ 2007, \cjaa, 7, 155 

\bibitem[de Grijs(2009)]{deg09} 
	de Grijs, R.\ 2009, \apss, 324, 283 

\bibitem[Ducati(2002)]{duc02} 
	Ducati, J.~R.\ 2002, VizieR Online Data Catalog, 2237,  

\bibitem[Elias et al.(2006a)]{eli06a}
   	Elias, J. H., Joyce, R. R., Liang, M., et al.\ 2006a, Proc. SPIE, 6269, 62694C 

\bibitem[Elias et al.(2006b)]{eli06b}
	Elias, J. H., Rodgers, B., Joyce, R. R., et al.\ 2006b, Proc. SPIE, 6269, 626914

\bibitem[Ford et al.(2001)]{for01} 
	Ford, A., Jeffries, R.~D., James, D.~J., \& Barnes, J.~R.\ 2001, \aap, 369, 871 

\bibitem[Fossati et al.(2008)]{fos08} 
	Fossati, L., Bagnulo, S., Landstreet, J., et al.\ 2008, \aap, 483, 891 

\bibitem[Friel \& Boesgaard(1992)]{fri92} 
	Friel, E.~D., \& Boesgaard, A.~M.\ 1992, \apj, 387, 170 

\bibitem[Gaia Collaboration et al.(2016)]{bro16} 
	Gaia Collaboration, Brown, A.~G.~A., Vallenari, A., et al.\ 2016, \aap, 595, A2

\bibitem[Gaia Collaboration et al.(2016)]{pru16} 
	Gaia Collaboration, Prusti, T., de Bruijne, J.~H.~J., et al.\ 2016, \aap, 595, A1 

\bibitem[Gaia Collaboration et al.(2017)]{van17}
       Gaia Collaboration, van Leeuwen, F., Vallenari, A., et al.\ 2017, \aap, 601, A19 


\bibitem[Gaia Collaboration et al.(2018)]{bro18} 
		Gaia Collaboration, Brown, A.~G.~A., Vallenari, A., et al.\ 2018, arXiv:1804.09365 


\bibitem[G{\'a}sp{\'a}r et al.(2009)]{gas09} 
	G{\'a}sp{\'a}r, A., Rieke, G.~H., Su, K.~Y.~L., et al.\ 2009, \apj, 697, 1578 

\bibitem[Gennaro et al.(2011)]{gen11} 
	Gennaro, M., Brandner, W., Stolte, A., \& Henning, T.\ 2011, \mnras, 412, 2469 

\bibitem[Goldman et al.(2013)]{gol13} 
	Goldman, B., R{\"o}ser, S., Schilbach, E., et al.\ 2013, \aap, 559, A43 

\bibitem[Gratton(2000)]{gra00} 
	Gratton, R.\ 2000, Stellar Clusters and Associations: Convection, Rotation, and Dynamos, 198, 225
	Binney, J., \& Tremaine, S.,\ 1987, Galactic Dynamics, Princeton University Press

\bibitem[Griffin \& Griffin(1986)]{gri86} 
	Griffin, R., \& Griffin, R.\ 1986, Journal of Astrophysics and Astronomy, 7, 195 

\bibitem[Griffin \& Griffin(2011)]{gri11}
	Griffin, R.~E.~M., \& Griffin, R.~F.\ 2011, Astronomische Nachrichten, 332, 105 

\bibitem[Hambly et al.(1995)]{ham95} 
	Hambly, N.~C., Steele, I.~A., Hawkins, M.~R.~S., \& Jameson, R.~F.\ 1995, \aaps, 109,  

\bibitem[Herter et al.(2008)]{her08}
  	Herter, T. L., Henderson, C. P., Wilson, J. C., et al. 2008, Proc. SPIE, 7014, 70140X

\bibitem[Hewett et al.(2006)]{hew06} 
	Hewett, P.~C., Warren, S.~J., Leggett, S.~K., \& Hodgkin, S.~T.\ 2006, \mnras, 367, 454 

\bibitem[Hillenbrand(1997)]{hil97} 
	Hillenbrand, L.~A.\ 1997, \aj, 113, 1733 
        
\bibitem[Hillenbrand \& White(2004)]{hil04} 
	Hillenbrand, L.~A., \& White, R.~J.\ 2004, \apj, 604, 741

\bibitem[H{\o}g et al.(2000)]{hog00} 
	H{\o}g, E., Fabricius, C., Makarov, V.~V., et al.\ 2000, \aap, 355, L27 



\bibitem[Jeffries(1999)]{jef99} Jeffries, R.~D.\ 1999, \mnras, 304, 821 

\bibitem[Jordi et al.(2005)]{jor05} 
	Jordi, K., Grebel, E.~K., \& Ammon, K.\ 2005, Astronomische Nachrichten, 326, 657 

\bibitem[Kippenhahn et al.(1970)]{kip70} 
	Kippenhahn, R., Meyer-Hofmeister, E., \& Thomas, H.~C.\ 1970, \aap, 5, 155 

\bibitem[Kirkpatrick et al.(2010)]{kir10} 
	Kirkpatrick, J.~D., Looper, D.~L., Burgasser, A.~J., et al.\ 2010, \apjs, 190, 100

\bibitem[Kirkpatrick et al.(2011)]{kir11} 
	Kirkpatrick, J.~D., Cushing, M.~C., Gelino, C.~R., et al.\ 2011, \apjs, 197, 19 

\bibitem[Kraus \& Hillenbrand(2007)]{kra07} 
	Kraus, A.~L., \& Hillenbrand, L.~A.\ 2007, \aj, 134, 2340 

\bibitem[Lada et al.(1984)]{lad84} 
	Lada, C.~J., Margulis, M., \& Dearborn, D.\ 1984, \apj, 285, 141 

\bibitem[Lada \& Lada(2003)]{lad03} 
	Lada, C.~J., \& Lada, E.~A.\ 2003, \araa, 41, 57 

\bibitem[Lan{\c c}on et al.(2007)]{lan07} 
	Lan{\c c}on, A., Hauschildt, P.~H., Ladjal, D., \& Mouhcine, M.\ 2007, \aap, 468, 205 

\bibitem[Lawrence et al.(2007)]{law07}
	Lawrence, A., et al.\ 2007, \mnras, 379, 1599

\bibitem[Lawrence et al.(2012)]{law12}
	Lawrence, A., Warren, S.~J., Almaini, O., et al.\ 2012, VizieR Online Data Catalog, 2314 



\bibitem[Lindegren et al.(2018)]{lin18} 
	Lindegren, L., Hernandez, J., Bombrun, A., et al.\ 2018, arXiv:1804.09366 

\bibitem[Lodieu et al.(2012)]{lod12} 
	Lodieu, N., Deacon, N.~R., \& Hambly, N.~C.\ 2012, \mnras, 422, 1495

\bibitem[Loidl et al.(2001)]{loi01} 
	Loidl, R., Lan{\c c}on, A., \& J{\o}rgensen, U.~G.\ 2001, \aap, 371, 1065

\bibitem[Luhman(2012)]{luh12} 
	Luhman, K.~L.\ 2012, \araa, 50, 65 

\bibitem[Luhman \& Rieke(1999)]{luh99} 
	Luhman, K.~L., \& Rieke, G.~H.\ 1999, \apj, 525, 440 

\bibitem[Mainzer et al.(2011)]{mai11} 
	Mainzer, A., Bauer, J., Grav, T., et al.\ 2011, \apj, 731, 53 

\bibitem[Mart{\'{\i}}n et al.(2018)]{mar18} 
	Mart{\'{\i}}n, E.~L., Lodieu, N., Pavlenko, Y., \& B{\'e}jar, V.~J.~S.\ 2018, \apj, 856, 40 

\bibitem[Massarotti et al.(2008)]{mas08}
	Massarotti, A., Latham, D., Stefanik, R.~P., et al.\ 2008, \aj, 135, 209

\bibitem[Mathieu(1984)]{mat84} 
	Mathieu, R.~D.\ 1984, \apj, 284, 643 

\bibitem[Melnikov \& Eisl{\"o}ffel(2012)]{mel12} 
	Melnikov, S., \& Eisl{\"o}ffel, J.\ 2012, \aap, 544, A111 
	
\bibitem[Melotte(1915)]{mel15} 
	Melotte, P.~J.\ 1915, \memras, 60, 175 

\bibitem[Mermilliod et al.(2008)]{mer08} 
	Mermilliod, J.-C., Grenon, M., \& Mayor, M.\ 2008, \aap, 491, 951 

\bibitem[Mermilliod, Mayor, \& Udry(2009)]{mer09}
	Mermilliod, J.~C., Mayor, M., \& Udry, S.\ 2009, \aap, 498, 949

\bibitem[Netopil et al.(2016)]{net16} 
	Netopil, M., Paunzen, E., Heiter, U., \& Soubiran, C.\ 2016, \aap, 585, A150 

\bibitem[Nicolet(1981)]{nic81} 
	Nicolet, B.\ 1981, \aap, 104, 185

\bibitem[Odenkirchen et al.(1998)]{ode98} 
	Odenkirchen, M., Soubiran, C., \& Colin, J.\ 1998, \na, 3, 583 

\bibitem[Pace et al.(2008)]{pac08} 
	Pace, G., Pasquini, L., \& Fran{\c c}ois, P.\ 2008, \aap, 489, 403 

\bibitem[Pecaut et al.(2012)]{pec12} 
	Pecaut, M.~J., Mamajek, E.~E., \& Bubar, E.~J.\ 2012, \apj, 746, 154 
	
\bibitem[Pecaut \& Mamajek(2013)]{pec13} 
	Pecaut, M.~J., \& Mamajek, E.~E.\ 2013, \apjs, 208, 9 

\bibitem[Perryman et al.(1998)]{per98} 
	Perryman, M.~A.~C., Brown, A.~G.~A., Lebreton, Y., et al.\ 1998, \aap, 331, 81 

\bibitem[Rayner et al.(2003)]{ray03} 
	Rayner, J.~T., Toomey, D.~W., Onaka, P.~M., et al.\ 2003, \pasp, 115, 362 

\bibitem[Rayner et al.(2009)]{ray09} 
	Rayner, J.~T., Cushing, M.~C., \& Vacca, W.~D.\ 2009, \apjs, 185, 289

\bibitem[Reid et al.(2002)]{rei02}
	Reid, I.~N., Gizis, J.~E., \& Hawley, S.~L.\ 2002, \aj, 124, 2721 

\bibitem[Reid et al.(2006)]{rei06} 
	Reid, I.~N., Lewitus, E., Burgasser, A.~J., \& Cruz, K.~L.\ 2006, \apj, 639, 1114

\bibitem[Rochau et al.(2010)]{roc10} 
	Rochau, B., Brandner, W., Stolte, A., et al.\ 2010, \apjl, 716, L90 

\bibitem[Roeser et al.(2010)]{roe10} 
	Roeser, S., Demleitner, M., \& Schilbach, E.\ 2010, \aj, 139, 2440

\bibitem[Salpeter(1955)]{sal55} 
	Salpeter, E.~E.\ 1955, \apj, 121, 161 

\bibitem[Scholz et al.(2011)]{sch11} 
	Scholz, R.-D., Bihain, G., Schnurr, O., et al.\ 2011, \aap, 532, L5

\bibitem[Shu(1982)]{shu82}
	Shu, F. H.,\ 1982, The Physical Universe, University Science Books

\bibitem[Skrutskie et al.(2006)]{skr06} 
	Skrutskie, M.~F., Cutri, R.~M., Stiening, R., et al.\ 2006, \aj, 131, 1163 

\bibitem[Skrzypek et al.(2015)]{skr15} 
	Skrzypek, N., Warren, S.~J., Faherty, J.~K., et al.\ 2015, \aap, 574, A78 

\bibitem[Skrzypek et al.(2016)]{skr16} 
	Skrzypek, N., Warren, S.~J., \& Faherty, J.~K.\ 2016, \aap, 589, A49 

\bibitem[Smith et al.(2014)]{smi14}
	Smith, L., Lucas, P.~W., Burningham, B., et al.\ 2014, \mnras, 437, 3603

\bibitem[Tang et al.(2014)]{tan14}
	 Tang, J., Bressan, A., Rosenfield, P., et al.\ 2014, \mnras, 445, 4287

\bibitem[Terrien et al.(2014)]{ter14} 
	Terrien, R.~C., Mahadevan, S., Deshpande, R., et al.\ 2014, \apj, 782, 61

\bibitem[Trumpler(1938)]{tru38} 
	Trumpler, R.~J.\ 1938, Lick Observatory Bulletin, 18, 167 

\bibitem[Tsvetkov(1989)]{tsv89} 
	Tsvetkov, T.~G.\ 1989, \apss, 151, 47 

\bibitem[Upgren(1962)]{upg62}
	Upgren, A. R.\ 1962, \aj, 67, 37

\bibitem[Vacca et al.(2003)]{vac03} 
	Vacca, W.~D., Cushing, M.~C., \& Rayner, J.~T.\ 2003, \pasp, 115, 389 

\bibitem[van Leeuwen(1999)]{van99} 
	van Leeuwen, F.\ 1999, \aap, 341, L71 

\bibitem[van Leeuwen(2009)]{van09} 
	van Leeuwen, F.\ 2009, \aap, 497, 209 

\bibitem[Wang et al.(2014)]{wan14} 
	Wang, P.~F., Chen, W.~P., Lin, C.~C., et al.\ 2014, \apj, 784, 57 

\bibitem[West et al.(2011)]{wes11} 
	West, A.~A., Morgan, D.~P., Bochanski, J.~J., et al.\ 2011, \aj, 141, 97 

\bibitem[Wright et al.(2010)]{wri10} 
	Wright, E.~L., Eisenhardt, P.~R.~M., Mainzer, A.~K., et al.\ 2010, \aj, 140, 1868

\bibitem[Zacharias \& Gaume(2011)]{zac11} 
	Zacharias, N., \& Gaume, R.\ 2011, Journ{\'e}es Syst{\`e}mes de R{\'e}f{\'e}rence Spatio-temporels 2010, 95
	
\bibitem[Zacharias et al.(2015)]{zac15}
	Zacharias, N., et al.\ 2015, \aj, 150, 101

\bibitem[Zuckerman \& Song(2004)]{zuc04} 
  	Zuckerman, B., \& Song, I.,\ 2004, \araa, 42, 685 
%
\end{thebibliography}
\end{document}